\documentclass[twocolumn]{emulateapj}
\pdfoutput=1

\usepackage{amssymb}
\usepackage{amsmath}
\usepackage{graphicx}
\usepackage{natbib}
\usepackage{bigints}
\usepackage{relsize}
\usepackage{multirow}
\usepackage{color}

\newcommand{\myemail}{radigan@stsci.edu}

\shorttitle{Brightness Variations Signal L/T Transition}
\shortauthors{Radigan et al.}

\begin{document}


\title{Strong Brightness Variations Signal Cloudy-to-Clear Transition of Brown Dwarfs}

\author{Jacqueline Radigan\altaffilmark{1,2},  David Lafreni{\`e}re\altaffilmark{3}, Ray Jayawardhana\altaffilmark{2,4}, Etienne Artigau\altaffilmark{3}}

\altaffiltext{1}{Space Telescope Science Institute, 3700 San Martin Drive Baltimore MD 21218; \myemail}
\altaffiltext{2}{Department of Astronomy \& Astrophysics, University of Toronto, Toronto, ON M5S~3H4, Canada } 
\altaffiltext{3}{D{\'e}partement de Physique, Universit{\'e} de Montr{\'e}al, C.P. 6128 Succ. Centre-Ville, Montr{\'e}al,  QC H3C~3J7, Canada}
\altaffiltext{4}{Department of Physics \& Astronomy, York University, Toronto, ON L3T~3R1, Canada}

\begin{abstract}
We report the results of a  $J$ band search for cloud-related variability in the atmospheres of 62 L4-T9 dwarfs using the Du Pont 2.5-m telescope at Las Campanas Observatory and the Canada France Hawaii Telescope on Mauna Kea.  We find 9 of 57 objects included in our final analysis to be significantly variable with $>$99\% confidence, 5 of which are new discoveries.  
In our study,  strong sinusoidal signals (peak-to-peak amplitudes $>$2\%) are confined to the L/T transition
(4/16 objects with L9-T3.5 spectral types and 0/41 objects for all other spectral types).  The probability that the observed occurrence rates for strong variability inside and outside the L/T transition originate from the same underlying true occurrence rate is excluded at $>$99.7\% confidence.  Based on a careful assessment of our sensitivity to astrophysical signals, we infer that 39$^{+16}_{-14}$\% of L9-T3.5 dwarfs are strong variables on rotational timescales.  If we consider only L9-T3.5 dwarfs with 0.8$<J-K_{\rm s}<$1.5, and assume an isotropic distribution of spin axes for our targets, we find that $80^{+18}_{-19}$\% would be strong variables if viewed edge-on; azimuthal symmetry and/or binarity may account for non-variable objects in this group.  These observations suggest that the settling of condensate clouds below the photosphere in brown dwarf atmospheres does not occur in a spatially uniform manner.  Rather, the formation and sedimentation of dust grains at the L/T transition is coupled to atmospheric dynamics, resulting in highly contrasting regions of thick and thin clouds and/or clearings. Outside the L/T transition we identify 5 weak variables (peak-to-peak amplitudes of 0.6\%-1.6\%).  Excluding L9-T3.5 spectral types, we infer that $60^{+22}_{-18}$\% of targets vary with amplitudes of 0.5\%$-$1.6\%, suggesting that surface heterogeneities are common among L and T dwarfs.  Our survey establishes a significant link between strong variability and L/T transition spectral types, providing evidence in support of the hypothesis that cloud holes contribute to the abrupt decline in condensate opacity and 1\,$\mu$m brightening observed in this regime.  More generally, fractional cloud coverage is an important model parameter for brown dwarfs and giant planets, especially those with L/T transition spectral types and colors. 
\end{abstract}

 \section{Introduction}
 \label{sect:intro}
Brown dwarfs (BDs) are objects thought to form similarly to stars, yet lack the required mass ($M \lesssim$ 0.07\,$M_{\odot}$) to burn hydrogen \citep{chabrier00}.  Without a sustained energy source they spend their lives cooling.  While young brown dwarfs may resemble the lowest mass stars, after the first $\sim$0.1-1 Gyr  their atmospheres have cooled to sub-stellar temperatures ($\lesssim$2200\,K).  The coolest brown dwarfs yet detected are reported to have temperatures as low as $\sim$300-400\,K \citep{liu11,cushing11,luhman12,dupuy13}, and represent the coolest atmospheres available to direct and detailed study outside of our solar system.  As such, a detailed understanding of brown dwarf atmospheres is an important stepping stone toward the understanding of giant planet atmospheres---including those recently discovered \citep[e.g. HR8799 system;][]{marois08} and cooler objects to be found in the near future through campaigns such as GPI \citep{gpi} and SPHERE \citep{sphere}---for which we will collect comparatively fewer data of significantly lower quality.

Direct spectra for hundreds of free-floating BDs in the solar neighborhood provide an unrivaled sample from which the majority of current knowledge about cool, cloudy atmospheres has been derived.   The standard picture is as follows.  As temperatures fall below $\sim$2200\,K refractory species---including iron, silicates, and metal oxide compounds---condense to form``dust'' clouds in substellar atmospheres \citep{burrows99,lodders99,burrows06}.  The formation and thickening of dust clouds leads to progressively redder spectral energy distributions, and characterizes the L spectral sequence.  However, at  temperatures of $\sim$1200\,K dust opacity is observed to diminish abruptly, signaling the transition to cloud-free and methane-rich T spectral types.  This transition from cloudy to cloud-free atmosphere is known as the ``L/T transition'' and is characterized by a dramatic spectral evolution (a blueward shift of $\sim$2 magnitudes in $J-K$, encompassing $\sim$L8-T5 spectral types), at near constant effective temperature \citep{golimowski04,stephens09}.

The disappearance of dust as a major opacity source is thought to occur as dust grains gravitationally settle below the photosphere.  For example, although we see prominent ammonia clouds in Jupiter's photosphere, thick iron and silicate clouds are thought to reside in its deep atmosphere, hidden from view \citep{lodders06}.  However, the detailed physics governing the dissipation and settling of condensates remains poorly understood, with models generally predicting a much more gradual disappearance of clouds over a wider range of effective temperatures \citep{tsuji03,marley02,allard03}.  The discrepancy between observations and models is perhaps best highlighted by the $\sim$1\,$\mu$m fluxes of L/T transition brown dwarfs, which counterintuitively brighten by a factor of approximately two-fold from L8 to T5 spectral types \citep{dahn02,tinney03,vrba04,dupuy12,faherty12}, whereas models predict monotonically decreasing fluxes. 
This discordant observation has led to the suggestion that, rather than uniformly settling below the photosphere, clouds are dynamically disrupted at L/T transition temperatures, opening up windows to the deep photosphere, which contribute to the abrupt decline in cloud opacity and resurgence of 1$\mu$m flux \citep{ackerman01,burgasser02_lt}.  
 The cloud disruption hypothesis makes a testable prediction:  patchy cloud coverage should produce rotationally modulated variability as cloud features rotate in and out of view \citep[typical rotation periods are $\sim$2-10\,hr;][]{reiners08}.

 Searches for cloud related variability have been ongoing for over a decade for a range of spectral types at both red optical \citep{tinney99,bailer-jones99,bailer-jones01,gelino02,koen03,koen05c,littlefair06,koen13} and infrared wavelengths \citep{artigau03,bailer-jones03,enoch03,koen04a,koen05a,morales-calderon06,lane07,clarke08,goldman08}, but have yielded mostly ambiguous results.

Observations in the red optical have targeted mainly late-M to early-L dwarfs, due to a significant drop-off in optical flux for later type dwarfs.  In the $I$-band, a rather high fraction of early L-dwarfs show some statistical evidence for variability \citep[as high as 30-80\%;][]{bailer-jones01,gelino02,koen03,koen04b,koen05b,koen05c}, although the fraction for which periodic variability has been claimed is at the lower end of this range ( $\sim$30\%).  Typical peak-to-peak amplitudes of periodic variables are of the order of a few percent, and are often comparable to the photometric noise level.   Complicating the interpretation of these results is that many of the claimed periodicities are inconsistent with rotation periods based on $v\sin{i}$ measurements \citep{kirkpatrick99,mohanty03,bailer-jones04,zapatero06,reiners08}.  Despite the existing ambiguity, there are cases where optical periods can be matched to periodic radio pulsation \citep{lane07} or periodic variations in $H\alpha$ \citep{clarke03}, and likely correspond to rotation periods.  Recent work by \citet{harding13} has shown that radio emitting late-M and L dwarfs are often periodic variables in the red optical, suggesting magnetism as an underlying cause for this subset of objects.  Recent work by \citet{koen13} reporting short timescale $I$-band variability for 125 ultra cool dwarfs found variability to be more common for early spectral types: of 24 objects found to be significantly variable on timescales of 2-3\,hr, 18 had spectral types earlier than L2.  Interestingly, this study included 5 objects with spectral types $>$L8 (and as late as T5.5) and 4 of them were found to show nightly changes in mean flux level, hinting at longer timescale variability in these objects.  Thus, based on red-optical studies, variability in late-M, and early-L dwarfs is relatively common and may be related to magnetic spot activity, dust meteorology, or a combination of both.  The work of \citet{koen13} hints at a re-emergence of $I$-band variability at later spectral types, possibly coincident with the L/T transition.

While optical studies have been mostly confined to early spectral types, the L/T transition is one of the most interesting regimes to test for cloud related variability and weather due to the ability of cloud holes to explain observed properties of the transition.  Due to a strong drop off in optical flux with increasing spectral type, a move to NIR wavelengths is required (wherein late-L and T-dwarfs are brightest).  Since the atmospheres of late-L and T dwarfs are increasingly neutral, they are less likely to support cool magnetic spots \citep{gelino02,mohanty02}, making the interpretation of detected variability in this regime less ambiguous.  However, in contrast to optical surveys, data obtained in the NIR is typically subject to larger amounts of correlated noise due to the bright IR sky, variable precipitable water vapor in Earth's atmosphere, and detector systematics.  This makes the interpretation of NIR time series a challenging task \citep[e.g.][]{bailer-jones03,artigau06}, and can be a source of false-positives if not accounted for.  There have been several surveys for variability of L and T dwarfs at NIR wavelengths.  In a study of 18 L and T dwarfs, \citet{koen04b} found no significant evidence of variability in the $J$ band above the $\sim$20 mmag level nor in the $H$ or $K_s$ bands above $\sim$40 mmag, but find marginal evidence of periodic variability at lower peak-to-peak amplitudes with periods of $0.8-1.5$\,hr, for a few objects in their sample.  Similarly, the $J$ band survey of \citet{clarke08} found variability to be confined to amplitudes $<15$\, mmag, reporting periodic variations for 2 of 8 late L and T dwarfs surveyed with amplitudes of 15 and 8 mmag and periods of 1.4~hr and 2~hr respectively.  The work of \citet{girardin13} echoes these conclusions, finding all but one target to be non-variable above $5-15$\,mmag, and find evidence for periodic variability of a T0.5 binary at the $\sim$25-60\,mmag  level, with a  $\sim$3\,hr periodicity.  In contrast to these studies, \citet{enoch03} found 3 of 9 L2-T5 dwarfs monitored in the $K_s$ band  to be variable at the 10\%-20\% level (2 of which had low-significance periodicities of 1.5\,hr and 3.0\,hr).  Similarly,  \citet{khandrika13} found 4 of 15 L and T dwarfs to be variable in $J$ and/or $K_s$ with peak-to-peak amplitudes of 10\%-60\%, and find a significant periodicity for the T1.5 dwarf 2M2139$+$02, which was previously reported to be variable by \citet{radigan12}.  Combined, these latter two studies find that 7 of 24 (or $\sim$30\%)  of L and T dwarfs are high-amplitude variables in the NIR.  It is notable that the detections in these latter studies are typically only 2-3 times the level of the photometric noise, and given the lack of detections in higher-precision surveys, this may suggest a large number of false positives due to correlated noise.  Alternatively, it is possible that differences in target selection, filter choice (i.e. the use of a $K_{\rm s}$ filter instead of or in addition to $J$), and observing strategy (i.e. observations of the same targets at multiple epochs) differentiate the high-yield \citet{enoch03} and \citet{khandrika13} studies from others.  The HST/WFC3 survey of \citet{buenzli14} report $\sim$40\,min spectral time series for 22 L5-T6 dwarfs and report significant variability (p$>$95\%) in at least one wavelength region from 1.1-1.7\,$\mu$m for six brown dwarf spanning the range of spectral types observed.  Periods and amplitudes of the variability are not well constrained due to the short observation window of this study.  Taken together, previous studies in the NIR do not find variability to be correlated with spectral type or color, and do not find evidence to support the hypothesis that variability may be more common at the L/T transition.

The most compelling detections of brown dwarf variability to date in the NIR---those whose amplitudes greatly surpass the photometric noise, and/or have been repeated at multiple epochs---have been mostly reported as single object detections.  The first such result was reported by by \citet{artigau09}, who found the T2.5 dwarf SIMP~J013656.57$+$093347.3 (SIMP0136+09) to be variable with a peak-to-peak amplitude of $\sim$50~mmag in $J$ and a period of 2.4\,hr (a 10-$\sigma$ detection).   Since this benchmark finding, there have been three additional reports of large-amplitude variables in the NIR: the T1.5 dwarf 2MASS~J21392676$+$0220226 \citep{radigan12} which varies with an amplitude as high as 26\% in $J$ on a 7.72\,hr timescale\footnote{Although reported as a single object, this target was first detected as part of the variability survey presented in this work},  the T0.5 binary SDSS~J105213.51$+$442255.7, which was found to be variable by \citet{girardin13} with an amplitude as high 6\% and a period of 3\,hr, and the T2 secondary component of the Luhman AB system \citep{gillon13,biller13} which has been observed to have an amplitude as high as 13\% in the $H$ band and a $\sim$5\,hr period.  These reports of high amplitude variables, all of which have early T spectral types, hint that large-amplitude variability may be more common in the L/T transition despite the lack of evidence in previous survey work.  Low sample sizes of early-T targets in previous studies may explain this discrepancy.

Here we present results of the largest, most sensitive search for NIR variability in brown dwarf atmospheres to date, with the specific goal of testing whether L/T transition dwarfs are variable at $\sim$1\,$\mu$m wavelengths. 
Our survey, conducted over 60 nights using the 2.5-m Du Pont telescope at the Las Campanas Observatory, and the Canada France Hawaii Telescope on Mauna Kea is described in section \ref{sect:obs}.  The analysis of light curves, detection limits, and our search sensitivity are described in section \ref{sect:lightcurves}.  In section \ref{sect:results} we present our results, and demonstrate a statistically significant increase in variability for early T-dwarfs.   Finally, in section \ref{sect:concl} we summarize the major conclusions of our study, and discuss their implications for our understanding of cloudy substellar atmospheres.


\section{A J-band Variability Survey of L and T dwarfs }
\label{sect:obs}
We have completed the largest and most sensitive survey for variability in L and T dwarf atmospheres to date, with the specific goal of testing the cloud disruption hypothesis of the L/T transition.
Our observations were carried out in the $J$-band, wherein contrast between cloud features and the underlying gaseous photosphere is thought to be largest \citep{ackerman01,marley02}.  Observations were were carried out over 60 nights (less time lost due to technical difficulties and poor weather) using the Wide field InfraRed Camera (WIRC) on the 2.5-m Du Pont telescope at Las Campanas,  divided into 5 12-night observing runs spanning July 2009 to May 2010.  These observations were supplemented with additional observations of northern targets using the Wide-field InfraRed Camera (WIRCam) on the 3.6-m Canada-France-Hawaii Telescope, observed in queue mode during the 2009A semester.  A log of all observations is provided in table \ref{tab:ch5_obslog}.

\subsection{Target Sample and Observing Strategy}
The majority of targets surveyed were selected from the DwarfArchives database of spectroscopically confirmed L and T dwarfs\footnote{ http://www.DwarfArchives.org, maintained by A. Burgasser, D. Kirkpatrick and C. Gelino}.  The T2 dwarf SIMP~J16291840$+$0335371 (SIMP1629$+$03), an unpublished discovery from the SIMP proper motion survey \citep{artigau09_simp} and the (at the time) recently discovered wide binary system SDSS J141624.08+134826.7AB \citep[SD1416+13;][]{schmidt10,bowler10,burningham10} are the only exceptions.  We note that  SIMP1629$+$03 was independently discovered and reported by the PANSTARRS collaboration \citep{deacon11}, where it is known as PSO~J247.3273$+$03.5932.   Targets were selected to span mid-L to late-T spectral types,  with special care taken to populate the L/T transition region of the color magnitude diagram.   Known binaries were avoided when possible, although some were observed when there were no other suitable targets.  From Las Campanas 57 unique targets were observed with $J<17$ and $\delta < 15$\,deg (corresponding to $\sim$3.5\,hr of visibility above an airmass of 1.6).  On a given night targets were selected based on visibility, and a weighing of observing conditions and target brightness.  Five additional targets were observed in the CFHT queue, and consisted of reasonably bright L9-T5 targets with right ascensions ranging from 21h to 8hr, and $\delta > 10$\,deg.

Due to a long history of ambiguous results in BD variability monitoring (see section \ref{sect:intro}), the present work was designed to (i) survey an unprecedented number of objects allowing for robust statistical analysis, (ii) achieve improved photometric precisions of $\sim$1\%, and (iii) better control for potentially confounding factors such as second order extinction and rapidly evolving light curve morphologies.  In order to observe a large number of objects, each target was typically observed only once, and monitored continuously over a $\sim$2-5\,hr interval with a $\sim$30-70\,s cadence.  Since ultra cool dwarfs are found to be rapid rotators with periods ranging from $\sim$2-10\,hr \citep[e.g.,][]{reiners08}, the chosen time baseline aims to cover a large fraction of a rotational period in order to detect rotationally modulated variability in a single epoch.  Continuous, high cadence monitoring was chosen instead of a sparser time sampling over weeks or months because (i) it leads to more precise differential photometry, and (ii) the evolution of surface features (e.g., clouds) on a BD's surface may prevent observations taken over multiple epochs from being phased together in a coherent way.   Finally, new $J$-band filters closely matching the Mauna Kea Observatory (MKO) system \citep{tokunaga05}, which cut off  precipitable water vapor bands redward of 1.35 $\mu$m \citep[e.g,][]{artigau06}, were purchased and installed on WIRC in order to minimize second order extinction effects (WIRCam on CFHT also uses an MKO $J$ filter).  For the same reason, targets were monitored almost exclusively at air masses $<$1.6.

The final sample consists of 62 unique target light curves.  A log of all observations is provided in table \ref{tab:ch5_obslog}.

\subsection{Las Campanas Observations}
Observations at Las Campanas utilized the NIR camera WIRC on the Du Pont 2.5~m telescope.  The camera consists of 4 HAWAII-I arrays, each with a 3.2\arcmin~field of view and a pixel scale of 0.2$\arcsec$.  The camera is intended as a wide-field survey camera, with 3\arcmin~gaps between detectors.  We did not use it as such, choosing to position our targets consistently on the south-west array, which we determined (from dark and flat field data) to be the least noisy of the four chips.   All exposures were read out using correlated double sampling.  Observations were obtained either in staring mode or by using a randomized dither pattern.  Both staring, and the use of a localized random dither pattern (rather than a wide N-point pattern) minimize the movement of stars on the detector, which serves to minimize position-dependent systematics such as residual flat-fielding errors.  Dome-flats (lamp on and off) and dark frames corresponding to each exposure time were typically taken either on the afternoon preceding, or the morning following each observation.   On several occasions we obtained twilight flat field frames, but did not find them to be an improvement over the dome flat fields.

 \subsubsection{Staring observations}
For staring observations the target centroid is kept fixed on the same pixel throughout the sequence.  This was accomplished using an $IDL$ routine to stream the incoming science images onto a standard laptop and compute real-time guiding corrections.  An alert was sounded, and manual closed-loop corrections to the guide-camera reference position were made each time the target strayed by more than 0.5 pixels from its initial position (approximately once every 5-15 minutes).  Without telescope offsets the efficiency of staring observations is significantly increased.  Individual exposures of 20-60\,s were used depending on target brightness, resulting in a cadence of 27-67\,s.  Nine-point dither sequences for the purpose of rough sky subtraction and centroiding were made at the beginning, end (and sometimes middle for long observations) of each contiguous staring sequence.  Staring observations were preferred in more crowded fields, and in poor seeing conditions.

 \subsubsection{Dithered observations}
 For targets significantly fainter than the sky, and in very good seeing conditions dithered observations were preferred.  A random dither pattern was employed wherein the telescope was offset by at least 3\arcsec~(15 pixels) after each exposure.  All pointings were typically contained in a 15\arcsec~(75 pixel) diameter circle, or box with the same side length, although slow drifts in pointing over long sequences sometimes caused the center of the dither pattern to drift.

\subsubsection{Reduction of WIRC Data}
All raw images were dark subtracted and corrected for non linearity using a linearity sequence of dome flats obtained on 27 Jul 2009.  Dome flat fields (constructed from median combined and differenced lamp-on and lamp-off images) were used to correct for inter-pixel variations in quantum efficiency.  
For dithered sequences, a running sky frame (constructed from dithered science images closest in time and sufficiently offset from the frame in question) was subtracted from each science image.  Faint stars were identified in an initial first-pass reduction, and subsequently masked for the second pass so as not to bias the sky frames.  Hot or dead pixels, as well as pixels having more than 35,000 counts ($>$3\% non-linear) were flagged and set to error values.  Isolated bad pixels were interpolated over using a gaussian fitting function in the vicinity of the bad pixel with the IDL Astronomy Library routine {\tt MASKINTERP}.  Except for sky subtraction, reduction of the staring sequences is almost identical to the procedure described above, however in these cases sky frames obtained before and after the staring sequence were used for rough sky subtraction.

\subsection{CFHT WIRCam Observations and Data Reduction}
For the WIRCam queue observations, photometric conditions, low airmass ($<$1.5-1.7), and seeing better than $\sim$1\arcsec~(for in-focus sequences) were provided as observing constraints.  A staring strategy was employed with a small defocus for bright targets (SDSS J075840.33+324723.4 and 2MASSI J2254188+312349).  Exposure times ranged from 15 to 40 seconds depending on target brightness.   In-focus sky sequences using a 9-point dither pattern and identical exposure times were taken before and after staring.  The raw data were automatically processed by the `I`iwi processing pipeline.\footnote{http://www.cfht.hawaii.edu/Instruments/Imaging/WIRCam/ IiwiVersion1Doc.html}  Manual sky subtraction was performed using the processed but  pre-sky-subtracted data products.  Sky images were median combined to create sky frames, then subtracted from the science sequences.   With a  20\arcmin$\times$20\arcmin~ field of view, and pixel scale of 0.3\,\arcsec\,pix$^{-1}$, WIRCam typically provides many tens of similar-brightness reference stars for differential photometry.

\section{Lightcurve Analysis}
\label{sect:lightcurves}

\subsection{Aperture Photometry}
\label{sect:phot}

For each monitoring sequence, aperture photometry was performed on the target and a set of reference stars.  For WIRC targets we used a circular aperture of radius 1.5 times the median full width at half maximum (FWHM) of all stars in each image.   For the in-focus WIRCam observations the seeing was usually quite good (median FWHM of 0.57 -0.74\arcsec ~or approximately 2 pixels), and we found that a larger aperture of 2.5 times the median FWHM of all stars in each image produced more stable photometry.  For the defocussed WIRCam observations source positions were determined from an iterative measurement of the center of light, and photometry was obtained within an aperture with radius 1.5 times  the width of the second moment of the distribution of light about the central position.  For all sequences residual sky levels in the vicinity of each star were measured inside an annulus centered on each source.   Targets and reference stars (i) affected by bad (i.e. interpolated) pixels in the photometry aperture, (ii) having failed flux extraction in over 25\% of images (i.e. where interpolation of bad pixels fails), or (iii) having an extended PSF compared to other stars in the frame (e.g., a galaxy or double star) were flagged for quality.\footnote{For the WIRCam sequences where a defocus was applied, the interpolation of bad pixels was unavoidable for most sources.  However, the interpolation of isolated bad pixels is more accurate for the wider, defocussed PSFs than for the in-focus PSFs.  Therefore, the two defocussed  WIRCam sequences in our sample are not flagged for quality due to bad/interpolated pixels in the photometry aperture.}

The raw light curves display fluctuations in brightness due to changing atmospheric transparency, seeing, airmass, and instrumental effects throughout the night.  To first order these changes are common to all stars, and can be removed.  This is done by dividing all raw light curves by a calibration curve formed from median-combining the relative-flux light curves of reference stars in the same field of view as each target.  First, light curves of all stars were converted from absolute to relative fluxes via division by their median brightness.  Next, for each reference star a calibration curve was created by median combining the light curves of other reference stars (excluding that of the brown dwarf target and star in question).  The raw light curve of each reference star was then divided by its calibration curve to obtain a corrected light curve.  For each corrected light curve the standard deviation, $\sigma$, was determined using the IDL {\tt ROBUST\_SIGMA} routine which calculates an outlier-resistant standard deviation as outlined in \citet{beers90}.  We also measure a second quantity $\sigma_{\rm pt}$, which is the standard deviation of the light curve subtracted from a shifted version of itself, $f_{i+1}-f_{i}$, divided by $\sqrt{2}$.  This latter quantity provides an estimate of the high frequency noise in the light curve, and is insensitive to low frequency trends.  After the first-pass corrections low signal-to-noise or poor-quality reference stars were identified and removed from the calibration curve if any of the following conditions were met:  (i) $\sigma_{\rm pt}>1.5\times$ that of the target star, (ii) $\sigma$ values that represent a 3-sigma departure from a smooth fit to the $\sigma$ vs. magnitude trend of all stars on the chip, or (iii)if the source was previously flagged for quality (i.e. due to bad pixels, failed flux extraction in over 25\% of images, or due to an extended PSF).  This procedure is designed to eliminate flagged and intrinsically variable sources, as well as sources that fall on unusually noisy regions of the array from the set of references used to construct the calibration curves.  The bulk properties of reference star light curves are discussed in detail in section \ref{sect:limits}.  The remaining subset of high-quality, high-signal-to-noise references, less the star in question (and always excluding the brown dwarf target), are then used to re-calibrate each raw light curve.  The above procedure is repeated for successive iterations until the number of good reference stars stabilizes (typically 1-3 iterations) and usually yields 3-10 calibration sources for the WIRC sequences, and more than ten calibration sources for the WIRCam sequences.

Example light curves of targets and simultaneously observed reference stars are provided in Appendix \ref{app:A}.

\subsection{Detection Limits for Variability}
\label{sect:limits}

\begin{figure}[ht!]
\includegraphics[width=1.0\hsize]{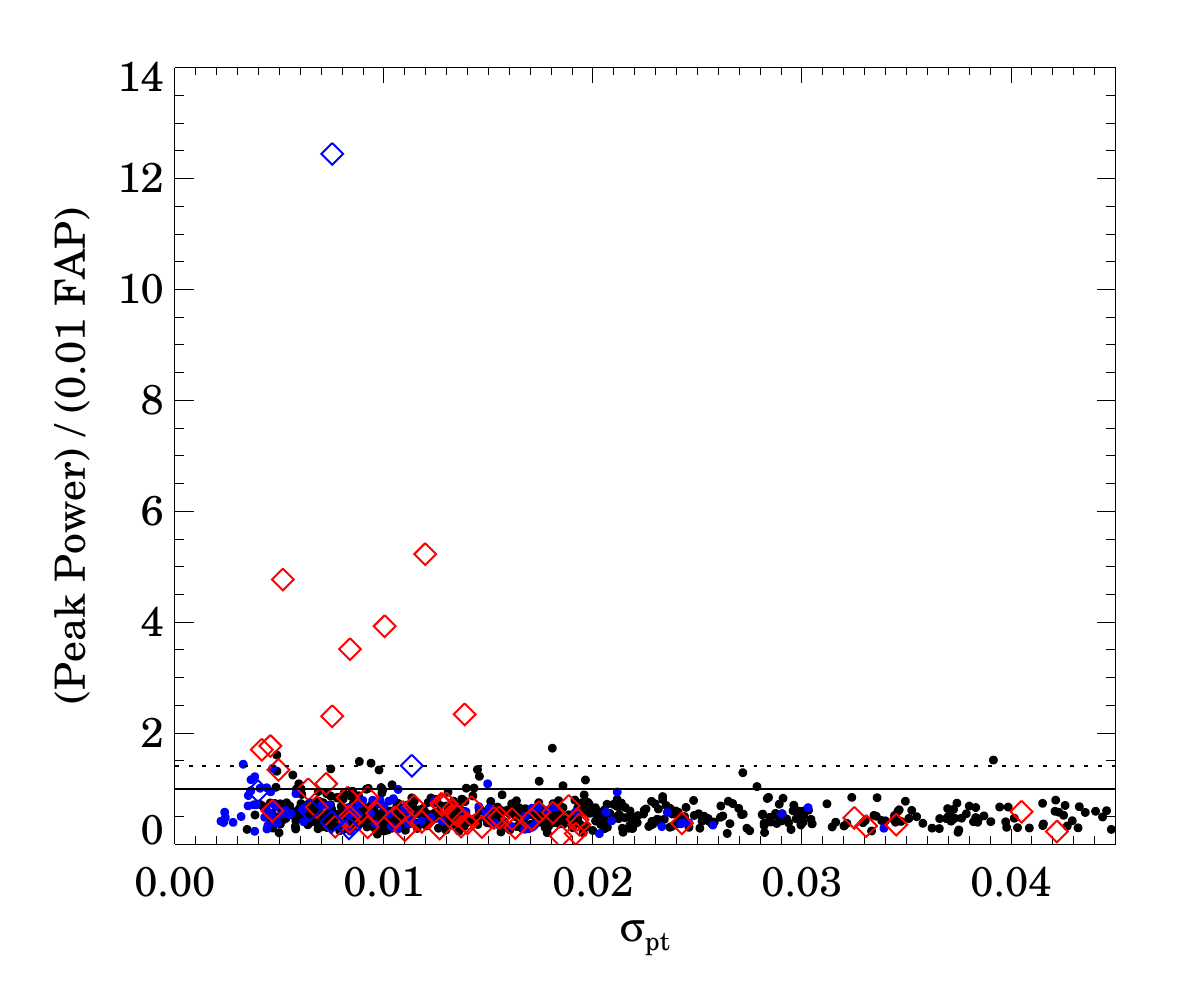}
\caption[Peak power in a Lomb-Scargle periodogram relative to the 0.01 FAP power for 742 reference stars]{Peak power in a Lomb-Scargle periodogram relative to the 0.01 FAP power for 742 reference stars observed with WIRC (black points) and 97 reference stars observed with WIRCam (blue points).  The ratio of peak power to 0.01 FAP level is over plotted for BD targets as red diamonds (WIRC) and blue diamonds (WIRCam). Horizontal lines are plotted for ratios of 1 (96\% detection probability) and 1.4 (99\% detection probability) as a visual aid.  \label{fig:peak_power}}
\end{figure}

The problem of detecting arbitrary variability in a light curve with both random and correlated noise contributions is challenging.  Even after dividing out common first order variations, there are typically a few reference stars in a given sequence that display residual low-frequency trends at the $\sim$1\% level.   We are able to make progress by noting that residual trends in the reduced light curves are often correlated in time with other observables.   In practice we found residual trends to be most strongly correlated with the calibration curve (e.g., due to second order differential extinction) or the seeing, while correlations with sky brightness and airmass were found to be negligible.  Weak correlations with the target's (${\bf x},{\bf y}$) position on the detector were found in some cases.  For the majority of reference star light curves we are able to remove correlated noise by fitting a function of the form ${\bf f}(t)= c_0+c_1{\bf F_0}(t)+c_2{\bf S}(t)+c_3{\bf x}(t)+c_4{\bf y}(t) + c_5{\bf x}(t)^2 + c_6{\bf y}(t)^2$ to the reduced light curves, and then dividing out the best fit.  Here, ${\bf F_0}$ is the calibration curve, ${\bf S}$ is the seeing, and ${\bf x}$, ${\bf y}$ are positions on the detector.  After dividing out correlated trends, we compute the remaining sinusoidal power in each light curve using the Lomb-Scargle method \citep{scargle82}, which is equivalent to least squares fitting of sinusoids.  Residual trends in the decorrelated light curves were  characterized using a control sample of 742 reference stars observed with WIRC throughout the course of our survey.    Light curves flagged for quality (due to extended PSFs, failed flux extraction, and bad pixels in the photometry aperture; see section \ref{sect:phot}) were excluded.  Other than placing a broad cut on $\sigma_{pt} < 0.05$, we did not eliminate any light curves based on their noise properties.  For each light curve we computed the power of sinusoidal signals in our data for a minimum period of 15 minutes up to a maximum period of twice the observation baseline.\footnote{While the maximum period that can be detected corresponds to the observation length,  the LS periodogram is still sensitive to the presence of lower frequency power.}   For each periodogram a 0.01 false alarm probability (FAP) was determined from 1000 simulated light curves produced by randomly permuting the indices of the originals.  We find that for 3.6\% of reference star light curves the ratio of peak power to the 0.01 FAP level is $>$1, as opposed to the expected 1\%, shown in figure \ref{fig:peak_power}.  Unfortunately we have no way of determining whether this power in the LS periodogram reflects intrinsic stellar variability in the near infrared, or residual systematic trends.  If we conservatively assume that none of our reference stars are variable, then we find the peak power due to correlated noise to be  $<$1.4 times the 0.01 FAP computed from simulations, 99\% of the time.  We therefore define $\beta$ as the ratio of the peak power in an LS periodogram to the 0.01 FAP computed from simulations, and set  our threshold for significant detections ($p>$99\%) to  $\beta>1.4$.   Similar results were found for 97 reference stars observed with WIRCam, also shown in figure  \ref{fig:peak_power}.

 As for the reference stars, the reduced target light curves were decorrelated against time varying observables, and $\beta$ values were computed.  We note that our procedure may remove or diminish intrinsic astrophysical variations that are, in part, coincidentally correlated with other observables that vary on similar timescales; we reason that such variations are not compelling and should not be ascribed high significance.    In other words, when assessing detections we only consider variability that is {\it not} correlated with other time-varying observables.
The $\beta$ values for BD targets are overplotted in figure \ref{fig:peak_power} for comparison.  It is immediately clear that the BD light curves are significantly more variable than those of the reference stars. Furthermore, our variability diagnostic of $\beta>1.4$ is in good agreement with a by-eye assessment of the light curves. Lomb-scargle periodograms for all targets and simultaneously monitored reference stars can be found in Appendix \ref{app:B}.

\subsection{Search Sensitivity}
\label{sect:sens}
In order to determine our completeness to intrinsic variability signals we injected sinusoidal signals of peak-to-peak amplitude $A$, period $P$, and random phase into our raw reference star light curves and re-analyzed them for detections, as described in section \ref{sect:limits}.  As opposed to white-noise simulations, using the the reference light curves themselves preserves the true noise properties of our data.  This procedure was completed for a grid of peak-to-peak amplitudes ranging from 1\%-10\%, and periods ranging from 1.5 to 20\,hr with 13 and 9 grid points respectively.  The recovery fraction of simulated signals at each grid point was computed as a function of the relative flux lightcurve RMS ($\sigma$) and the observing baseline ($\Delta t$).  Recovery rates for the injected variable signals were then determined as a function of observed light curve properties:
\begin{itemize}
\item $\sigma_{a}=\sigma_{\rm pt} \sqrt{\delta t / 67 {\rm s}}$  where $\delta t$ is the average cadence of the observation.  Since not all sequences use the same exposure times, $\sigma_a$ is the noise estimate {\em adjusted} for a 67\,s cadence.
\item the observation time baseline, $\Delta t$
\end{itemize}

\begin{figure}[ht!]
\centering
\includegraphics[width=0.99\hsize]{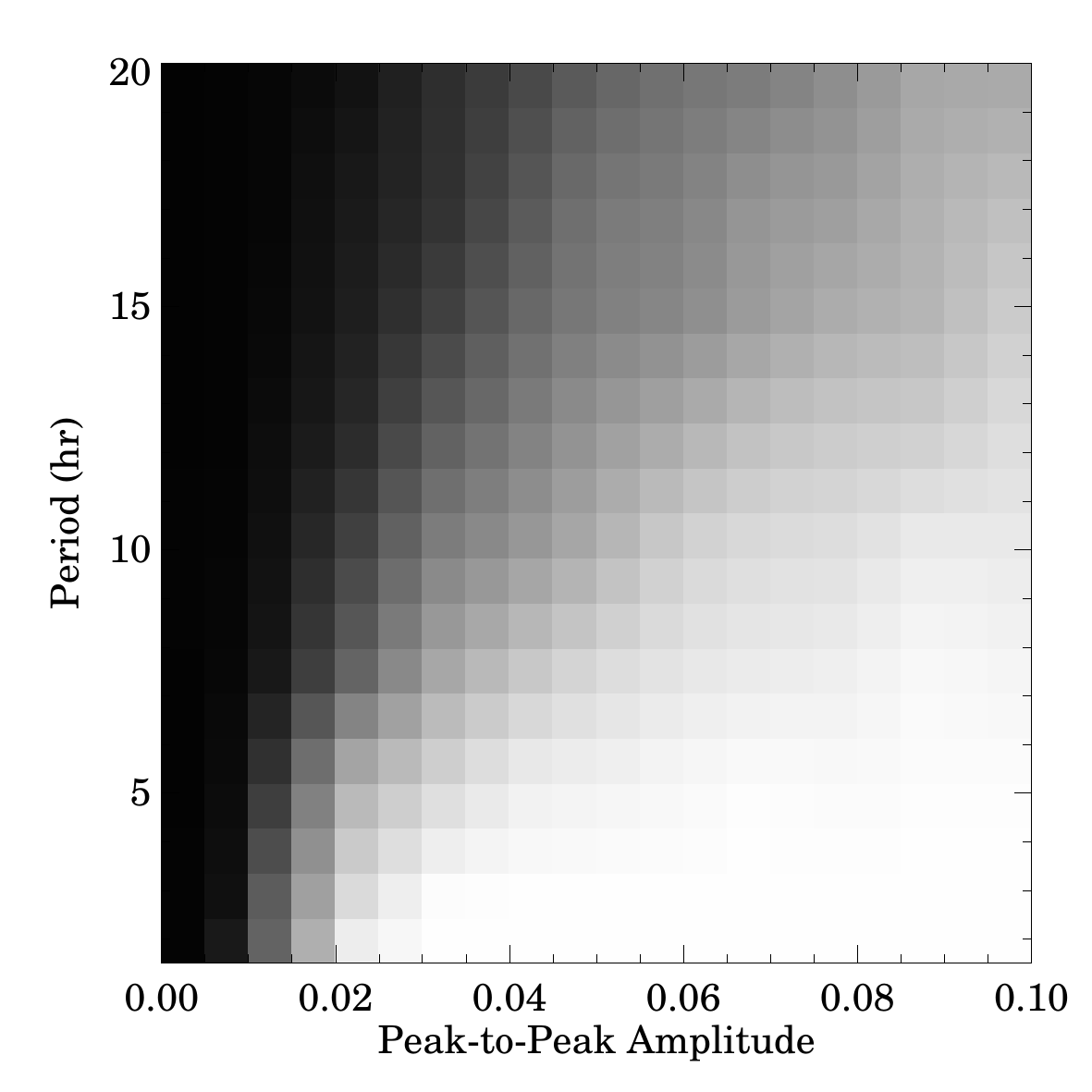}
\caption{Recovery rates of simulated sinusoidal signals as a function of peak-to-peak amplitude and period.  This slice in the 4-dimensional grid corresponds to a light curve with $\sigma_{a}=0.01$, and a time baseline of 3.5\,hr.  The shading varies linearly from recovery rates of 0\% (black) to 100\% (white). \label{fig:sens_slice}}
\end{figure}

For every point on the grid of amplitudes and periods we divided the simulated light curves into 4 bins in $\Delta t$ centered on 1.75, 3., 4.25, and 5.5\,hr (we seldom have an observing sequence longer than 5.5\,hr).  For each bin in $\Delta t$ we determined the detection fraction as a function of $\sigma_{a}$. This was done for 20 points in $\sigma_{a}$ ranging from 0.01 to 0.1 using a sliding bin of width 0.02.  To further smooth over noise we then fit the detection fraction with a complementary error function of the form

$$  
f_{\rm det}=c_0+c_1 {\rm erfc}[(\sigma_{\rm a}-c_2))/c_3)]
$$

with parameter constraints of c$_1 >$0 and 0$<c_2<$0.03.  The above functional form was chosen because it provides a very good match to the shape of the $f_{\rm det}$ vs. $\sigma_{\rm a}$ curves, and was more stable than a high-order polynomial fit.  All fitting was visually verified.  This procedure generated a 4-dimensional grid in $A$, $P$, $\sigma_{a}$, and $\Delta t$, varying smoothly (and for the most part monotonically) in all dimensions.  An image of a grid slice for $\sigma_{a}=0.015$, and $\Delta$t=3.5\,hr is shown in figure \ref{fig:sens_slice}.  Thus, the recovery fraction of a given signal in a given light curve, denoted here as $f_{\rm det}(\sigma_{a},\Delta t;A,P)$, is given from a 4-dimensional linear interpolation on this grid.

\begin{figure*}[ht!]
 \begin{tabular}{cc}
\includegraphics[width=0.45\hsize]{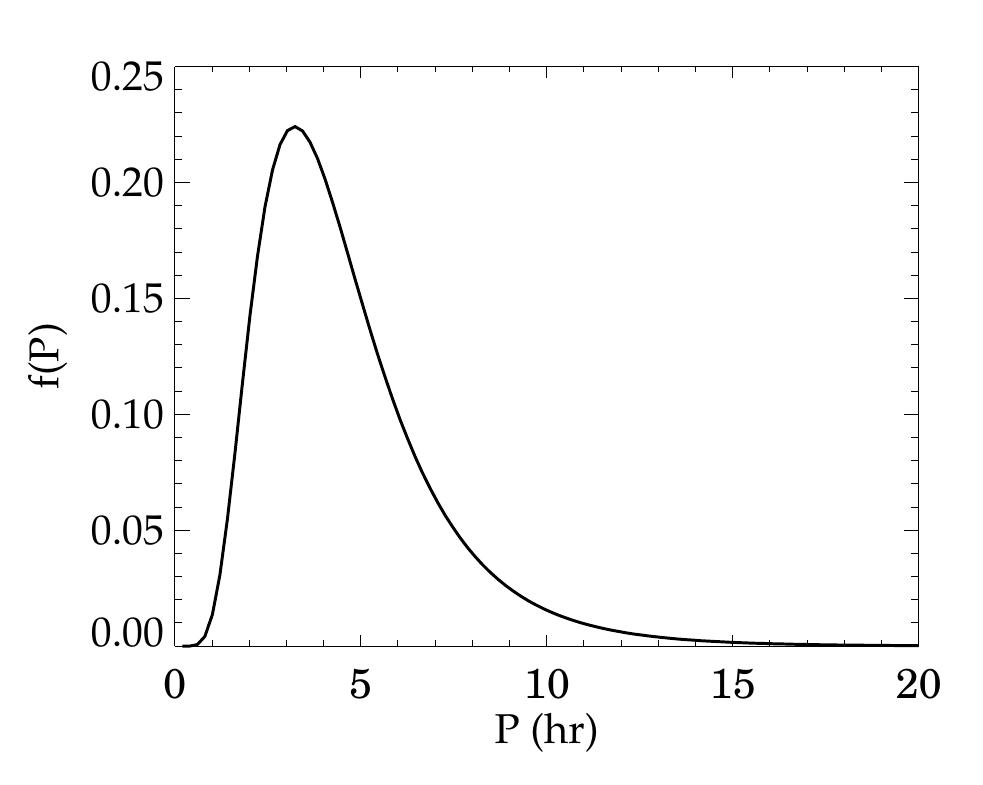} &
\includegraphics[width=0.45\hsize]{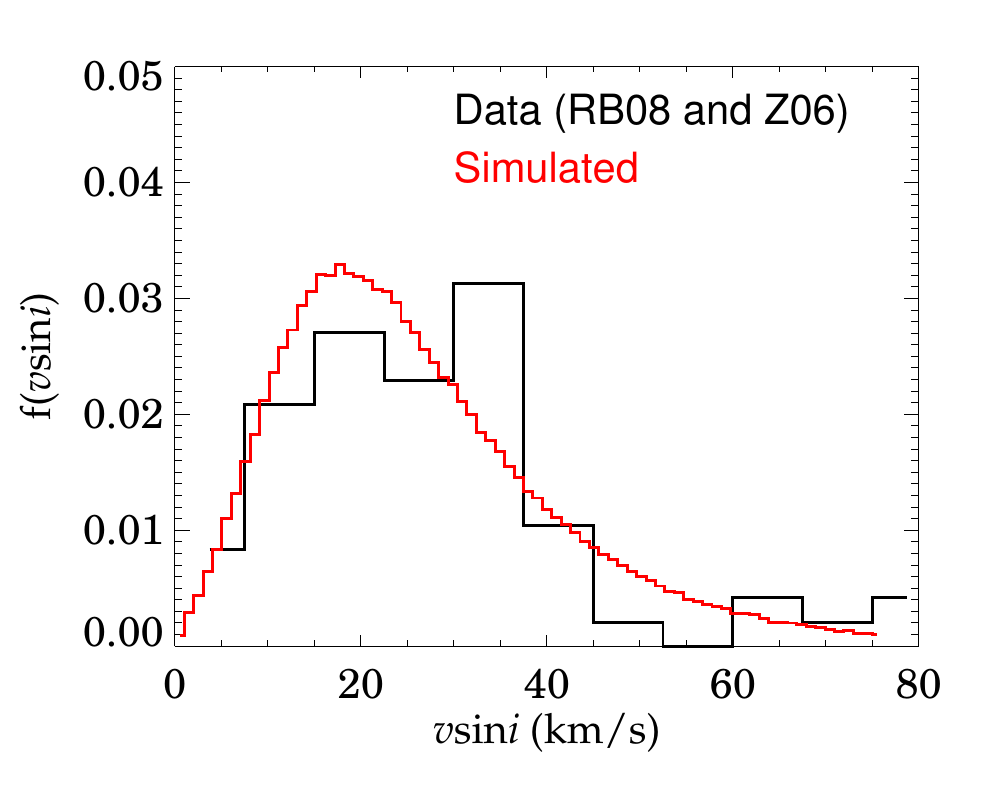}
\end{tabular}
\caption{The lognormal period distribution of L and T dwarfs inferred from  published $v\sin{i}$ values provided in  \citet{reiners08} and \citet{zapatero06} is shown on the left.  The right panel shows a comparison between the published $v\sin{i}$ values, and a simulated $v\sin{i}$ distribution.  The simulated distribution was produced in Monte Carlo fashion, using the inferred period distribution $f(P)$ and a distribution of inclinations $f(i)\propto \sin{i}$ as inputs.  Targets were assumed to have radii of 1\,$R_{\rm Jup}$. \label{fig:per}}
\end{figure*}

In addition to the instrumental detection limits, the period distribution will affect the recovery rate of variable signals. For a single BD target our sensitivity to a particular variability amplitude is given by 

\begin{equation}
f_{\rm sens}[\sigma_{a},\Delta t; A]=\int_{0}^{\infty} f_{\rm det}[\sigma_{a},\Delta t; A,P] f(P)dP
\end{equation}

where $f(P)$ is the distribution of periods for our sample.   Previous $v\sin{i}$ studies of L and T dwarfs provide some clue as to the the period distribution $f(P)$.  Under the assumption of a lognormal period distribution given by

\begin{equation}
f(P)=\frac{1}{P\sigma\sqrt{2\pi}}{\rm e}{^{-\frac{(\ln{P}-\mu)^2}{2\sigma^2}}}
\end{equation}

we determined maximum likelihood values of $\mu=1.41$ and $\sigma=0.48$ via comparison with $v\sin{i}$ data for L and T dwarfs from  \citet{reiners08} and \citet{zapatero06}.  The adopted period distribution is shown and compared to existing data in figure \ref{fig:per}.

\begin{figure*}[ht!]
\begin{tabular}{cc}
\includegraphics[width=0.45\hsize]{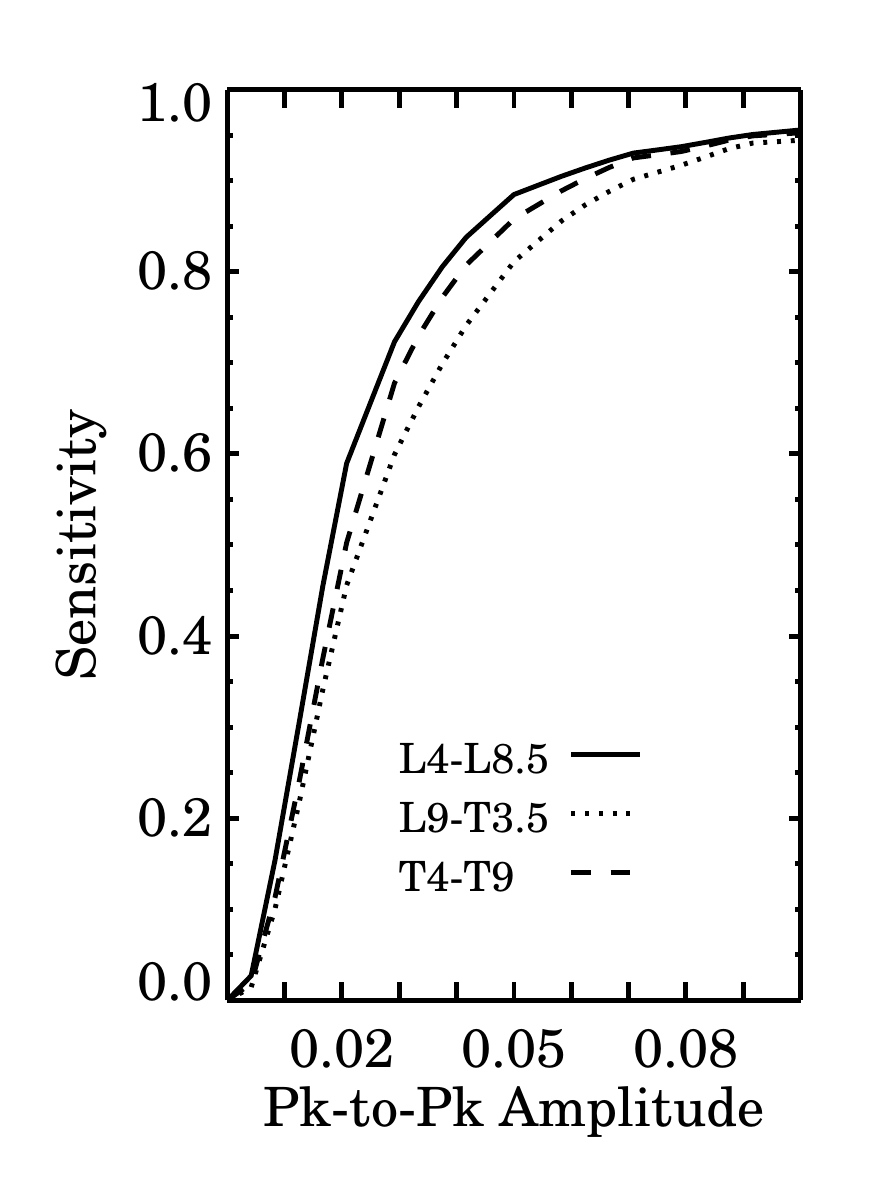} &
\includegraphics[width=0.45\hsize]{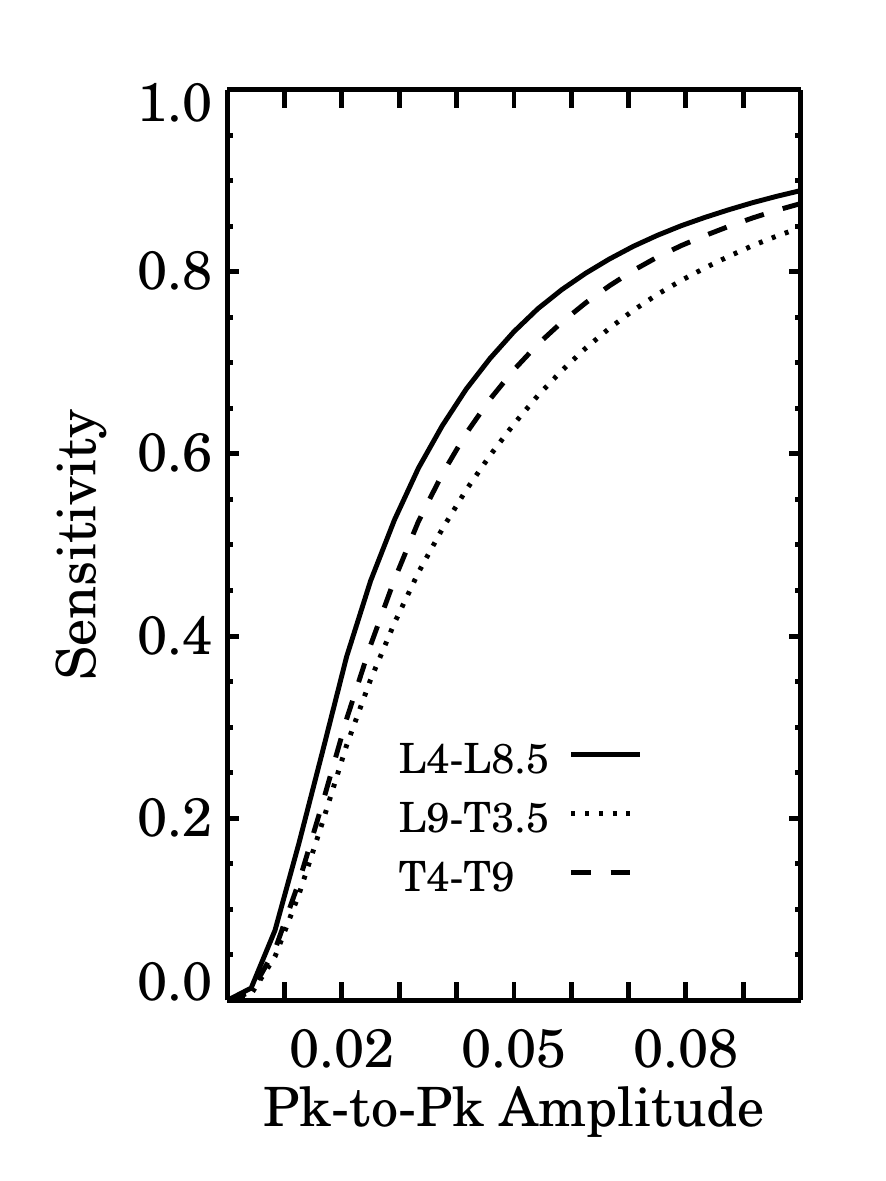}
\end{tabular}
\caption{{\it Left panel:} Sensitivity to sinusoidal signals of a given peak-to-peak amplitude, $A$, averaged over all targets in each of three spectral type bins: L4-L8.5, L9-T3.5, T4-T9.  We find approximately equal sensitivities in all spectral type bins.  These sensitivities reflect recovery rates of sinusoidal signals with a distribution of periods inferred from $v\sin{i}$ data for brown dwarfs (see section\ref{sect:sens}).   Our sensitivity to brightness changes of equivalent amplitude over the span of a monitoring sequence is typically much larger (see $\sigma_a$ values in table \ref{tab:lc_info} for an idea of the precision achieved at 67\,s cadence for individual targets).  {\it Right panel:} Same as left panel, but showing reduced sensitivity to a maximum observable amplitude for a target viewed edge-on (i.e., $i=90^{\circ}$), assuming an isotropic distribution of spin axes for our targets.  \label{fig:sens}}
\end{figure*}

The $\beta$ values for each target, as well as the amplitudes for which $f_{\rm sens}[\sigma_{a},\Delta t; A]$=0.5  are provided in table \ref{tab:lc_info}.  Our sensitivity to signals as a function of amplitude, averaged over all survey targets (i.e. averaging over the $f_{\rm sens}[\sigma_{a},\Delta t; A]$ curves for individual targets), is shown in figure \ref{fig:sens}.  When examining recovery rates in different spectral type bins (L4-L8.5,  L9-T3.5, and T4-T9) we find no significant biases as a function of spectral type.

\section{Results}
\label{sect:results}
\subsection{Detections of Variability}
\label{sect:detections}
\begin{figure*}[ht!]
\includegraphics[width=0.95\hsize]{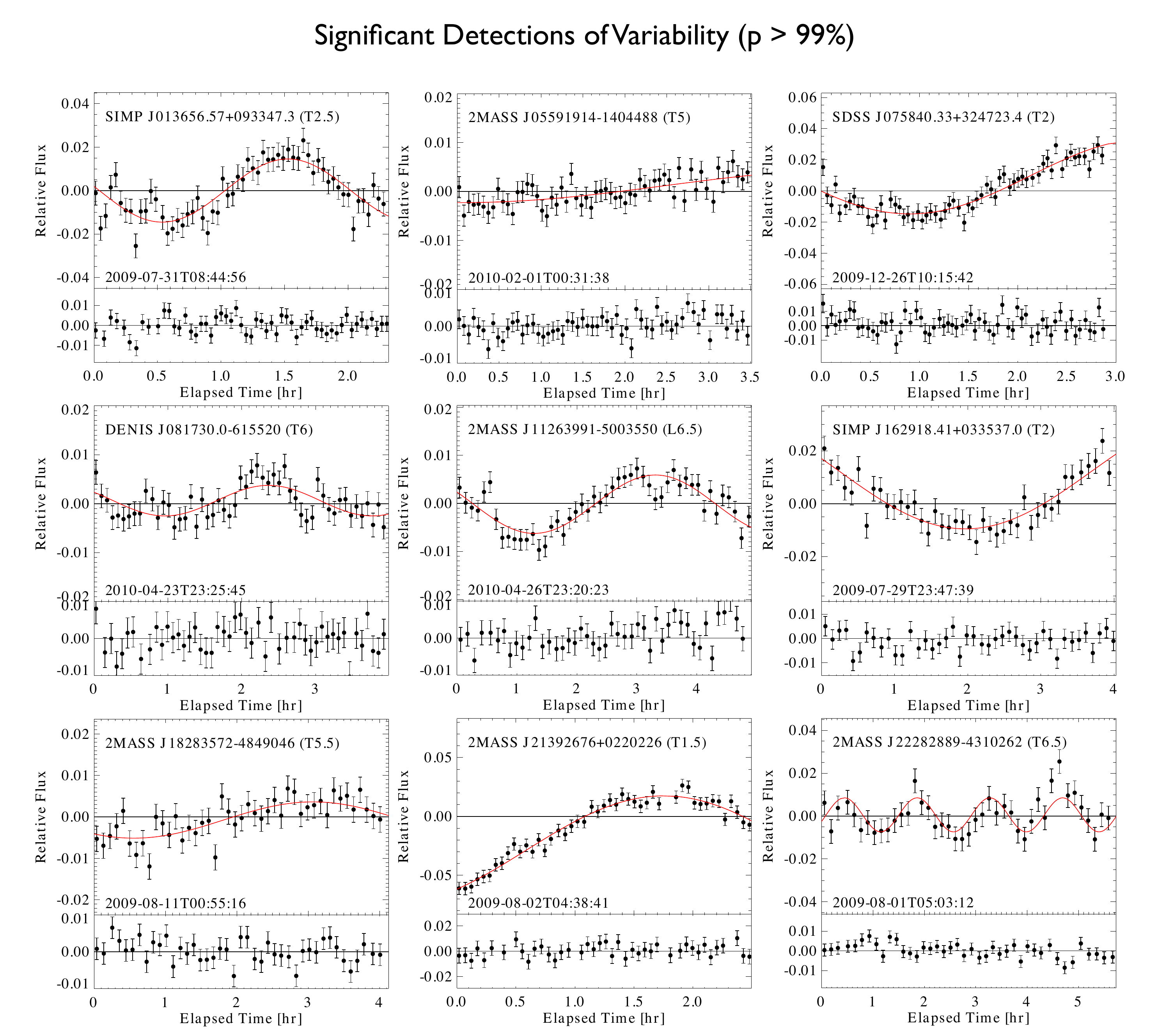} 
\caption{Light curves of significantly variable  (p$>$99\%) targets in our sample. Light curves are shown in the upper panels, and the light curve of a similar-brightness comparison star observed simultaneously is shown in the lower panels.  All data shown were obtained using WIRC on the Du Pont 2.5-m telescope except for those of 2M0758+32, which were obtained using WIRCam on the CFHT.  All light curves have been binned down from their original cadence by factors of 3-7. \label{fig:1}  }
\end{figure*}

\begin{figure*}[ht!]
\includegraphics[width=0.95\hsize]{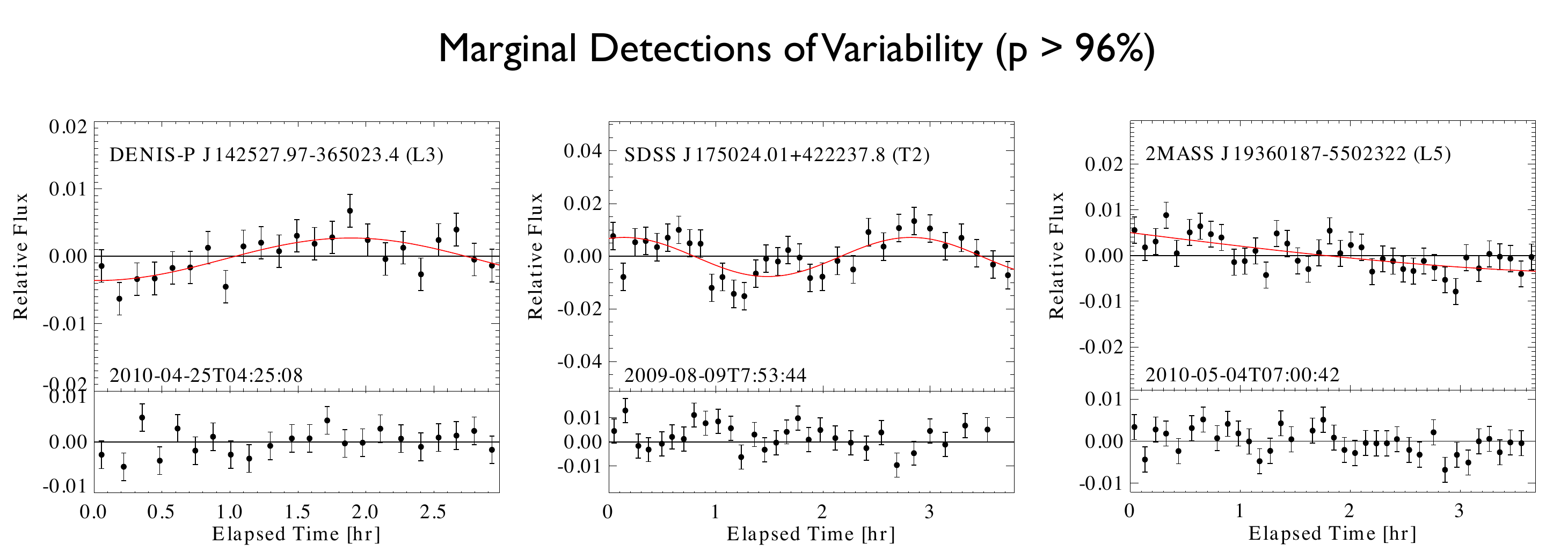} 
\caption{Light curves of marginally variable targets  ($p>$96\%) in our sample.  With a 4\% false positive rate, we might expect $\sim$2 false positives at this significance level, and we therefore do not consider these detections to be reliable. Light curves are shown in the upper panels, and the light curve of a similar-brightness comparison star observed simultaneously is shown in the lower panels.  All light curves have been binned down from their original cadence by factors of 3-7. \label{fig:2} }
\end{figure*}

\begin{figure*}[ht!]
\includegraphics[width=0.95\hsize]{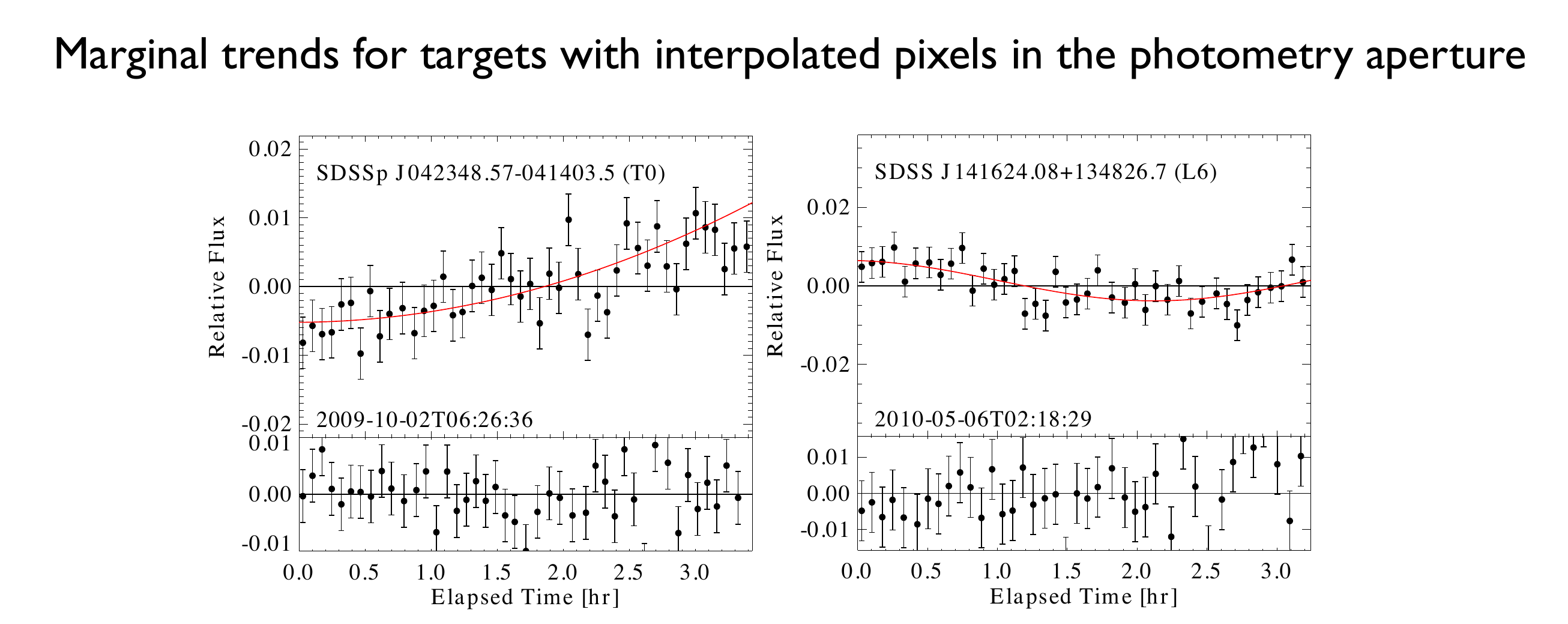} 
\caption{Light curves of targets flagged for quality by our reduction pipeline due to bad/interpolated pixels in the photometry aperture.  These targets have been excluded from the statistical sample considered in section \ref{sect:stats}.   Light curves have been binned down from their original cadence by factors of 3-7. \label{fig:3}  }
\end{figure*}

In the previous section we determined a detection limit of $\beta > 1.4$ ($p>99$\%) for which we expect less than one false positive in the entire sample.  This limit is conservative because some reference stars, all of which were assumed to be non-variable, may in fact be intrinsically variable.  Light curves for significantly variable targets are shown in figure \ref{fig:1}.
Targets for which $1 < \beta < 1.4$ are noted as marginal detections ($p>$96\%), and their light curves are shown in figure \ref{fig:2}.  Two targets, SDSSp~J042348.57$-$041403.5AB (SD0423$-$04) and SD1416$+$13 have been flagged for quality for having one or more bad pixels in the photometry aperture, and would not have passed our cuts for inclusion in the reference star library used to determine the false-positive rate.  Therefore, these targets have been disqualified from our statistical sample, and subsequent analyses.  We note for completeness that both targets have significant or marginally significant trends in their light curves, which we show in figure \ref{fig:3}.

The 9 targets for which we have detected significant variability ($p>99$\%) are listed in table \ref{tab:detections}.  Five of these, including three new high-amplitude variables at the L/T transition, are reported for the first time as part of this survey, with detailed follow-up observations of the largest variable in our sample, the T1.5 dwarf 2M2139$+$02, already reported by \citet{radigan12}.  
The three additional marginal detections (p$>$96\%) are noted in table \ref{tab:detections} for completeness, however this number is consistent with expected number of false-positives ($\sim$2), and they are treated as non-detections in subsequent analysis.

Each highly significant detection ($p>99$\%) is described in further detail below.

\paragraph{SIMP~J013656.57+093347.3}

The T2.5 dwarf SIMP0136+09 \citep{artigau06} was previously found to be variable with a peak-to-peak amplitude as high as 8\% by \citet{artigau09} and with a period of 2.4\,hr.  Its light curve is known to change in amplitude and shape from epoch to epoch \citep{metchev13_proc}.   The observations from Las Campanas captured an amplitude of only 2.9\%, and were not long enough to capture a full period.  A sinusoidal fit to the light curve yields a period of $\sim$2\,hr which underestimates the known value.

\paragraph{2MASS J05591914-1404488}
The T4.5 dwarf 2M0559-14 \citep{burgasser00b} was found to be variable, with a steady increase in brightness of $\sim$0.6\% observed over the 3.5\,hr observing sequence. Because we observe an approximately linear trend, the estimated amplitude (0.7\% peak-to-peak) and period ($\sim$10\,hr) derived from a sinusoidal fit are highly uncertain. Nonetheless, since the amplitude is close to that observed over the course of the observation, we adopt this value for the purpose of our subsequent analysis of the variability occurrence rate in section \ref{sect:stats}.   If the observed variability is rotationally modulated, we can conclude that 2M0559-14 has a period much larger than 3.5\,hr.  This target has been noted in the literature for being overluminous for its spectral type at both NIR and MIR wavelengths \citep{dahn02,dupuy12}.  It is therefore a suspected binary despite remaining unresolved in high angular resolution imaging \citep{burgasser03} and radial velocity observations \citep{zapatero07}.  Ground based monitoring of this target in the $J$ band by \citet{clarke08}, found no evidence for variability above $\sim$5\,mmag.  Conversely, in a $\sim$40 minute spectroscopic observation with HST/WFC3 \citet{buenzli14} detected significant variability (p$>$95\%) of this target in the $H$-band with a slope of $\sim$1.2-1.4\%\,hr$^{-1}$, and  tentative variability (p$>$68\%) in the $J$-band with a slope of $\sim$-0.7\%\,hr$^{-1}$.  Although our observation corresponds to a much smaller slope of $\sim$0.17\%\,hr$^{-1}$, the high significance of the detection suggests that this source is indeed a weak variable in the $J$ band.

\paragraph{SDSS~J075840.33+324723.4}
Variability of the T2 dwarf SD0758$+$32 \citep{knapp04} is presented for the first time here.  The peak-to-peak amplitude as estimated from a sinusoidal fit is 4.8\%, with an estimated period of  $\sim$5\,hr.  This source was found to be non-variable above $\sim$5-10\,mmag in the monitoring program of \citet{girardin13}.

\paragraph{DENIS~J081730.0-615520}
The nearby T6 dwarf DE0817$-$61 \citep{artigau10} was observed to vary with a peak-to-peak amplitude of 0.6\% and a period of 2.8$\pm$0.2\,hr as estimated by a sinusoidal fit to its light curve.   In a $\sim$40 minute spectroscopic observation with HST/WFC3 \citet{buenzli14} also found this source to vary significantly in the $J$ band (p$>$95\%) with a slope of $\sim$0.7\%\,hr$^{-1}$.

\paragraph{2MASS~J11263991-5003550}
The target 2M1126$-$50 \citep{folkes07} is a peculiar L dwarf with $J-K_s$ colors that are unusually blue for its L4.5 optical or L6.5 NIR spectral type, most likely the result of thin, large-grained condensates \citep{burgasser08}.  Given the variability presented here we suggest that the blue colors and peculiar spectral morphology for this target are due to holes in the cloud layer.  A peak-to-peak amplitude of 1.2\%, and a tentative period of $\sim$4\,hr is inferred from a sinusoidal fit.

\paragraph{SIMP~J162918.41+033537.0}
The target SIMP1629$+$03 was included in our survey as an unpublished T2 dwarf discovered in the SIMP proper motion survey \citep{artigau09}, and was independently discovered in PANSTARRS as PSO J247.3273+03.5932 \citep{deacon11}.  We estimate a peak-to-peak amplitude of $\sim$4.3\% and a period of $\sim$6.9\,hr from a sinusoidal fit to the 4\,hr light curve.  These estimates are highly uncertain since our observations encompass only the trough of the light curve.

\paragraph{2MASS~J18283572$-$4849046} 
The T5.5 dwarf 2M1828$-$48 \citep{burgasser04} was observed to vary with a peak-to-peak amplitude of 0.9\% and a period of $\sim$5\,hr, inferred from a sinusoidal fit to its light curve.  We note that one of the reference stars observed simultaneously shows a similar trend at a less significant level, which can been seen from figure \ref{fig:lsper} in Appendix \ref{app:B}.    We expect $\sim$0.57 false positives in our sample, so it would not be unreasonable to find a single false positive.  This source was previously monitored in the $J$ band by \citet{clarke08} who do not find any significant variability, but note a weak trend in its light curve, at the same level as a trend seen in nearby ``check'' stars.

\paragraph{2MASS~J21392676+0220226}
The target 2M2139$+$02 \citep{reid08} was observed to vary with a peak-to-peak amplitude of $\sim$9\% in a 2.5\,hr observation.  Follow-up observations of this target by \citet{radigan12} reveal a $J$ band peak-to-peak amplitude as high as 26\% and a period of $\sim$7.72\,hr.  This is the largest variability reported for a BD to date.

\paragraph{2MASS~J22282889$-$4310262}
Variability of 2M2228$-$43 \citep{burgasser03_2mass} was first detected by \citet{clarke08} who find a peak-to-peak $J$-band amplitude of 1.5\% and a period of 1.4\,hr.  This target was confirmed to vary by \citet{buenzli12} who reported pressure/wavelength dependent phase shifts of the light curve.   We confirm this target's variability, measuring a peak-to-peak amplitude of 1.6\% in the $J$ band, and a period of $1.42\pm0.03$\,hr based on a sinusoidal fit to this target's light curve over 4 continuous rotations.  We note that this target is rather red for a T6.5 dwarf, and given the observed variability we suggest that its red color may be attributed to residual cloud patches in its atmosphere.  \

\subsection{Non-detections, and marginal cases:  comparisons with previous work}
For completeness we describe null and marginal detections in our dataset for targets that have been previously monitored for variability at similar (i.e. NIR) wavelengths.

\paragraph{2MASSW~J0030300$-$145033}
The L7 target 2M0030$-$14 \citep{kirkpatrick00} was monitored in the $K_s$ band by \citet{enoch03} and found to be variable with an amplitude of 0.19$\pm$0.11\,mag and a period of 1.4\,hr.  We observed a flat light curve for this target in the $J$ band  ($\sigma_a$=0.011) over a time period of 3.2\,hr.

\paragraph{SDSSJ015141.69$+$124429.6} The T1 target  SD0151$+$12 \citep{geballe02} was monitored in the $K_{\rm s}$ band by \citet{enoch03} and found to be variable with an amplitude of 0.42$\pm$0.14\,mag and a period of 3\,hr.  We observed a flat light curve for this target in the $J$ band  ($\sigma_a$=0.006) over a time period of 3.7\,hr.

\paragraph{2MASSI~J0243137$-$245329} 
The T6 target 2M0243$-$24 \citep{burgasser02_spt1} was monitored for variability by \citet{buenzli14} with HST/WFC3, who do not detect any significant trends within a $\sim$40\,min observation window.  We observed a flat light curve for this target in the $J$ band  ($\sigma_a$=0.011) over time period of 3.7\,hr.

\paragraph{2MASSI~J0328426$+$230205} The L9.5 target 2M0328$+$23 \citep{kirkpatrick00} was monitored in the $K_{\rm s}$ band by \citet{enoch03} and found to be variable with an amplitude of 0.43$\pm$0.16\,mag and no identifiable periodicity. We did not detect any significant variability in this target ($\sigma_a$=0.006) in our observation spanning 3.7\,hr.

\paragraph{SDSSp~J042348.57$-$041403.5}
The T0 binary SD0423$-$04 \citep{geballe02} was monitored in the $K_{\rm s}$ band by \citet{enoch03}, who find some evidence for variability with an amplitude of 0.3$\pm$0.18\,mag, but caution that the signal may be caused by a variable reference star rather than being intrinsic to the brown dwarf.  \citet{clarke08} observed this source in the $J$ band, and detected variability with an amplitude of 0.8\,mmag and period of $\sim$2\,hr.   Our observation of this source shows a small brightening over 3.6\,hr, however this target is flagged for quality due to bad pixels in the photometry aperture, and the significance of this trend is therefore highly uncertain.

\paragraph{2MASS~J09393548$-$2448279}
The T8 target 2M0939$-$24 \citep{tinney05} was monitored in the $J$ and $K_s$ bands by \citet{khandrika13}, who found evidence for variability with an amplitude of 0.31\,mag in the $K_s$ band.  These authors report only a marginal detection due to the low signal to noise of the light curve. No periodicity is reported.  We did not detect significant variability in our $J$ band monitoring of this target spanning 5.9\,hr ($\sigma_a$=0.015\,mag).   Although it has not been resolved by high angular resolution imaging \citep{goldman08}, this source is a suspected binary due to over-luminosity \citep{burgasser08_0939}.

\paragraph{SDSSp~J125453.90$-$012247.4} 
The T2 NIR spectral standard SD1254$-$01 \citep{leggett00} was monitored in the $J$ band by \citet{koen04a} and \citet{girardin13}, who find no evidence for variability above the $\sim$5-10\,mmag level. No variability is detected in our observations spanning 3.1\,hr ($\sigma_a$=0.021\,mag).

\paragraph{SDSS~J141624.08$+$134826.7}
The unusually blue L6 dwarf SD1416+13 was monitored in the $J$ and $K_s$ bands by \citet{khandrika13}, who find marginal evidence for variability with an amplitude of 5.4\%.  We found this target to be marginally variable in our observations spanning 3.2\,hr, albeit with a much smaller amplitude of  $\sim$1\% ($\sigma_a$=0.007).  However, this target was flagged for quality in our reduction pipeline due to bad pixels in the photometry aperture.  Follow up will be required to determine whether the observed variability of this target is real and persistent.

\paragraph{2MASSI~J1546291$-$332511}
{
The T5.5 dwarf 2M1546$-$33 \citep{burgasser02_spt1} was monitored by \citet{koen04a}, who found no evidence of variability above the $\sim$5\,mmag level based on 7 data points spanning 4.4\,hr.  Our continuous monitoring of this target over 3.8\,hr did not detect any variability ($\sigma_a$=0.008).
}

\paragraph{SDSSp~J162414.37$+$002915.6}
The T6 dwarf SD1624$+$00 \citep{strauss99} was monitored by \citet{koen04a}, who find no evidence of variability above the $\sim$5\,mmag level.  The HST/WFC3 spectroscopic study of \citet{buenzli14} identify tentative variability in the water band of this target, but not in the continuum.  Our observation of this target over 3.6\,hr did not detect any variability ($\sigma_a$=0.017).

\paragraph{2MASS~J17502484$-$0016151} The L5.5 dwarf 2M1750$-$00 \citep{kendall07} was found to be significantly variable in the $J$ band by \citet{buenzli14}, who detected a slope of $\sim$0.7\%\,hr$^{-1}$ over a $\sim$40\,min observation.   We did not detect any significant variability in our 2.5\,hr observation of this target ($\sigma_a$=0.005), but note a relatively large value of $\beta$=0.99, perhaps hinting at low-level variability just below our detection threshold.

\paragraph{2MASSI~J2254188$+$312349} The T4 dwarf 2M2254$+$31 \citep{burgasser02_spt1} was monitored in the $K_{\rm s}$ band by \citet{enoch03} and found to be variable with an amplitude of 0.48$\pm$0.20\,mag and no identifiable periodicity.  We observed a flat light curve for this target in the $J$ band  ($\sigma_a$=0.008) over time period of 3.4\,hr.\\

 Several targets found to be non-variable in our survey (2M0030$-$14, SD0151$+$12, 2M0328$+$23, and 2M0939$-$24, 2M1750$-$00) have previous detections reported in the literature.  In addition, two of our targets which show evidence for variability, but were flagged for quality (SD0423$-$04 and SD1416+13) have been previously reported as marginal variables.  Due to differences in sensitivity, cadence, and wavelength, as well as the transient nature of variability \citep[e.g.][]{metchev13_proc}, it is hard to draw conclusions from comparisons between observations of a given target with different instrumental setups and at different epochs.  Taken as a whole, however, our survey finds no evidence for significant $J$-band variability with amplitudes above $\sim$1.6\% outside the L/T transition.  This implies that large amplitude variability in this regime is rare, in agreement with  the $J$-band surveys of \citet{koen04a,koen05a}, \citet{clarke08} and \citet{girardin13}, and in contrast to the studies of \citet{enoch03} and \citet{khandrika13} who find large amplitude variability at all spectral types.  We note that many of the large amplitude detections from the latter two surveys are in the $K_{\rm s}$ band, which may account for part of this discrepancy.

\subsection{Properties of variable targets}
\label{sect:properties}
\begin{figure*}[ht!]
\centering
\includegraphics[width=0.8\hsize]{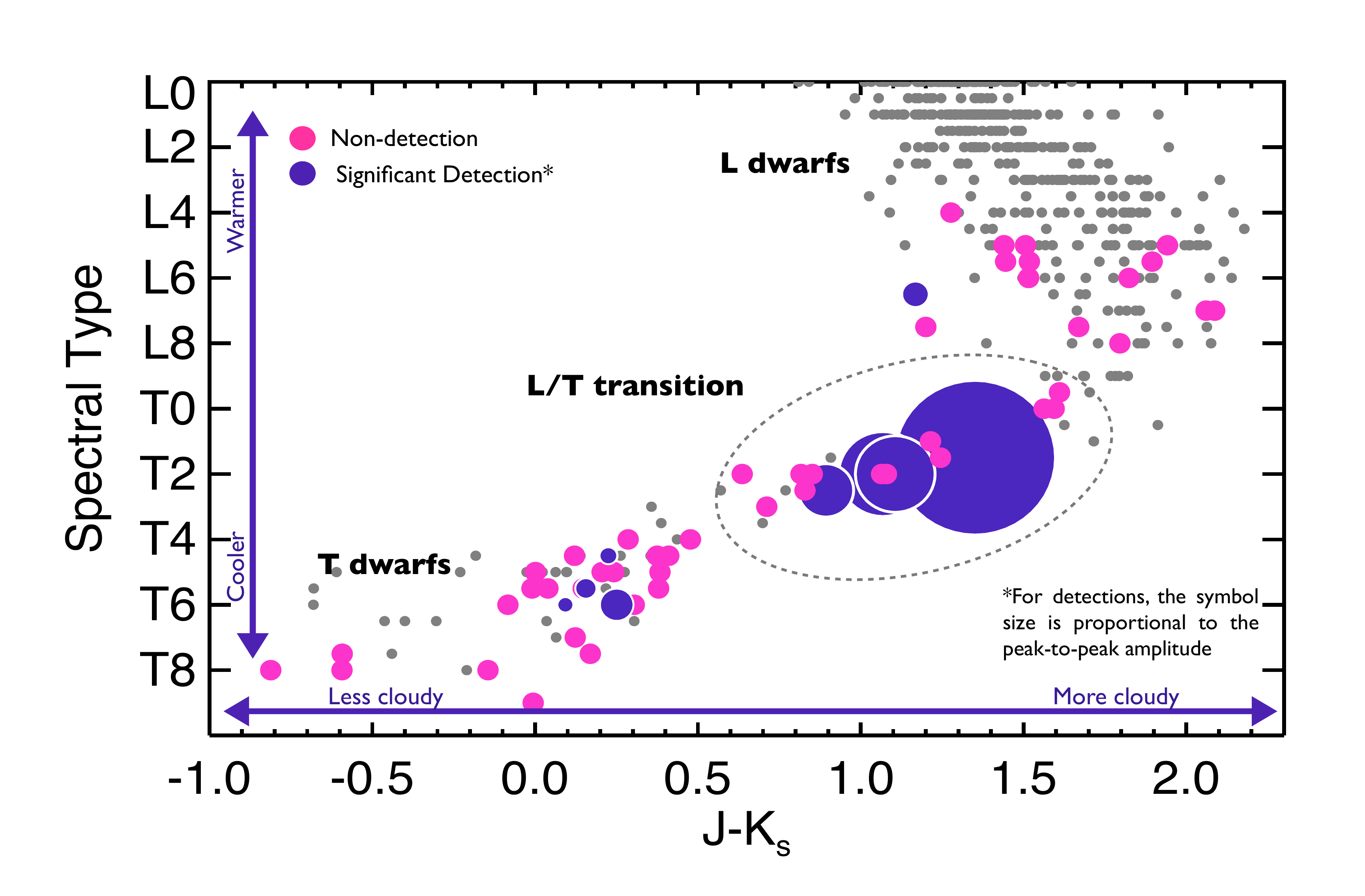} 
\caption{NIR spectral type versus 2MASS $J-K_s$ color for all targets observed in our program.  Grey points show the population of known field L and T dwarfs with $J<$16.5.  Purple circles show detections, with the linear symbol size drawn proportional to the peak-to-peak amplitude of variability detected (ranging from 0.9-9\% for the smallest to largest symbols).  A grey dashed line encircles the objects considered part of the L/T transition sample.  Our average sensitivity or completeness to sinusoidal signals of a given peak-to-peak amplitude is shown in figure \ref{fig:sens}, where we find approximately equal sensitivities in all spectral type bins, demonstrating that a fair comparison is made between populations. \label{fig:spt_col}}
\end{figure*}

For the 57 targets that pass our quality requirements (i.e. targets that were not flagged for bad pixels in our reduction pipeline, and are not resolved equal mass binaries---the composition of out statistical sample is described in section \ref{sect:stats}), detections and non-detections are identified as a function of NIR spectral type versus color in figure \ref{fig:spt_col}.  
On this plot, large amplitude variability ($\gtrsim$2\%) occurs exclusively for early-T spectral types with $J-K_s$ colors between 0.8\,mag and 1.3\,mag.   This is the first correlation between NIR variability and spectral type/color ever reported for brown dwarfs.  The statistical significance of this trend is discussed in section \ref{sect:stats}.

\begin{figure*}[ht!]
\includegraphics[width=0.99\hsize]{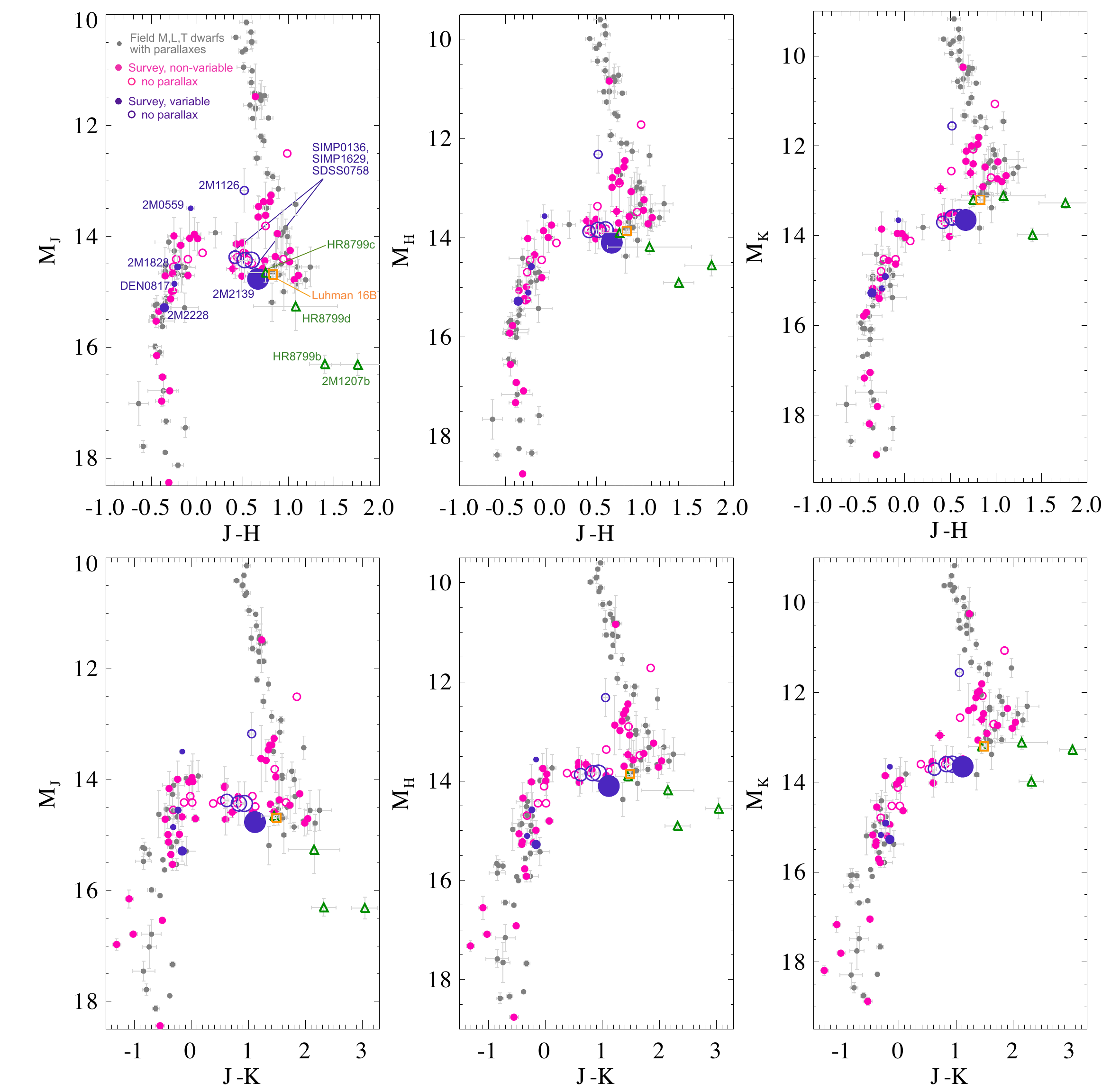} 
\caption{ Color magnitude diagrams for variable and non-variable survey targets (pink and purple points respectively).  Open circles mark targets without parallax data, for which spectroscopic parallaxes were found using the spectral type-absolute magnitude relationship defined by \citet{dupuy12}.  Field ultra cool dwarfs with parallaxes from the database of Trent J. Dupuy \citep{dupuy12} are shown as grey circles. \label{fig:cmds}}
\end{figure*}

Our survey targets are identified on a series of NIR color-magnitude diagrams (CMDs) in figure \ref{fig:cmds}, where we used the spectral-type versus absolute magnitude relationship of \citet{dupuy12} to estimate absolute magnitudes for targets without parallaxes.  Parallaxes, spectral types and MKO system photometry for our sample is provided  in table \ref{tab:ch5_targ}.  In cases where no published MKO photometry was available we used the correction formulae provided as a function of spectral type by \citet{stephens04} to convert from 2MASS to MKO magnitudes.   Positions of the HR8799bcd directly imaged planets \citep{marois08,currie11}, as well as the highly variable secondary component of the Luhman 16 system \citep{burgasser13} are shown for reference.  Clustering of the high-amplitude variables, including Luhman 16B,  in the lower portion of the ``$J$-band brightening sequence'' is evident, suggesting that the temperature contrast between cloud tops and clearings peaks in this region of the diagram, as clouds are first disrupted.  Further along the brightening sequence, temperature contrasts may decline due to a cooling of the underlying atmosphere, and cloud tops sitting progressively lower in the photosphere.  The HR8799bd planets, and 2M1207b appear as an extension of the L-dwarf sequence, becoming progressively redder with increasing magnitude.  Due to the ability of low-gravity atmospheres to retain clouds down to lower effective temperatures, cloud disruption occurs at much lower temperatures (if at all) in low-gravity atmospheres \citep[e.g.][]{metchev06,luhman07,marley12}.  The HR8799c planet is an interesting exception, sitting directly in the L/T transition region of the CMD occupied by highly varying field brown dwarfs, making it a prime candidate for variability monitoring.  The NIR spectrum of HR8799c obtained by \citet{oppenheimer13} is also consistent with an early T spectral type for this planet.

Outside the L/T transition, the highest amplitude detections are for the L6.5 dwarf 2M1126$-$50 (A=1.2\%) which has unusually blue NIR colors for its spectral type, and the T6 dwarf 2M2228$-$43 (A=1.6\%), which has a relatively red $J-K_{\rm s}$ color for its spectral type.  The presence of breaks in the cloud layer in the former case, and residual photospheric cloud features in the latter case may provide an explanation for the respective colors of these objects.  However, we detect variability with smaller amplitudes (ranging from 0.6-0.9\%) in three other mid-T dwarfs with unremarkable NIR colors.  Further study will be required to determine the significance of any trends between variability and $J-K_{\rm s}$ color for objects outside the L/T transition.  It is notable that $J$-band variability is found in several mid-T dwarfs, for which silicate clouds are expected to have settled below the photosphere.  These observations may provide non-spectroscopic evidence for the persistence of photospheric clouds---potentially formed from salts and sulfides (e.g. KCl and Na$_2$S)---in late-type atmospheres as suggested by \citet{morley12}.  Alternatively, \citet{robinson14} argue that thermal fluctuations can also give rise to flux variations.  Multi-wavelength studies will be required to determine the most likely cause of variability in late-type objects.

\begin{figure*}[ht!]
\includegraphics[width=0.99\hsize]{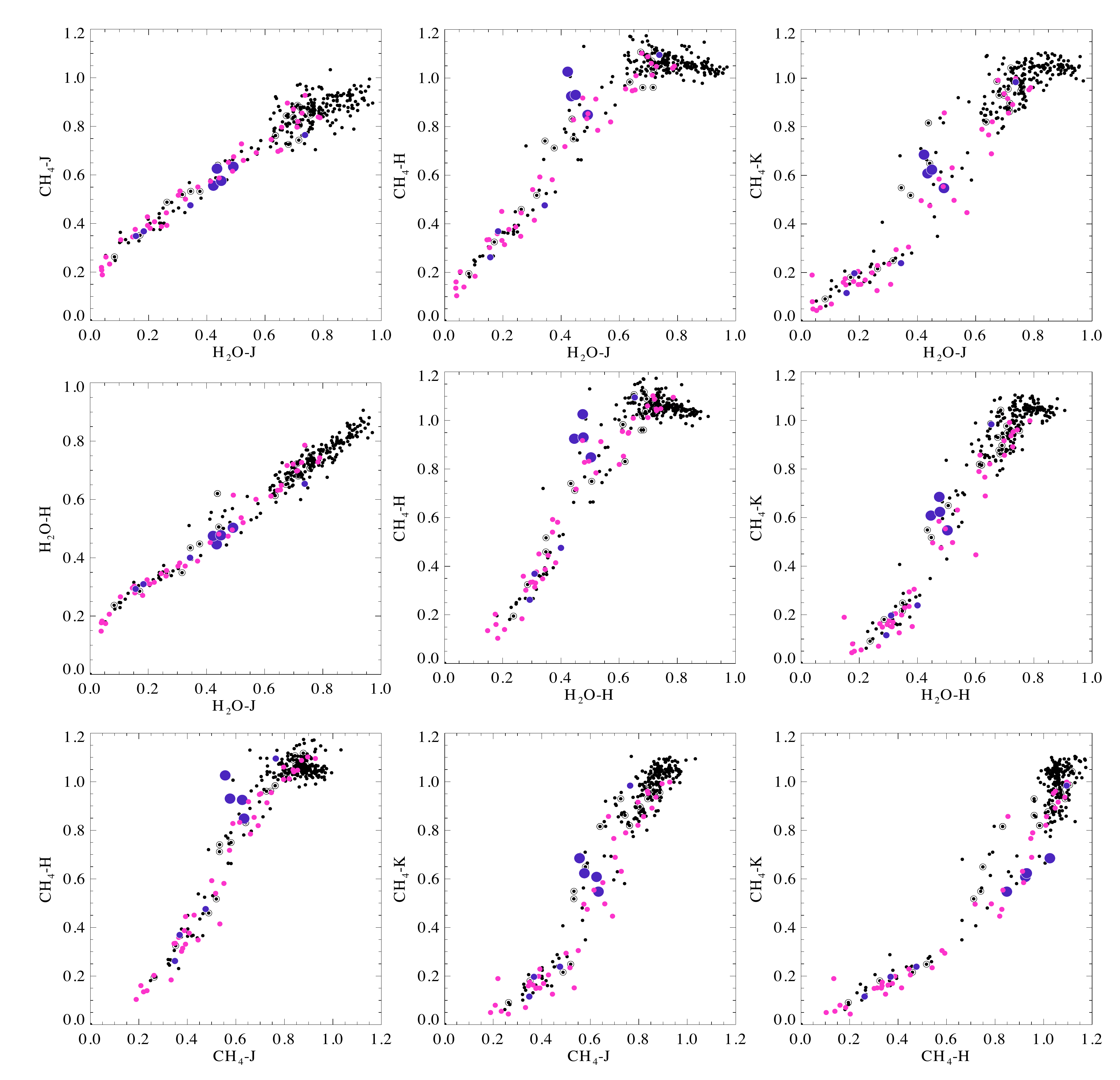} 
\caption{ Spectral indices as defined in \citet{burgasser06} for L5-T9 dwarfs with spectra available in the SpeX Prism Library (black dots).  Symbol size has been scaled with amplitude for variable targets.  Non-variable survey targets are indicated by filled pink circles.  Large purple symbols mark the large-amplitude variables inside the L/T transition, while small purple symbols show the low-amplitude variables outside the transition.   Resolved binaries are circled.  \label{fig:indices}}
\end{figure*}

In order to examine spectral trends in our sample, we downloaded all publicly available L and T dwarf spectra from the SpeX Prism Spectral Libraries\footnote{{ http://pono.ucsd.edu/$\sim$adam/browndwarfs/spexprism/} }, and computed spectral indices defined in \citet{burgasser06}.    All high-amplitude variables in our sample have spectra in the archive, with the exception of SIMP1629$+$03, for which we used the indices reported in \citet{deacon11}.   Plots of different combinations of spectral indices are shown in figure \ref{fig:indices}.  For the most part, the high amplitude variables have unremarkable spectral features.  The only exception is $H$-band CH$_4$ absorption (measured using the CH$_4-H$ index), which appears to be somewhat red on average for the early-T dwarf variables.   Weak methane absorption in the $H$-band of 2M2139$+$02 (our strongest variable, and largest CH$_4-H$ outlier) was noted by \citet{radigan12} who attribute it to low gravity and/or thick clouds, but caution that in the absence of any other indicators of youth, unusual cloud properties rather than surface gravity are likely responsible.    
Alternatively, since atypically strong CH$_4$ absorption in the $H$-band can be a signature of spectral binaries \citep{burgasser10}, this trend could in part be due to a selection bias for single rather than binary BDs.

\subsection{Variability Statistics}
\label{sect:stats} 

Here we examine occurrence rates of variability within our sample.   Of 62 targets monitored for variability, only 57 are considered in the statistical analysis:  16 L9-T3.5 dwarfs spanning the L/T transition color shift, bracketed by 15 L4-L8 dwarfs and 26 T4-T9 dwarfs.  While the L/T transition is traditionally defined to span $\sim$L8-T5 spectral types (e.g. see section \ref{sect:intro}), we find that in our sample this range includes both L dwarfs at the bottom of the dimming/reddening branch of the CMD that are predominantly cloudy, as well as T dwarfs that are predominantly clear based on relatively red and blue $J-K_{\rm s}$ colors respectively.  Since we wish to test whether variability is associated with the dissipation of clouds at the L/T transition, we have limited our ``L/T transition" bin to L9-T3.5 spectral types.  This more narrow grouping contains objects with truly intermediate $J-K_{\rm s}$ colors, indicative of intermediate levels of condensate opacity (see figure \ref{fig:spt_col}).  We adopt this more narrow definition of L/T transition spectral types in our analysis hereafter.

 We have excluded targets whose light curves were flagged for quality in the reduction pipeline (SD0423-04 and SD1416+13), as well as resolved binaries 2M0518$-$28 \citep{burgasser06_tbin}, 2M1404$-$31 \citep{looper08}, and 2M2052$-$16 \citep{stumpf11}.\footnote{SD0423$-$04 would have also qualified for exclusion due to binarity if not already flagged for quality.}   Binaries contaminate the sample for two reasons.  First, since all targets are unresolved in our observations, if variability were observed for a known binary it would be unclear which component were responsible.  Second, our sensitivity to a given amplitude for the (unknown) variable component is not well defined due to the constant flux contribution of the non-variable component.    An exception was made for the highly unequal mass T2/T8 binary 2M1209$-$10 \citep{liu10}, which has an unresolved spectral type of T3.  This target was  kept in the statistical sample since the T2 primary component contributes 80\% of the total flux in the $J$ band.  Therefore, our sensitivity is not greatly affected by the companion and the primary would be the likely source of any observed variability.  Similarly, using kernel phase interferometry, \citet{pope13} found the L5 target 2MASS~J19360187$-$5502322 to be a candidate binary with a contrast ratio of $\sim$18 in the $J$ band.  In this case the flux contribution of the possible companion is nearly negligible for the purpose of our survey.

Since resolved systems do not account for all binaries, we expect some remaining and unavoidable contamination of our sample.  We have not excluded targets that are candidate or suspected binaries based on spectral analysis \citep{burgasser10} or over luminosity \citep[2M0939$-$24 and 2M0559$-$14;][]{burgasser08_0939,dahn02}, as this would potentially bias our sample by excluding targets whose unusual characteristics are due to peculiar atmospheric properties, rather then unresolved binarity.

According to Baye's Theorem, we can write the probability of some true occurrence rate of variability, $\nu$, as:

\begin{equation}
P(\nu | D)  \propto   P(D|\nu) P (\nu)
\label{eq:eq3}
\end{equation}

where $D$ is shorthand for data, and $P(\nu)$ is the prior probability of $\nu$, which we assume to be flat on $[0,1]$, and zero everywhere else.  This leaves us to calculate the probability of our data given a true occurrence rate, $P(D | \nu)$.   Since our observed sensitivity differs significantly from target to target we must determine the individual probabilities for each target light curve, and take the product

\begin{equation}
P(D|\nu) = \prod_i P(D_i | \nu)
\label{eq:eq4}
\end{equation}

For the $i^{\rm th}$ light curve, the probability of a detection is equal to the probability that the source is variable, $P({\rm variable}|\nu)$, multiplied by the probability that the signal is detected, $f_{\rm sens}(\sigma_i,\Delta t_i)$, where the sensitivity is a function of measurable light curve properties.

The probability of a source being variable is simply given by the binomial distribution for $k=1$ and $N=1$, where $k$ is the number of occurrences, and $N$ is the number of events, which is given by  $P({\rm variable}|\nu)=\nu$.  Now let us define a detection parameter, $x_i$, that is equal to one in the case of a detection and zero otherwise. We can then write an expression for the probability corresponding to the $i$th light curve as

\begin{equation}
\label{eq:eq5}
P(D_i|\nu) =
\begin{cases}
\nu f_{\rm sens}(\sigma_i,\Delta t_i) &  \text{if }  x_i=1 \\
1-\nu f_{\rm sens}(\sigma_i, \Delta t_i) & \text{if } x_i=0
\end{cases}
\end{equation}

Determining the probability distribution of the true occurrence rate, $\nu$ is then a simple matter of determining $x_i$  and $f_{\rm sens}(\sigma_i,\Delta t_i)$ for each target, and substituting these values in equations \ref{eq:eq3}-\ref{eq:eq5}.

In the above method, we have neglected one important step.  In section \ref{sect:sens}, detection sensitivities are computed as a function of the peak-to-peak sinusoidal amplitude: $f_{\rm sens}(\sigma_i,\Delta t_i, A)$.  Therefore, in order to obtain   $f_{\rm sens}(\sigma_i,\Delta t_i)$ we must marginalize over amplitude in some reasonable way.  In practice, we will restrict our analysis to specified ranges in amplitude and assume that within these finite ranges, the amplitude distribution is flat.  This is described in further detail for the specific cases explored below.

\subsubsection{A statistically significant increase in variability at the L/T transition}
The results of our survey are summarized in figure \ref{fig:spt_col} where detections of variability are shown on a diagram of NIR spectral type versus $J-K_{\rm s}$ color.  This figure makes a compelling case for an increase in large-amplitude variability (peak to peak amplitudes $\ge$2\%) within the L/T transition region of the diagram (taken here to mean L9-T3.5 spectral types).   More formally, we can obtain probability distributions for the true occurrence rate of large-amplitude variability inside and outside of the L/T transition using equations \ref{eq:eq3}-\ref{eq:eq5}.   If we consider only large-amplitude detections (i.e. taking $x_i$=1 only for detections with peak-to-peak amplitudes $\ge$2\%), then variability is detected exclusively at the L/T transition:  4/16 objects with L9-T3.5 spectral types, compared to 0/41 objects at all other spectral types. 
We assume that our sensitivity to peak-to-peak amplitudes $\ge$0.02 is approximated by the average sensitivity to amplitudes between 0.02 and 0.05, given by  

$$
f_{\rm sens}(\sigma_i,\Delta t_i)= \frac{\int_{0.02}^{0.05} f_{\rm sens}(\sigma_i,\Delta t_i,A)dA }{ \int_{0.02}^{0.05}dA}
$$ 

We do not know the true distribution of amplitudes, $f(A)$, and this marginalization assumes that $f(A)$ is flat within the specified amplitude range.  The true distribution likely peaks at low values in this range, and drops off to zero beyond this range; therefore our assumption of a flat distribution from 0.02 to 0.05 (where the majority of variables were found) is probably a reasonable one.  
More importantly, it turns out that the choice of marginalization does not strongly influence our final result, or the significance of our findings.

\begin{figure*}[ht!]
 \begin{tabular}{cc}
\includegraphics[width=0.45\hsize]{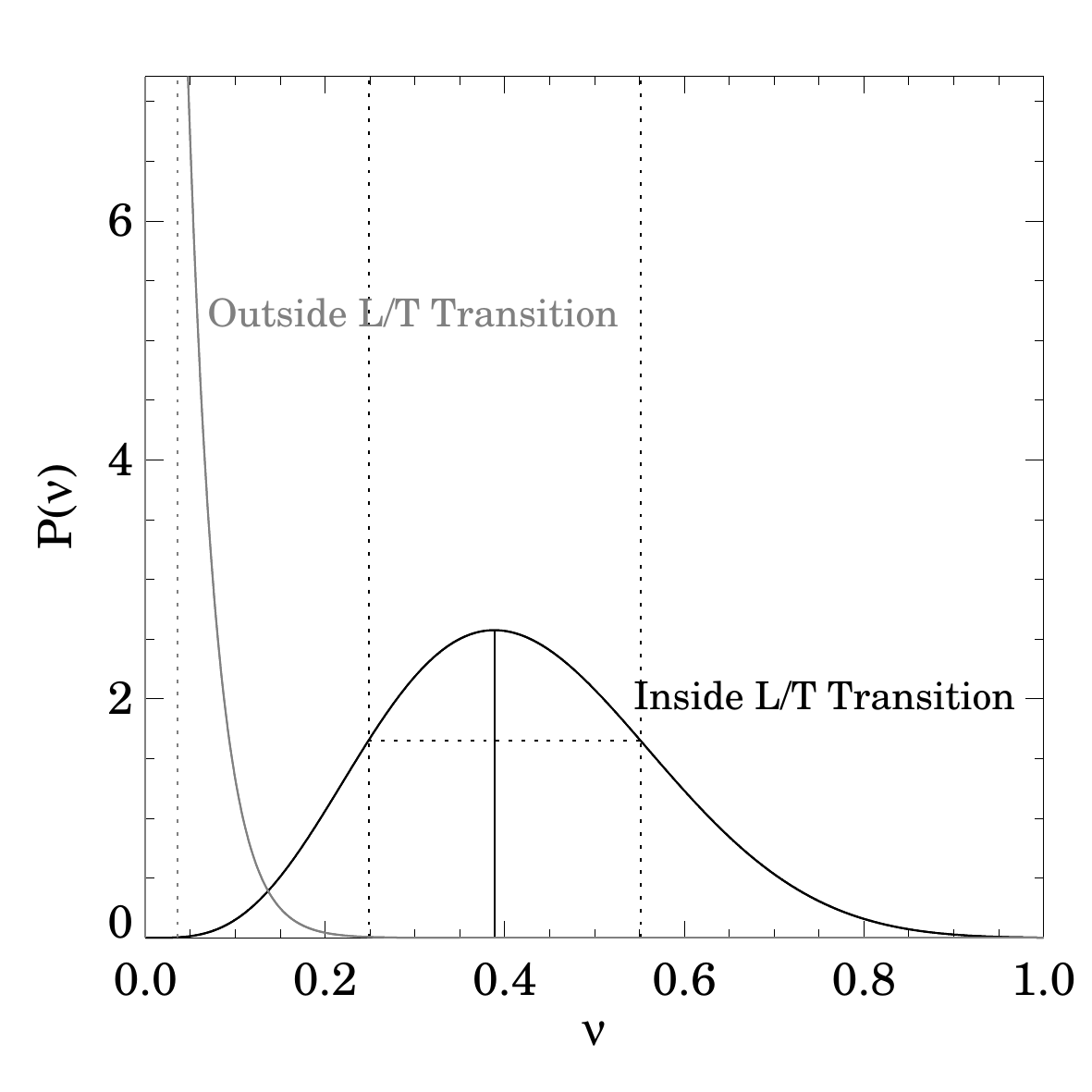} &
\includegraphics[width=0.45\hsize]{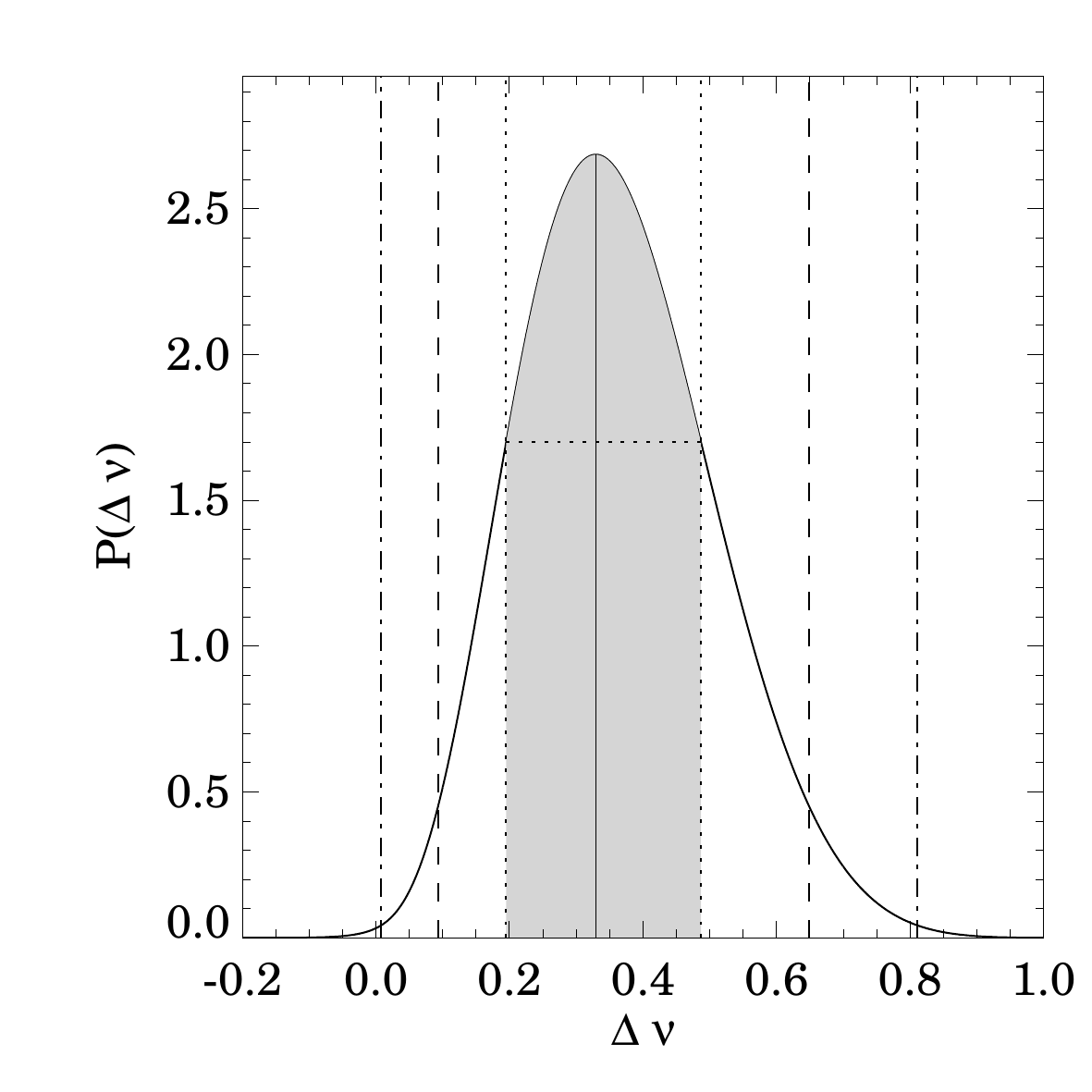}
\end{tabular}
\caption[Probability distributions for the observed frequencies of high amplitude ($>$2\% peak-to-peak) variability for objects inside and out of the L/T transition ]{{\em Left:}Probability distributions for the observed frequencies of high amplitude ($>$2\% peak-to-peak) variability for objects inside (solid black line) and out of (solid grey line) the L/T transition (defined here as L9-T3.5 spectral types).  {\em Right:} Probability distribution for the difference in observed frequency inside and outside of the transition.  This distribution is obtained by integrating over the two distributions on the left according to equation \ref{eq:eq_prob}.  Dotted, dashed and dash-dotted lines indicate the shortest 68\%, 95\% and 99.7\% credible intervals respectively.  The ``shortest'' credible interval (as opposed to an equal-tail interval) is constructed such that probabilities inside the intervals are uniformly larger than those outside. \label{fig:prob}}
\end{figure*}

Probability distributions for the occurrence rate of high-amplitude variability both inside and outside of the L/T transition are shown in the left panel of figure \ref{fig:prob}.  In order to quantify any difference in $P(\nu)$ between these populations, a new distribution $P(\Delta \nu)$ was constructed according to

\begin{equation}
P(\Delta \nu)=\bigintssss_{-1}^{+1} P_{\rm out}(\nu) P_{\rm in}(\nu+\Delta\nu)d\nu 
\label{eq:eq_prob}
\end{equation}

where $P_{\rm out}$ and $P_{\rm in}$ are the distributions for $\nu$ outside and inside the L/T transition respectively.  The distribution, $P(\Delta \nu)$,  is shown in the right panel of figure \ref{fig:prob}.   We are able to exclude the null hypothesis that $\Delta \nu=0$ with $>$99.7\% confidence.  If we compare only the L4-L8 sample and the L9-T3.5 sample, we can exclude $\Delta \nu=0$ at $>$95.4\% confidence.  Therefore, the probability that the observed occurrence rates for large-amplitude variability inside and outside the L/T transition originate from the same underlying true occurrence rate, $\nu$, is excluded with high confidence. This work establishes for the first time a statistically significant increase in (high amplitude) variability for L9-T3.5 spectral types.

\subsubsection{The frequency of strong variability inside the L/T transition}

In our sample, we found 4/16 or 25\%  of L9-T3.5 
dwarfs to be significantly variable with peak-to-peak amplitudes $>$2\%, with a probability distribution for the occurrence rate of variability shown in figure \ref{fig:prob}.  From this distribution we find an occurrence rate of $39^{+16}_{-14}$\% 
for high amplitude variability within the transition.  
This raises an important question:  if patchy clouds are responsible for observed properties of the L/T transition such as $J$-band brightening, then why aren't all L/T transition objects variable?  There are several factors that could lead us to underestimate the true rate of variability, which we discuss in the sections below.

\subsubsection{The inclination distribution of targets}
 If a target is inclined by an angle $i$ to the observer, the integrated light spectrum composed of cloudy and clear regions is not greatly affected, yet we might expect the observed variability amplitude to be smaller by a factor of $\sin{i}$.  If we assume an isotropic distribution of spin axes, then the distribution of inclinations is $f(i)\propto \sin{i}$ which yields a  distribution of projected amplitudes of $f(\sin{i})\propto \tan{i}$.  We can integrate over this distribution of projected amplitudes to find our reduced sensitivity to the maximum (i.e. $i$=90$^\circ$) amplitude
 
\begin{equation}
f_{\rm sens}[\sigma_{a},\Delta t,A_{i=0}]=\int_{0}^{1} f_{\rm det}[\sigma_{a},\Delta t, A\sin{i}] f(\sin{i})d(\sin{i})
\label{eq:eq6}
\end{equation}
 
Our reduced sensitivity due to projection effects is shown in the right panel of figure \ref{fig:sens}.  The occurrence rate for variability $\ge$2\% adjusted for inclination angle is $53^{+21}_{-18}$\% (figure \ref{fig:prob2}), which suggests that the majority of L/T transition targets would be variable with peak-to-peak amplitudes above $2\%$ if observed edge-on.

\subsubsection{Large variability is limited to a sub-range of the L/T transition}
Within the L9-T3.5 sample, variability is confined to objects with $J-K_s$ colors between 0.9 and 1.3.   With a $J-K_s$ color of 1.46$\pm$0.04  \citep{kniazev13}, the variable T0.5 dwarf Luhman~16B suggests that variability extends to redder objects than captured in our sample.  Nonetheless, it is possible that large-amplitude $J$-band variability due to patchy clouds peaks in a very narrow color range. 

\begin{figure*}[ht!]
 \begin{tabular}{cc}
\includegraphics[width=0.45\hsize]{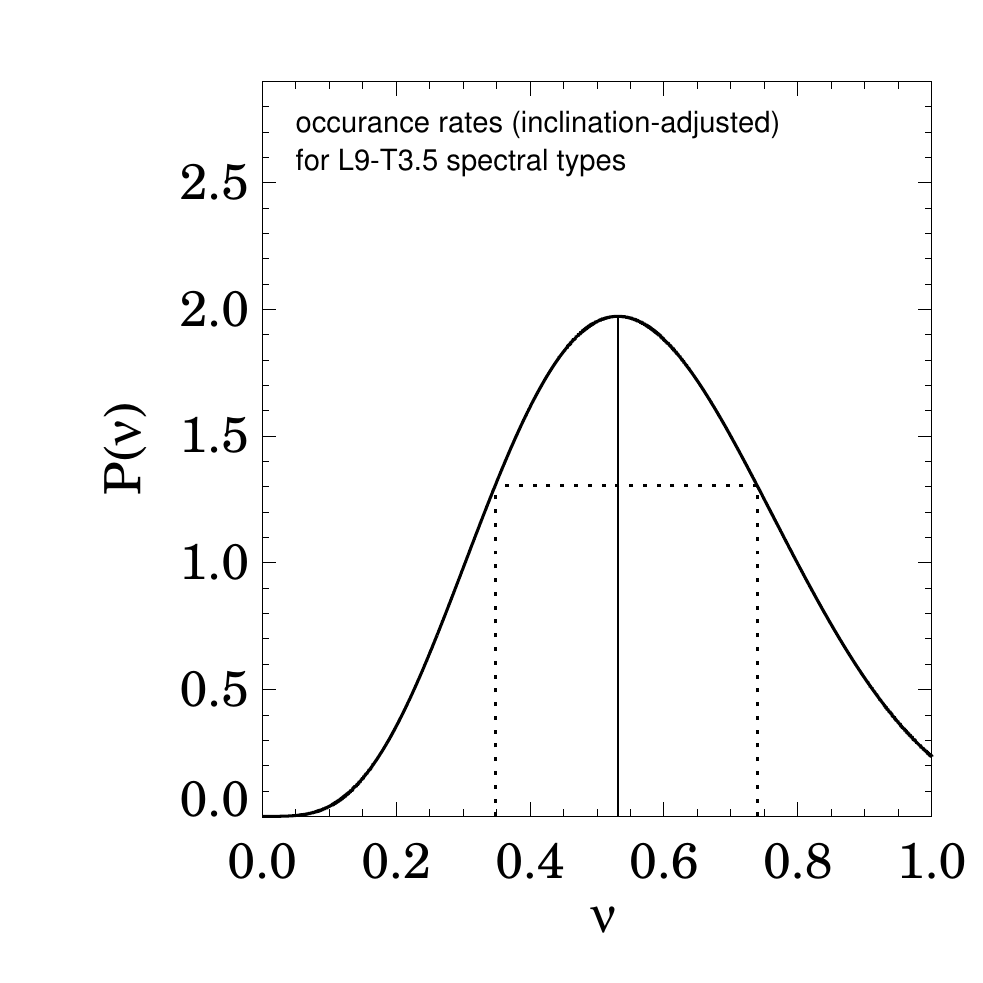} &
\includegraphics[width=0.45\hsize]{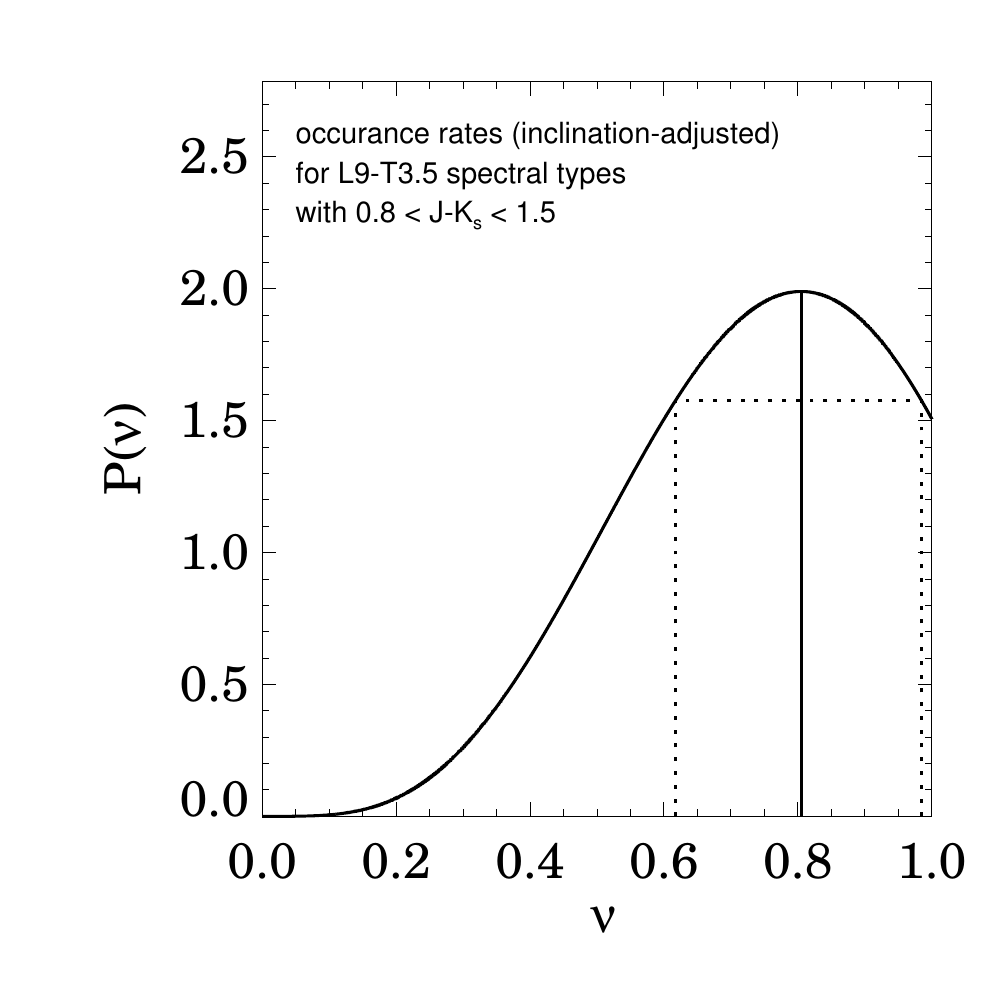}
\end{tabular}
\caption{Probability distributions for the occurrence rate of high-amplitude ($>2$\%) variability, when a target is observed at a 90 degree inclination, within the L/T transition.  {\em Left:} Occurrence rates for the L9-T3.5 sample.  {\em Right:} Occurrence rates for the L9-T3.5 subsample wherein $0.8>J-K_s>1.5$.   Dotted lines indicate the shortest 68\% credible intervals. \label{fig:prob2}}
\end{figure*}

If we confine ourselves to the subsample of L9-T3.5 dwarfs with $0.8<J-K_s<1.5$, we have an observed variability fraction of 4/11.  From equations \ref{eq:eq3}-\ref{eq:eq5}, and adjusting for target inclinations using equation \ref{eq:eq6}, we find an occurrence rate for large-amplitude variability (for targets viewed edge-on) of $80^{+18}_{-19}$\%.  The inferred distribution is shown in figure \ref{fig:prob2}, and is not incompatible with a 100\% occurrence rate of variability in this sub-range.

\subsubsection{Binarity, azimuthal symmetry, and duty cycle}
Binaries at the L/T transition can potentially dilute the sample with individual components that fall outside the transition region. While we have excluded known binaries from our statistical sample, it is likely that unresolved multiples remain.  Just over half of our targets have  reports of high resolution imaging in the literature, indicated in table \ref{tab:lc_info}.   Several high angular resolution  imaging studies of the brown dwarf binary fraction conclude that  $\sim10-15$\% of brown dwarfs are binaries with separations $\gtrsim$2-3\,AU \citep[e.g.][]{burgasser03,burgasser06,reid08,bouy03}.  Radial  velocity studies of young brown dwarfs have demonstrated that there may be just as many unresolved binaries at small separation as are found at large separation, meaning that up to 30\% of objects are binaries. In addition, there have been suggestions that the binary fraction may be even larger (up to 67\%) inside the L/T transition due to the rapid evolution of objects through these spectral types \citep{burgasser07_ltbin,burgasser10}, although similar numbers of binaries have been {\em resolved} at L9$-$T4 spectral types as are found outside this range \citep{goldman08,radigan13}.    Therefore, we might guess that up to $\sim$15-30\% of remaining L9-T3.5 targets are unresolved multiples (and more than this if the largest estimates of the L/T transition binary frequency are correct).

\citet{burgasser10} found three of our non-variable L/T transition targets to be candidate binaries based on spectral template fitting.  The T0 dwarf 2M1511+0607 was deemed to be a strong candidate, meeting all 6 of the authors' selection criteria, and was fit significantly better by a composite L5/T5 template compared to any single templates.  This object has a high probability of being a binary, and likely contaminates our sample.  The T2 dwarf 2M0949-15 and the T0 dwarf 2M1207+02 were flagged as weak binary candidates.  No high angular resolution imaging has been reported for these targets to date.  While these sources are potential binary contaminants of our sample, we point out that two of our variable targets, SD0758$+$32 and 2M2139$+$02, were also flagged as weak and strong binary candidates respectively by \citet{burgasser10}.  High angular resolution imaging of SD0758+32 in \citet{radigan13} and archival HST imaging of 2M2139$+$02 reported in \citet{radigan12} failed to resolve any companions.  Furthermore, a preliminary parallax for 2M2139$+$02 by \citet{smart13} implies an absolute magnitude that is consistent with this target being a single brown dwarf.  Given that these sources remain unresolved in high angular resolution images, and display large variability (which would be washed out by the presence of an equal luminosity companion), we suggest they are most likely single early T-dwarfs.

We further note that some objects with patchy clouds may not be variable simply due to a high degree of azimuthal symmetry in their atmospheres (e.g. the banding of clouds). And finally,  as cloud patterns have been shown to evolve \citep{artigau09,radigan12,metchev13_proc}, the variable fraction may provide an indication of the duty cycle for large asymmetries in cloud patches, rather than a frequency for cloud patchiness.  One of our targets in particular, SD0758+32, was observed by \citet{girardin13} and found to be non-variable above $\sim$5-10\,mmag over the course of their observations.

In summary, we find that target inclinations, azimuthal symmetry and binarity, as well as the spectral type and color range considered may explain the low observed occurrence rate for strong variability at the L/T transition, even if patchy clouds can explain the spectral morphologies for all single objects in this regime.

\subsection{Variability Outside the L/T transition}
We report 5 detections of low-amplitude variability outside the L/T transition ($<$2\% peak-to-peak).  
The fraction of low-amplitude variables in our sample, excluding the L/T transition, is 5/41.  If we assume a flat distribution of amplitudes from $0.5-1.6$\%, we can use equations \ref{eq:eq3}-\ref{eq:eq5} to obtain a probability distribution for the true occurrence rate of variability at these amplitudes.\footnote{The top of this range corresponds to the largest amplitude observed outside the L/T transition.  The bottom of this amplitude range corresponds to an average detection sensitivity of 3.7\% (see figure \ref{fig:sens}), which means we could have detected such a low amplitude in approximately one of the 41 targets in the non-L/T transition sample.  We do not consider smaller amplitudes, for which we would expect less than one detection in the entire sample.}   We estimate that $60^{+22}_{-18}$\% of mid-to-late  L  and T dwarfs (excluding L9-T3.5 spectral types) will vary with amplitudes between 0.5\% and 1.6\% in the $J$-band.  If we account for target inclinations using equation \ref{eq:eq6}, the inferred variability fraction of targets viewed edge-on peaks at 100\%, with lower limits of 70\% (44\%) corresponding to the 68\% (95\%) credible intervals.   Due to the low number of detections and our low sensitivity to small signals, the distribution we find is quite broad, but nonetheless suggests that most brown dwarfs are in fact variable if observed with high enough precision, over long enough time baselines.  Space-based observations using the Spitzer Space telescope by Metchev et al. 2014 (in prep) probe the frequency of low-level variability outside the L/T transition with greater sensitivity, and reach similar conclusions.

For objects outside of the L/T transition, only one of the detected variables is an L dwarf, while the other four variables are T-dwarfs.  This might suggest that variability is more common or of higher amplitude in T-dwarfs.  However, our sample size of mid-to-late T dwarfs was approximately double that of the L dwarfs, and two additional marginal detections of variability outside the L/T transition (see table \ref{tab:detections}) both occur for L-dwarfs, which (if real) would balance out the relative numbers.  Further observations will be required to determine if T-dwarfs are indeed a more variable population.

Finally, we find that strong $J$-band variability (peak to peak amplitudes $>$2\%) outside of L9-T3.5 spectral types must be comparatively rare.  While our conclusion echoes those of \citet{koen04b} and \citet{clarke08} who find no evidence for $J$-band variability $\gtrsim$20\,mmag in a combined sample of 24 early L to mid T dwarfs falling predominantly outside of the L/T transition, it differs from a recent study by \citet{wilson14} who find strong variability to be more evenly distributed as a function of spectral type in a sample of 69 L0-T8 dwarfs.\footnote{The variability study of \citet{wilson14} was published during the late stages of the review of our manuscript, and a more comprehensive comparison is beyond the scope of this work.}

\section{Summary and Conclusions}
\label{sect:concl} 

This work reports the most extensive and sensitive survey of brown dwarf variability to date at NIR wavelengths.  Of 57 targets included in our final statistical sample---monitored for variability in continuous sequences of $\sim$2-6\,hr---we detected significant variability ($p>$99\%) in 9 objects.  Of these, 5 objects (including 2M2139$+$02, for which follow-up observations were reported in \citet{radigan12}) are reported to be variable for the first time here.  For L9-T3.5 spectral types, 4/16 objects were found to be variable with peak-to-peak amplitudes ranging from 2.9\% to 9\%, while 5/41 targets outside the transition exhibit low-level variations with amplitudes ranging from 0.6\%$-$1.6\%.  The major conclusions of this work can  be summarized as follows:

\begin{enumerate}

\item We find a statistically significant increase in strong $J$-band variability (peak-to-peak amplitudes larger than 2\%) within the L/T transition (L9-T3.5 spectral types) at $>$99.7\% confidence. 
\item We infer that $39^{+16}_{-14}$\% of L9-T3.5 dwarfs are periodic variables with peak-to-peak amplitudes $>$2\%.

\item In our sample, sinusoidal signals with peak-to-peak amplitudes $>$2\% are confined to a narrow region of the L/T transition consisting of early T-dwarfs with  intermediate $J-K_{\rm s}$ colors.  
The flux contrast  between clouds and clearings (or thick and thin cloud patches) may peak within this narrow range.  
\item Correcting for target inclinations, we infer that $80^{+18}_{-19}$\% of L9-T3.5 dwarfs with $0.8>J-K_{\rm s}>1.5$ would be variable if viewed edge-on.  Azimuthal symmetry of cloud patches or binary contaminants may account for remaining non-variable objects. This is consistent with the majority of early L/T transition dwarfs having high contrast cloud patches in their atmospheres (see section \ref{sect:stats}).  
\item It follows from conclusion 4, that the development of spatially heterogeneous clearings or thin-cloud regions may contribute to the abrupt decline in cloud opacity and $J$-band brightening observed at the L/T transition.
\item We identify a tentative correlation between strong variability and weak $H$-band CH$_4$ absorption for early T-dwarfs.   
Otherwise, the variable targets have rather unremarkable NIR spectra (see discussion in section \ref{sect:properties}).
\item Outside the L/T transition we estimate that $60^{+22}_{-18}$\% of targets may vary with amplitudes of 0.5-1.6\%, suggesting that surface heterogeneities (e.g. cloud patches of lower contrast) are common among L and T dwarfs.   
\item The largest detections of variability outside the L/T transition were made for an unusually blue L6.5 dwarf (1.2\% peak-to-peak) and a somewhat red T6.5 dwarf (1.6\% peak-to-peak), suggesting that cloud patchiness (clearings in the former case, and residual clouds in the latter case) may influence emergent spectra for a wide range of spectral types.  However, we also detect lower-level variability (0.6\%-0.9\%) in 3 T-dwarfs with unremarkable NIR colors.  Further monitoring of will be required to determine the significance of any color trends outside the L/T transition.   
\item Finally, the detection of $J$-band variability in four mid-T dwarfs provides  non-spectroscopic evidence for the persistence of clouds---potentially composed of salts and sulfides--- in late-type objects.  Temperature perturbations may provide an alternate explanation, and multi-wavelength monitoring will be required to distinguish between variability mechanisms for late-type brown dwarfs for which silicate clouds are expected to have settled below the photosphere. 
\end{enumerate}

For the first time here, we report a statistically significant correlation between high amplitude variability and L/T transition spectral types.  This result suggests that cloud dispersal at the L/T transition proceeds in a spatially non-uniform manner, leading to localized regions of clouds and clearings (or alternatively thick and thin cloud patches).   Our observations therefore support the longstanding hypothesis of \citet{ackerman01} and \citet{burgasser02_lt} that the development of cloud holes plays an important role in the transition from cloudy to clear spectral types, and may contribute to puzzling properties of the transition such as $J$-band brightening.  This implies that substellar models of the L/T transition must include at minimum a multi-component surface model, and in the idealized case a fully 3D radiative hydrodynamical treatment.   First steps in this direction have been made by \citet{freytag10}, \citet{marley10}, \citet{showman13}, and \citet{xi14}.   More generally, fractional cloud coverage has been established as an additional parameter that may influence the emergent colors of brown dwarfs, and by extension directly imaged planets, especially those with L/T transition spectral types such as HR8799c \citep{marois08,oppenheimer13}.

Variable brown dwarfs, and notably the new population of high-amplitude variables at the substellar L/T transition, provide novel opportunities to constrain cloud properties and dynamics in cool atmospheres.   Surface variations in cloud thickness produce a chromatic variability signature, providing an unprecedented opportunity to probe cloud structure via multi-wavelength monitoring as in \citet{radigan12}, \citet{buenzli12}, \citet{apai13}, \citet{heinze13}, and \citet{biller13}.  And finally, mapping the evolution of features over multiple rotations will provide a way to study atmospheric dynamics in the high-gravity, non-irradiated regime.

\clearpage

\appendix

\section{A. Example light curves for targets and reference stars}
\label{app:A}
Example light curves for two non-variable targets (plus reference stars) with $J=$16.6 and $J=$15.6 are provided.  In addition, example light curves for a low-amplitude variable with $J$=14, and a sequence taken in bad weather conditions (variable extinction and seeing) are also provided.  In each example a series of plots are shown with the following layout.  Light curves for the target (top left) and up to 7 closest-brightness reference stars are shown directly below.  Light curves have been binned by a factor of 3-5.  A finding chart is shown in the top right with the target and reference stars labelled by number (associated with the ``ID'' field given with each light curve).  The target always has an ID of 0 and is circled on the finding chart.  In the lower right are diagnostic plots from top to bottom of: the sky brightness, median FWHM of all reference stars on-chip, first order (i.e. global) variations of all stars on the chip (e.g. due to changing atmosphere, instrumental, and/or procedural effects) and the relative light curve RMS as a function of magnitude of all stars on-chip.  In the latter plot the target is indicated by a black filled circle, while reference stars used to build the 1st order calibration curve are shown as open circles.  Reference stars excluded from the calibration curve are indicated as filled grey circles.

\clearpage
\begin{figure}[ht!]
\caption{Non-variable example (J=16.6): SDSS J104829.21+091937.8 and reference stars \label{fig:targ39}}
\begin{tabular}{cc}
\multirow{2}{*}[0in]{\includegraphics[width=0.41\hsize]{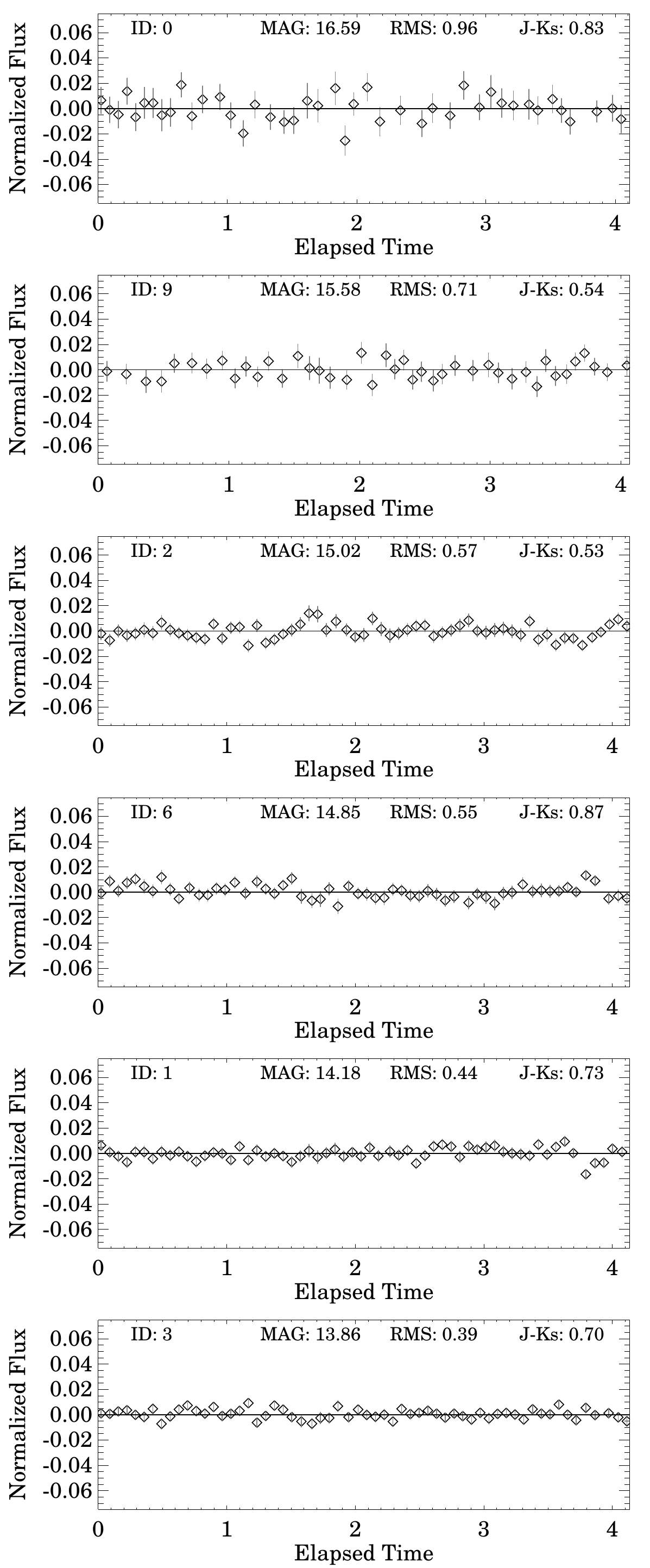}} &
\multirow{1}{*}[0.1in]{\includegraphics[width=0.5\hsize]{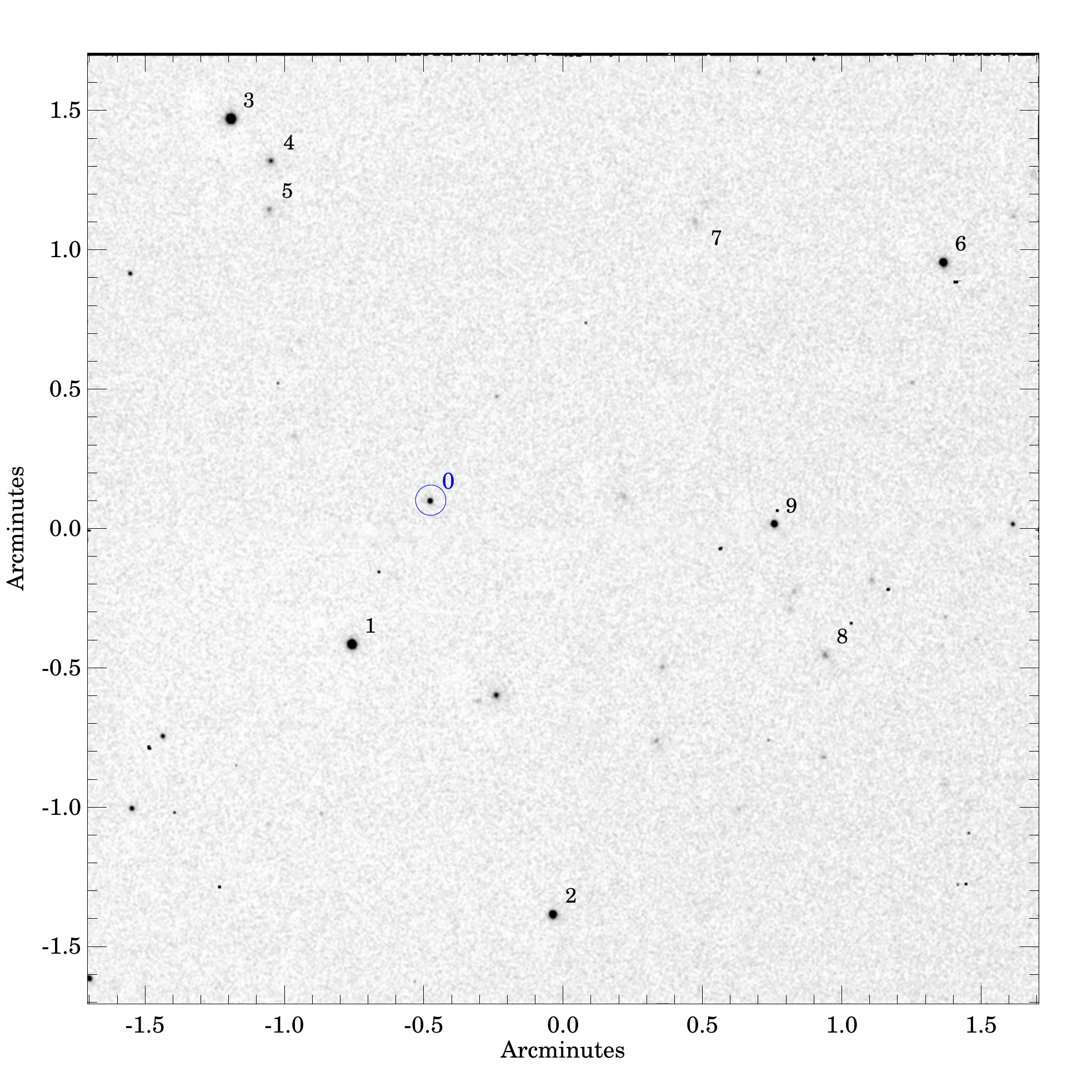}} \\
& \multirow{1}{*}[-3.4in]{\includegraphics[width=0.41\hsize]{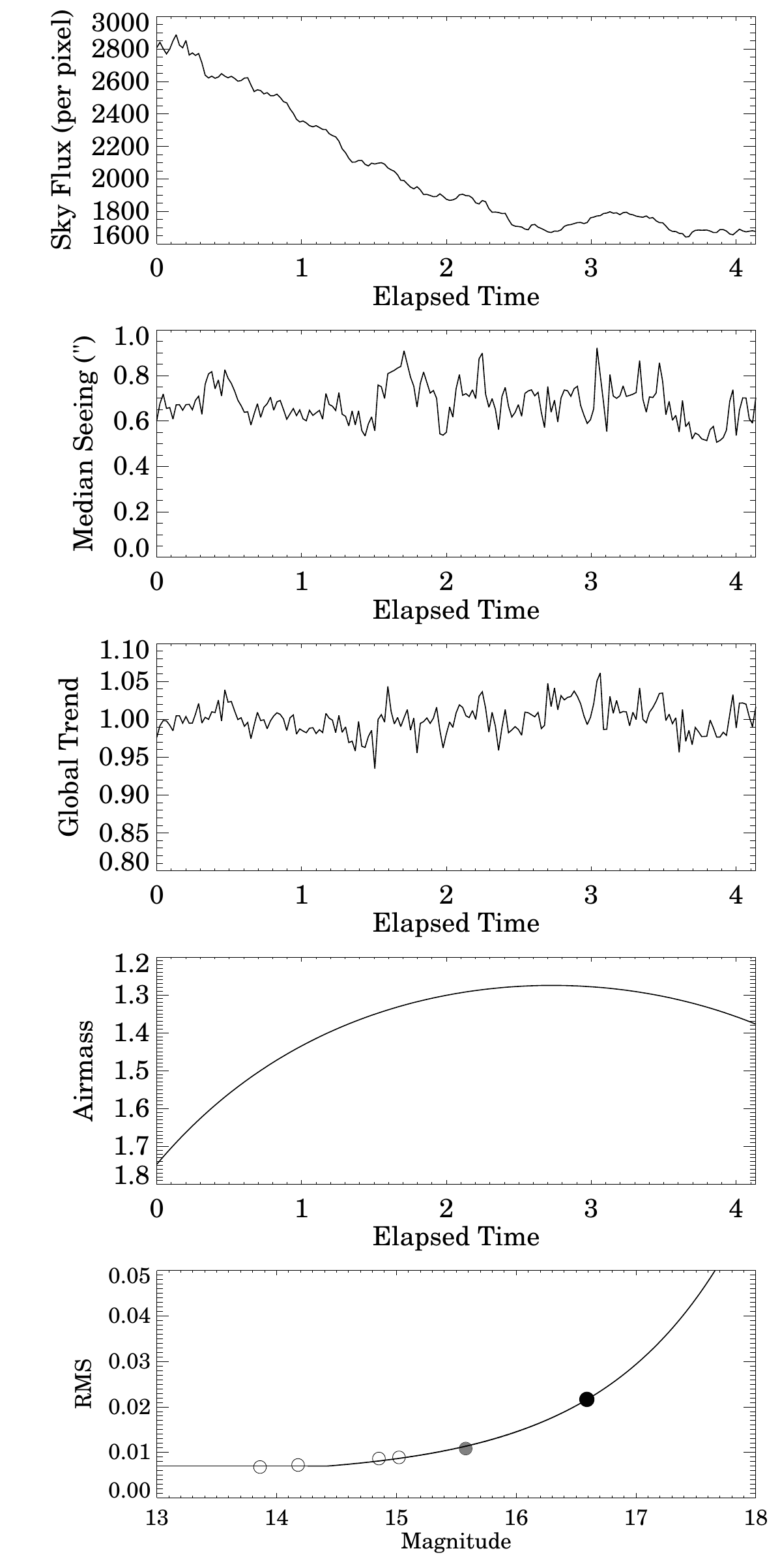}} \\
\end{tabular}
\end{figure}

 \clearpage
\begin{figure}[ht!]
\caption{Non-variable example (J=15.6)  2MASSI J1546271-332511 and references stars \label{fig:aplc2}}
\begin{tabular}{cc}
\multirow{2}{*}[0in]{\includegraphics[width=0.41\hsize]{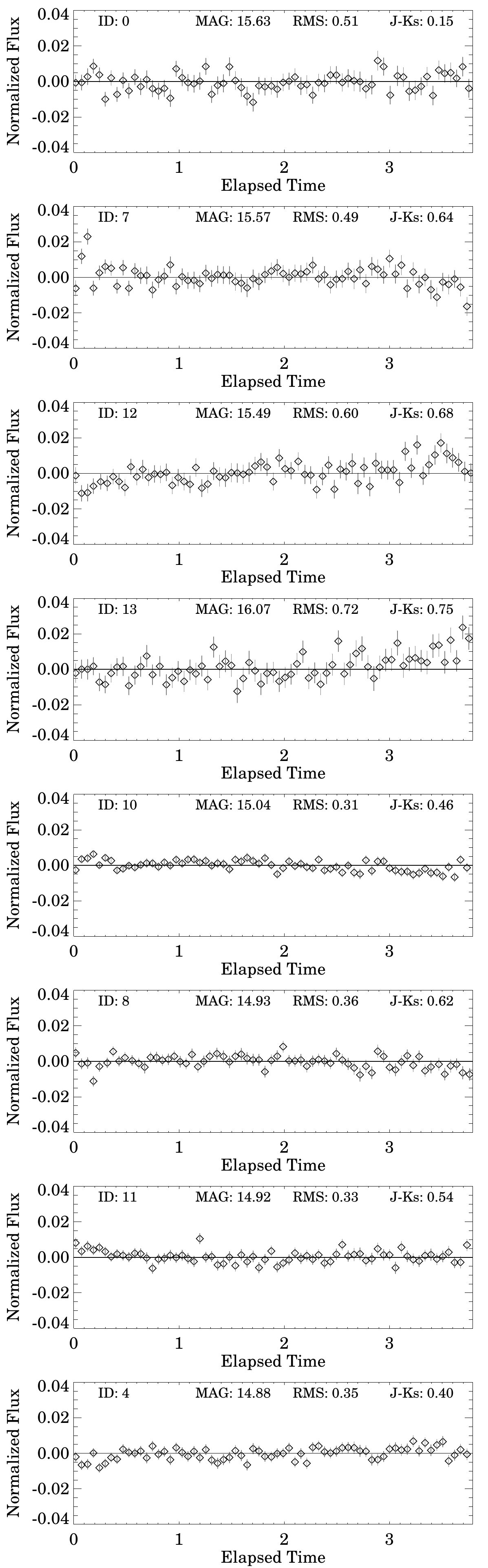}} &
\multirow{1}{*}[0.1in]{\includegraphics[width=0.5\hsize]{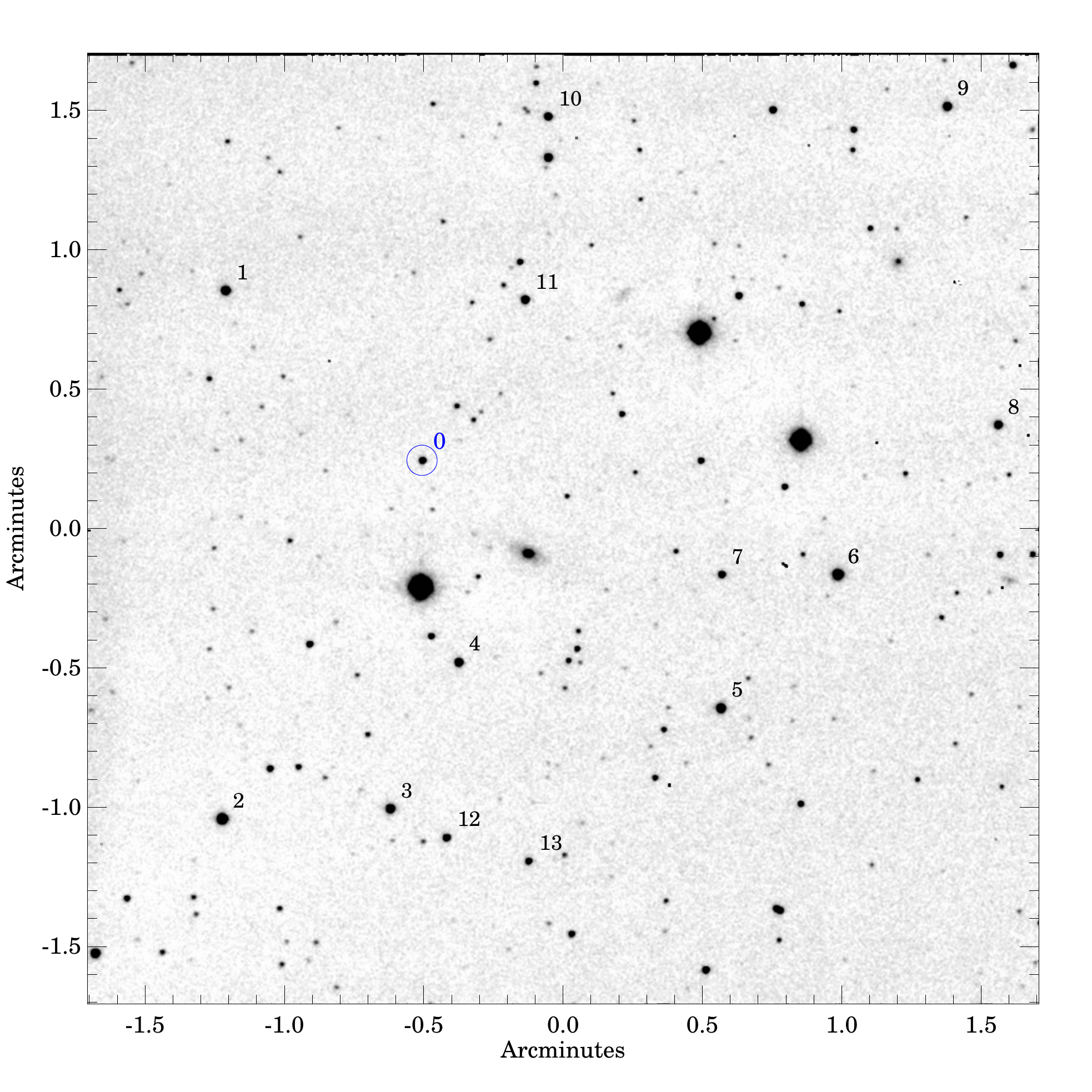}} \\
& \multirow{1}{*}[-3.4in]{\includegraphics[width=0.41\hsize]{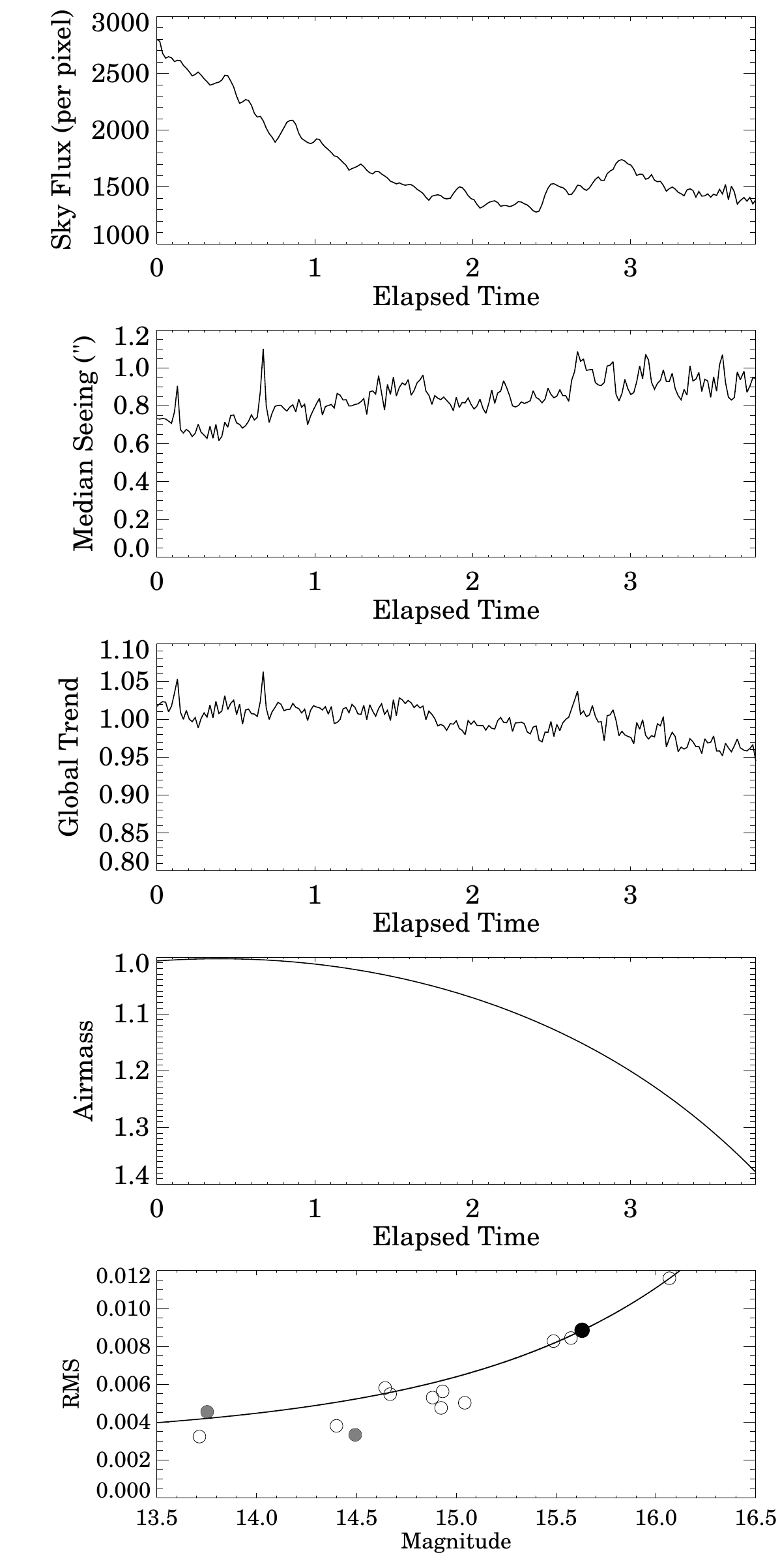}} \\
\end{tabular}
\end{figure}
\clearpage

\clearpage
\begin{figure}[ht!]
\caption{Low-amplitude example (J=14.0): 2MASS J11263991-5003550 and reference stars \label{fig:aplc2}}
\begin{tabular}{cc}
\multirow{2}{*}[0in]{\includegraphics[width=0.41\hsize]{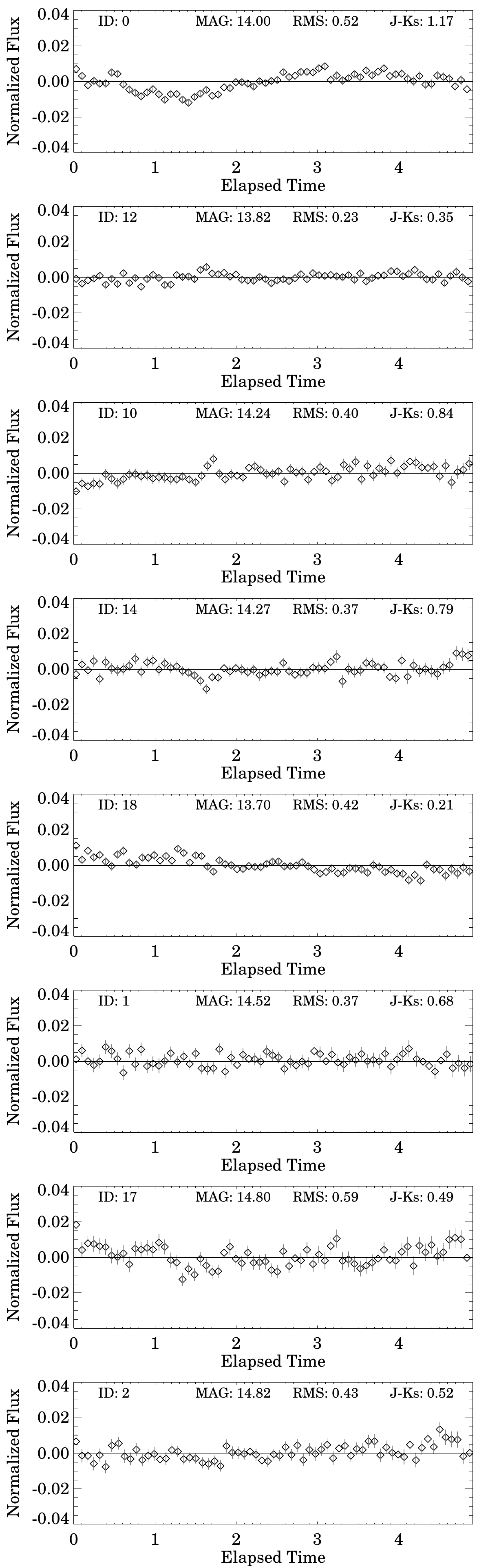}} &
\multirow{1}{*}[0.1in]{\includegraphics[width=0.5\hsize]{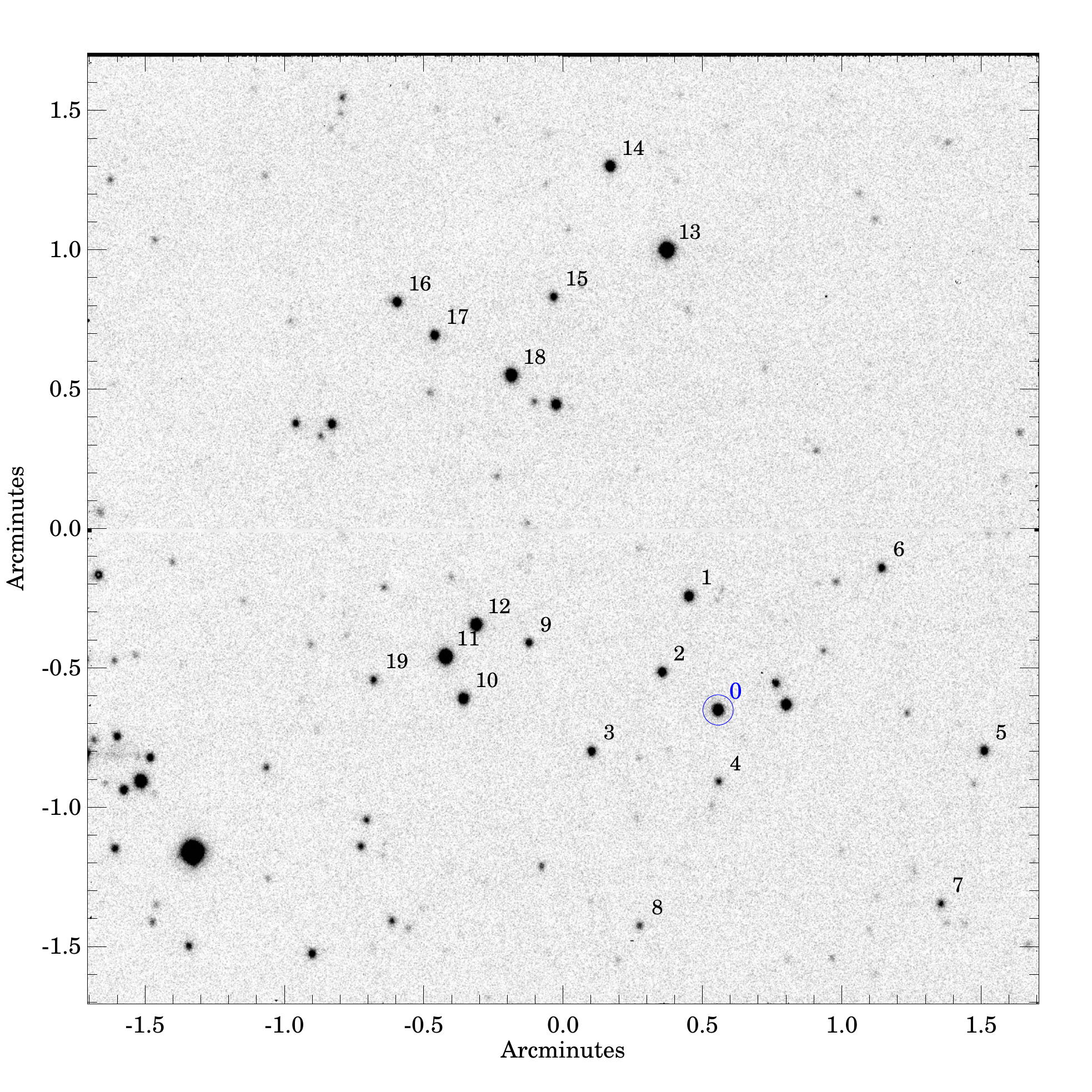}} \\
& \multirow{1}{*}[-3.4in]{\includegraphics[width=0.41\hsize]{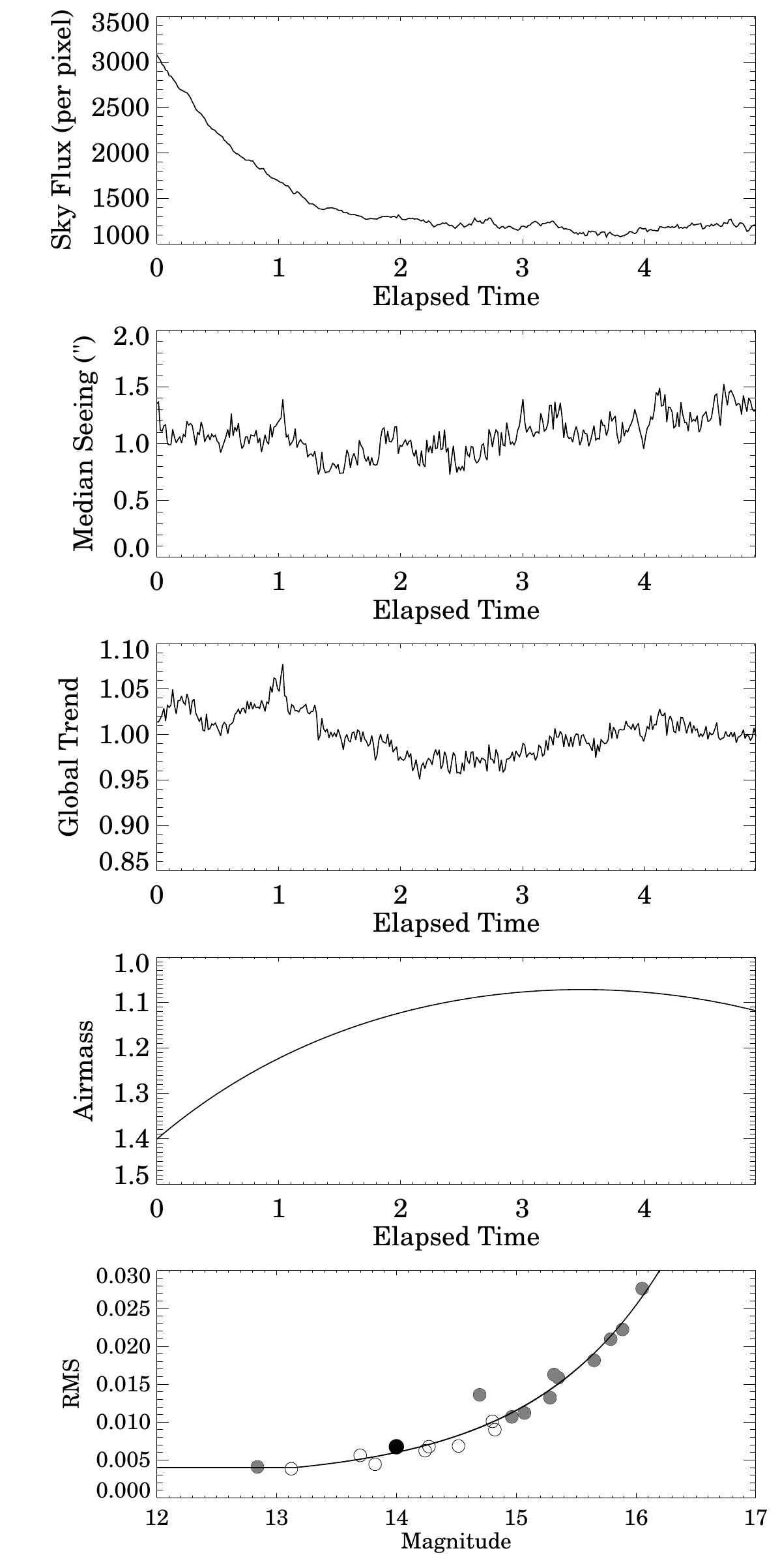}} \\
\end{tabular}
\end{figure}

\clearpage
\begin{figure}[ht!]
\caption{Bad weather example: SDSSp J125453.90-012247.4 (J=14.9) and reference stars \label{fig:aplc3}}
\begin{tabular}{cc}
\multirow{2}{*}[0in]{\includegraphics[width=0.41\hsize]{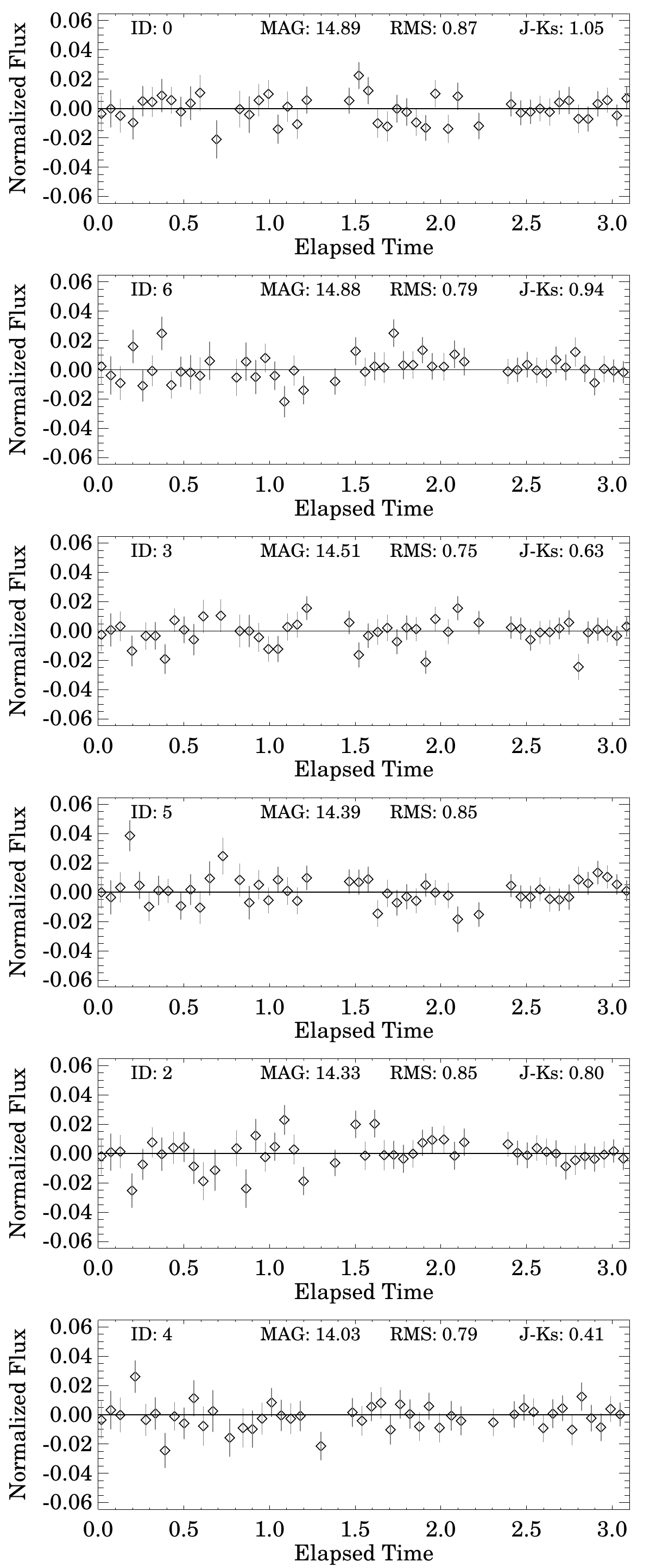}} &
\multirow{1}{*}[0.1in]{\includegraphics[width=0.5\hsize]{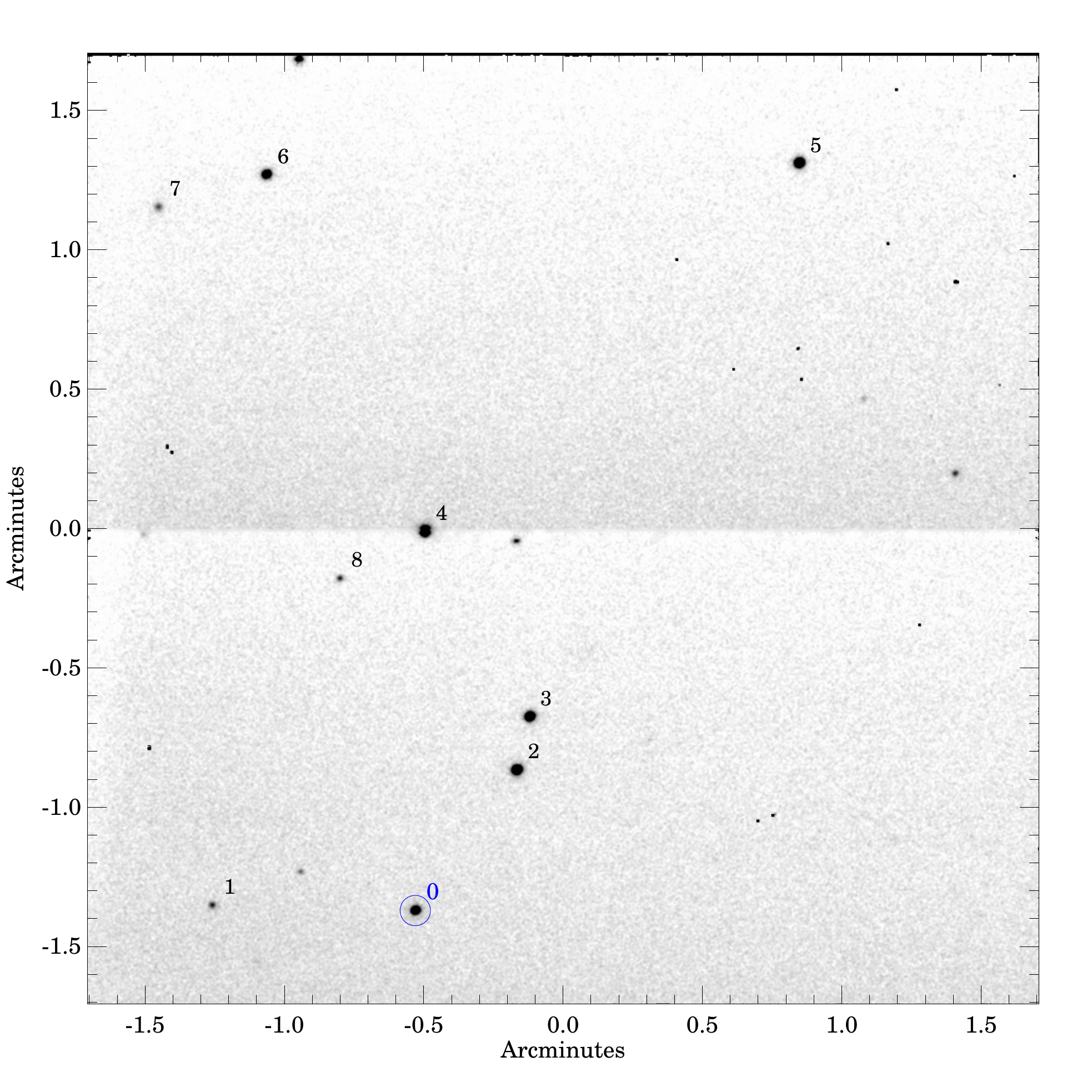}} \\
& \multirow{1}{*}[-3.4in]{\includegraphics[width=0.41\hsize]{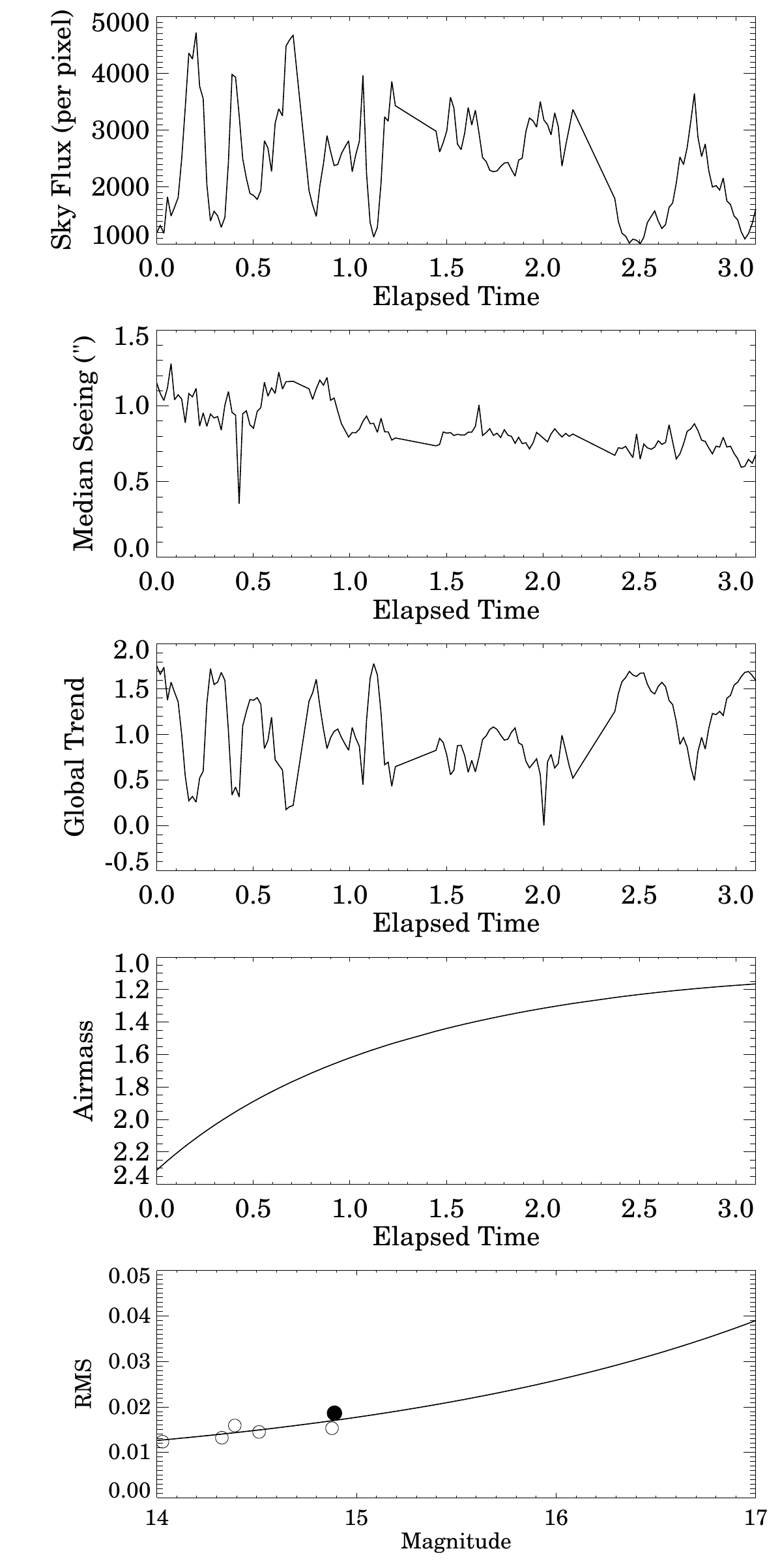}} \\
\end{tabular}
\end{figure}
\clearpage

\section{B. Periodograms for targets and reference stars}
 \label{app:B}
 
  \begin{figure*}[ht!]
\begin{tabular}{cccc}
\includegraphics[width=0.24\hsize]{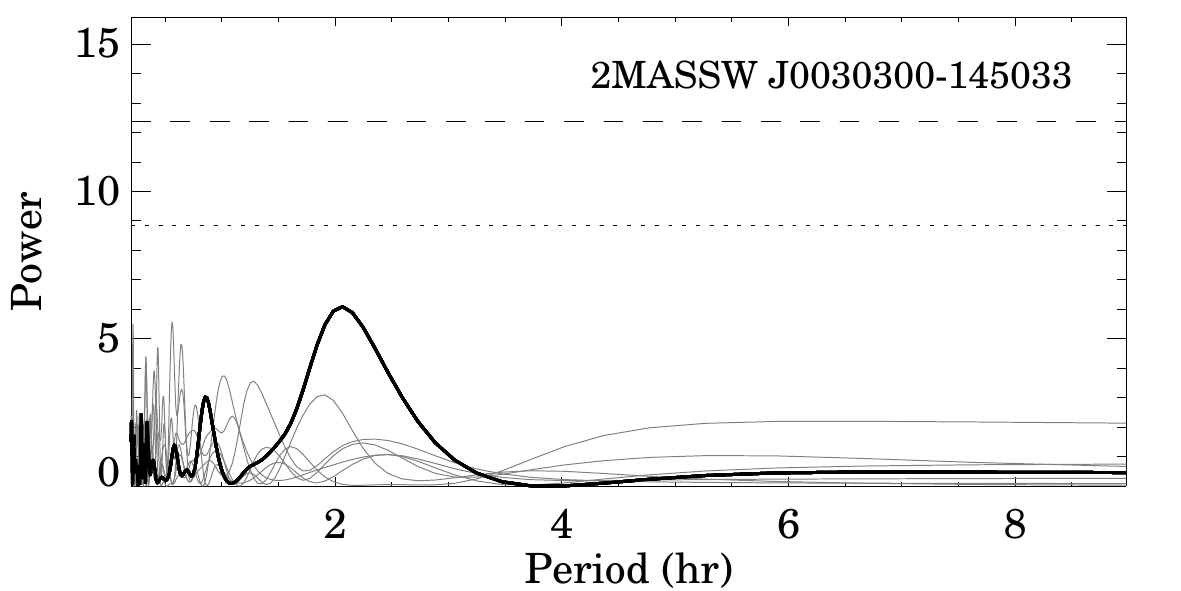} &
\includegraphics[width=0.24\hsize]{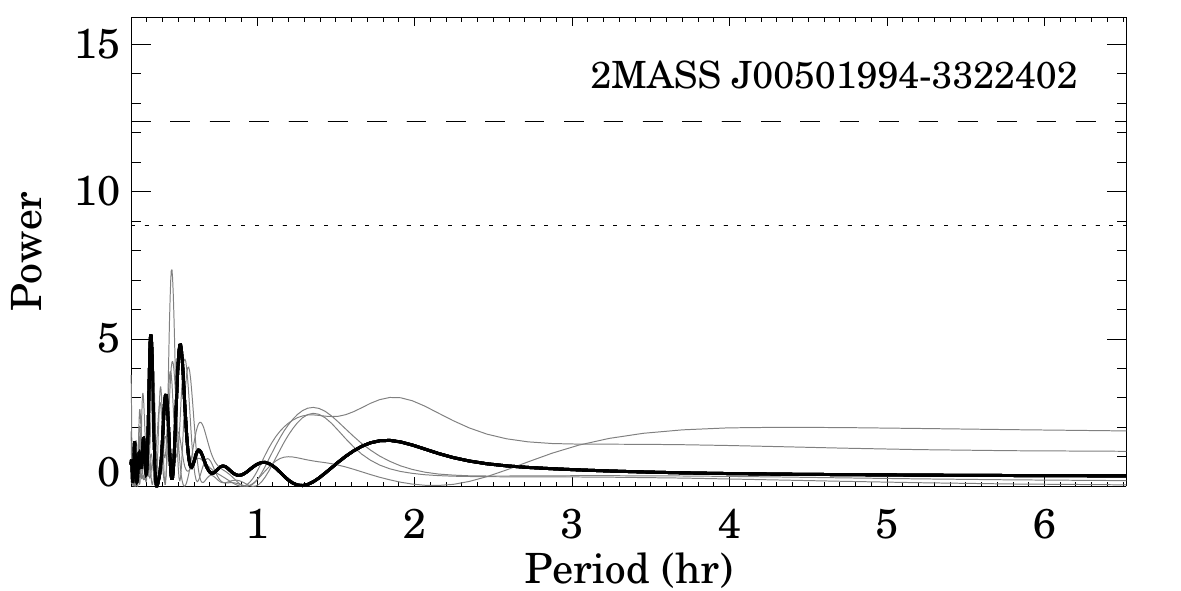} &
\includegraphics[width=0.24\hsize]{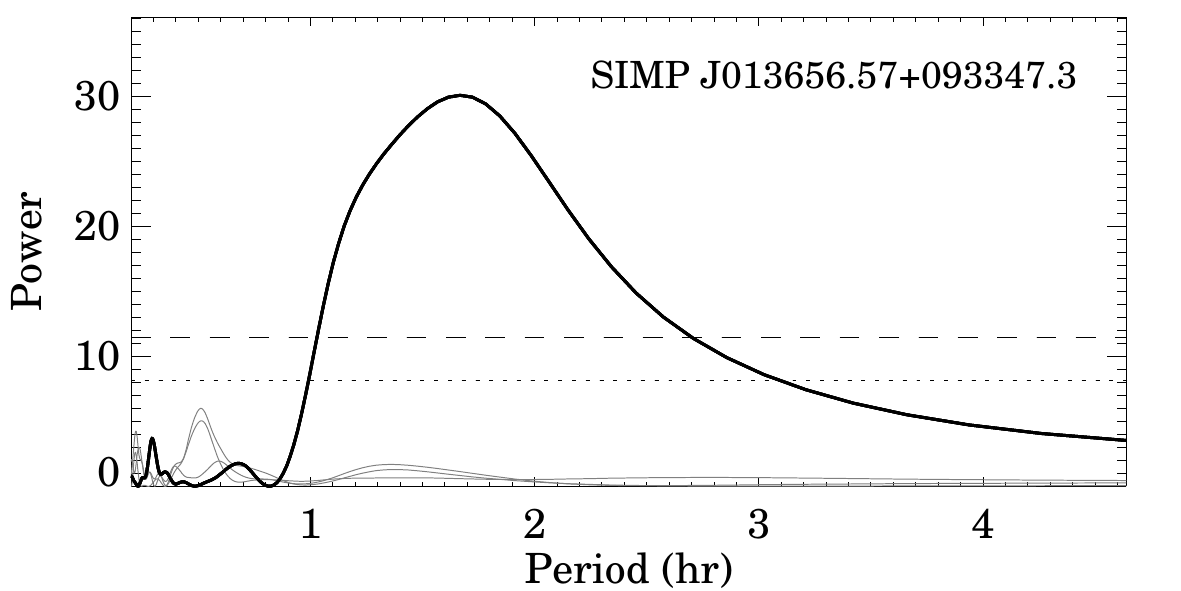} &
\includegraphics[width=0.24\hsize]{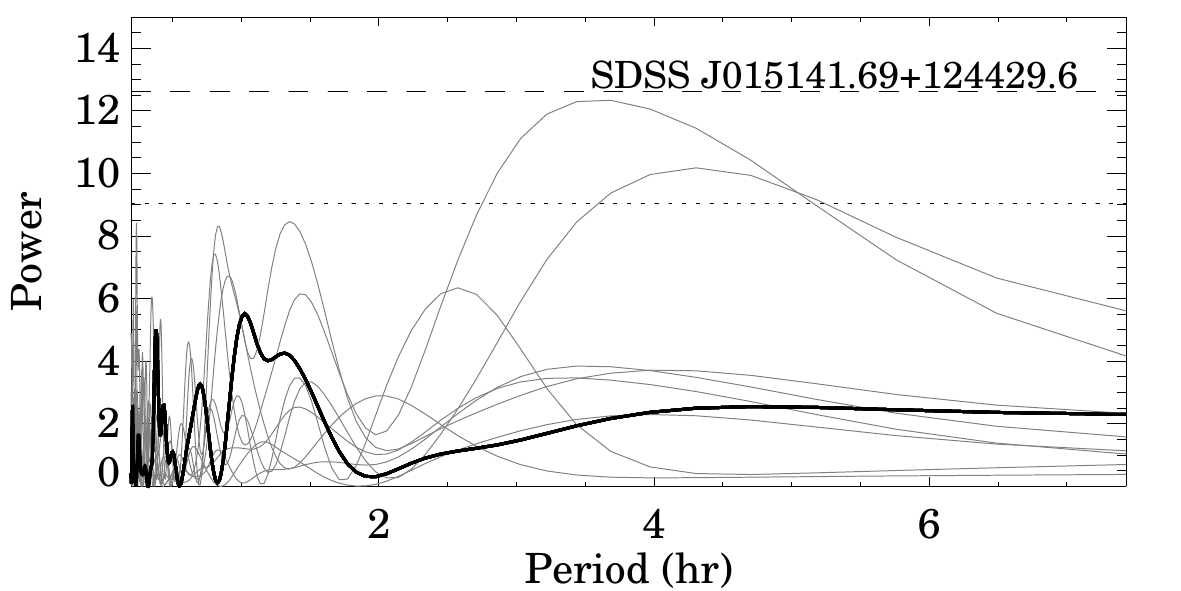} \\

\includegraphics[width=0.24\hsize]{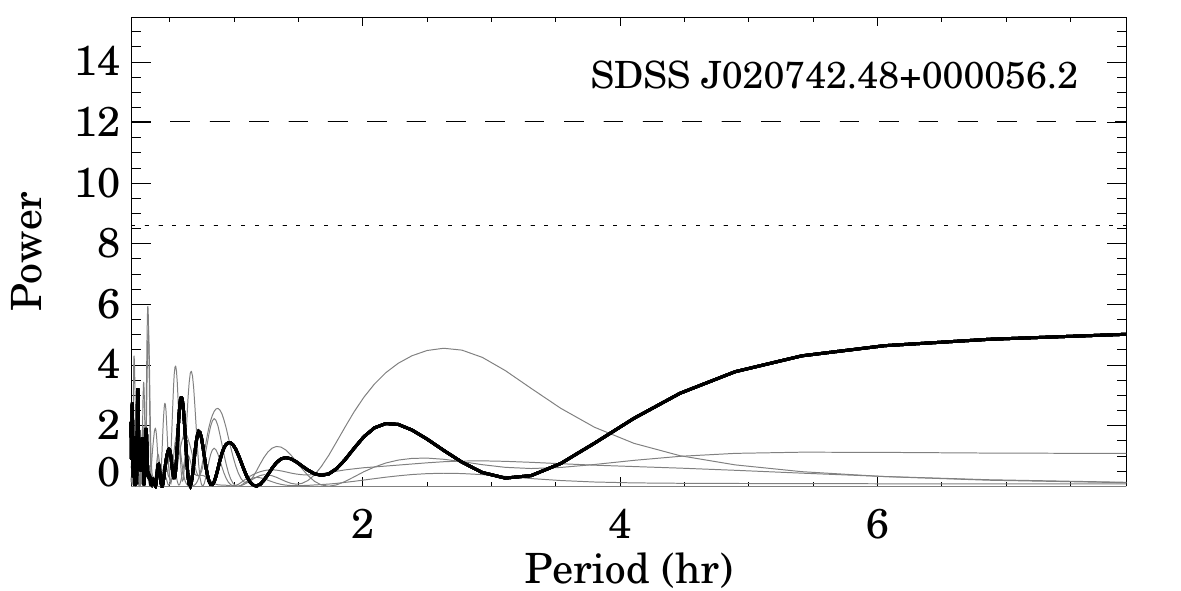} &
\includegraphics[width=0.24\hsize]{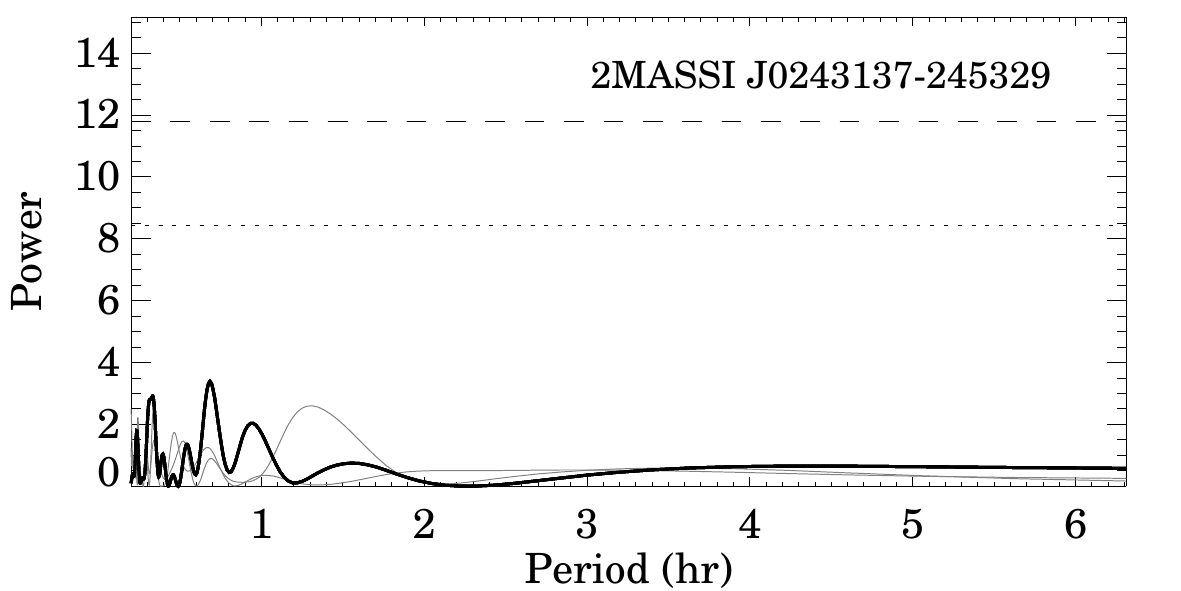} &
\includegraphics[width=0.24\hsize]{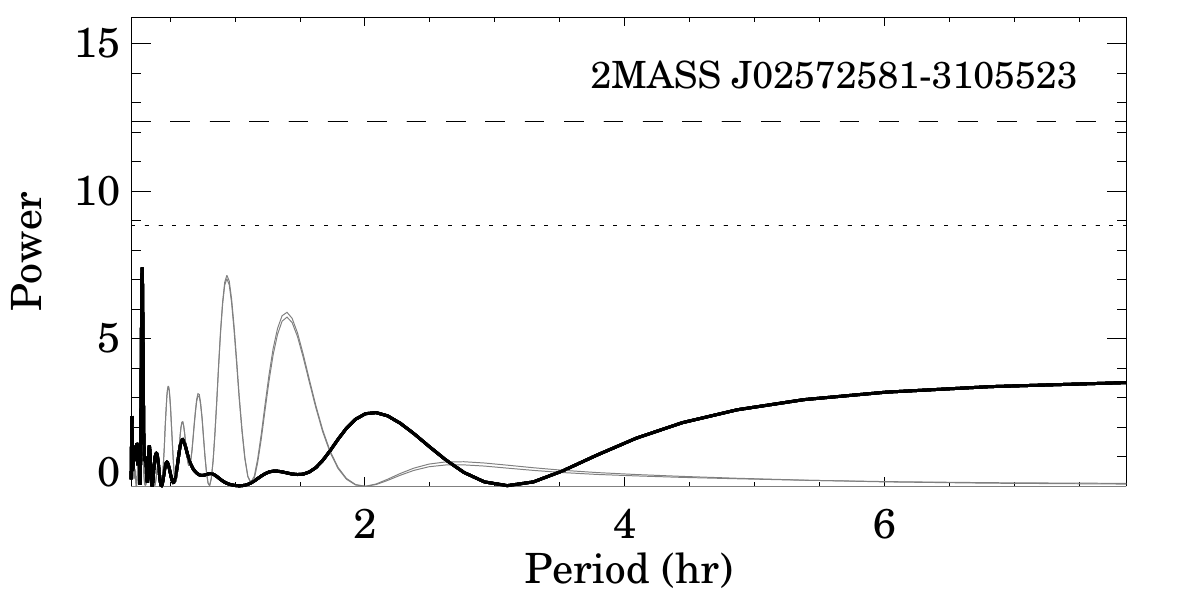} &
\includegraphics[width=0.24\hsize]{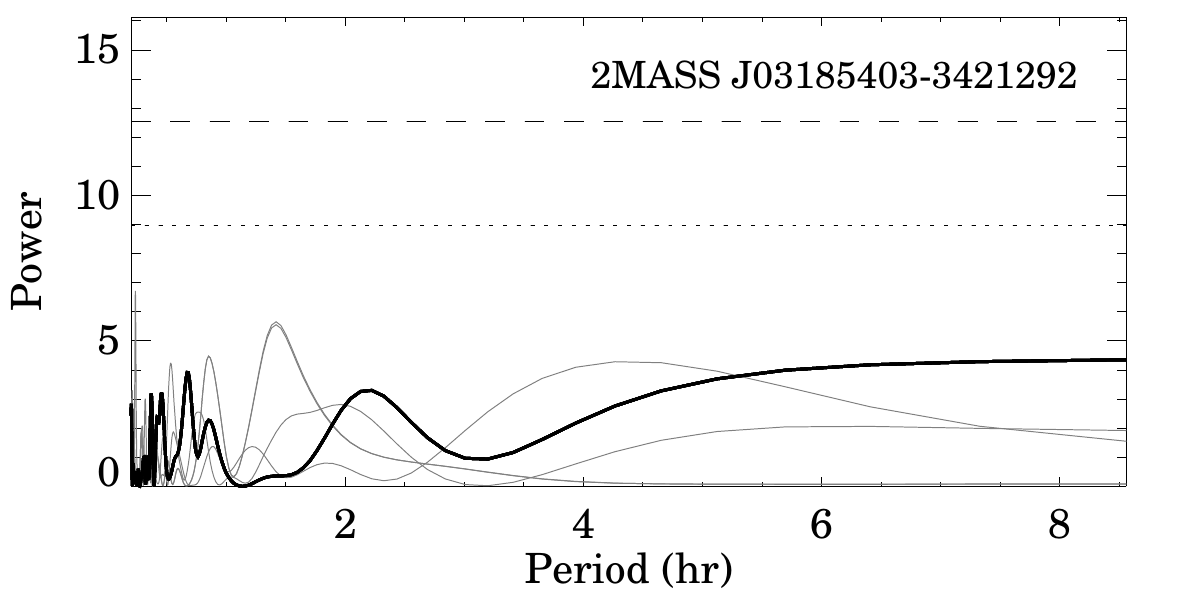} \\

\includegraphics[width=0.24\hsize]{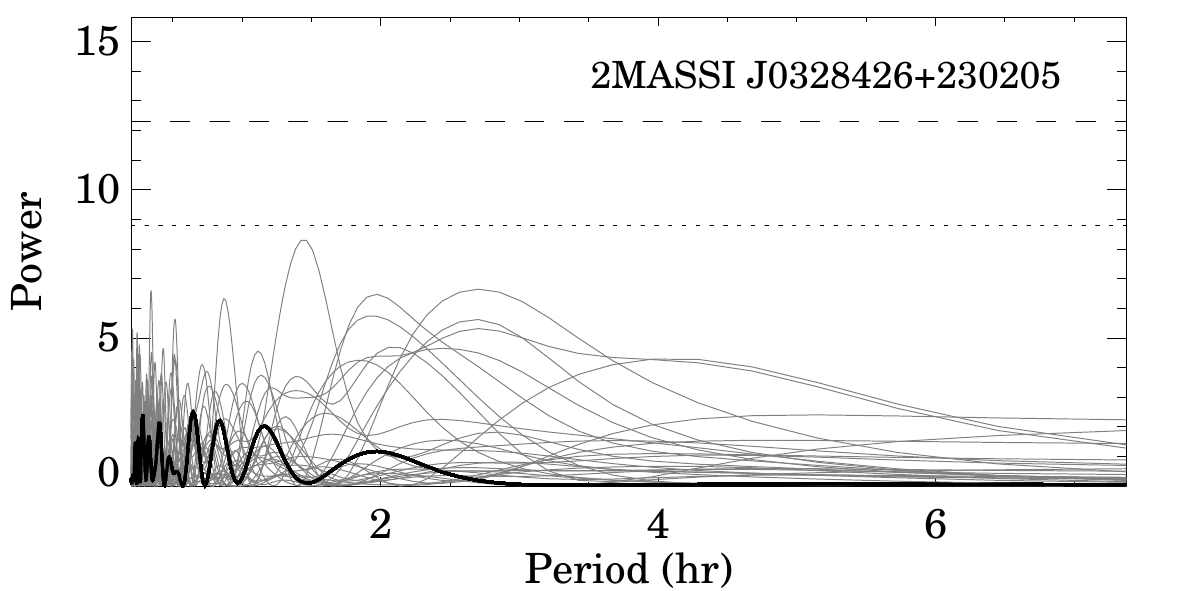} &
\includegraphics[width=0.24\hsize]{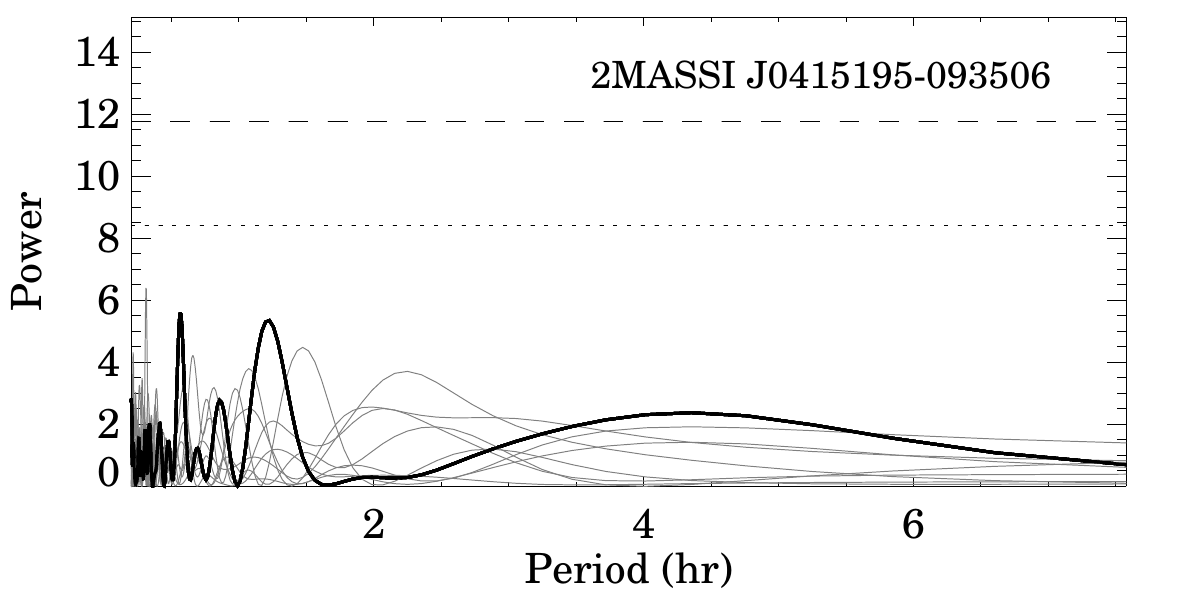} &
\includegraphics[width=0.24\hsize]{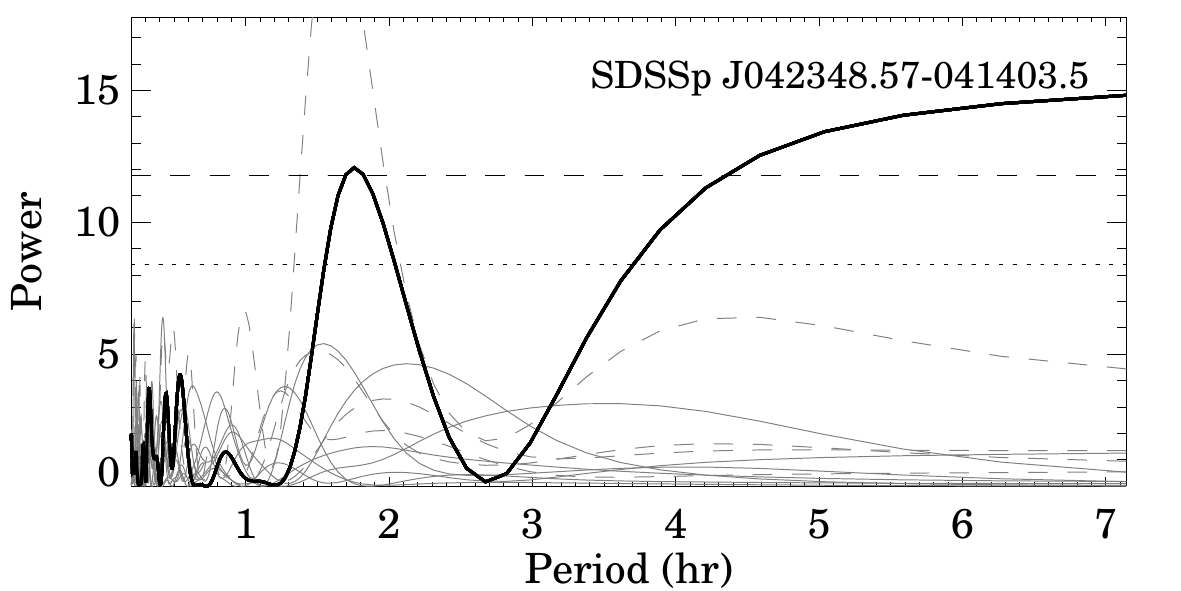} &
\includegraphics[width=0.24\hsize]{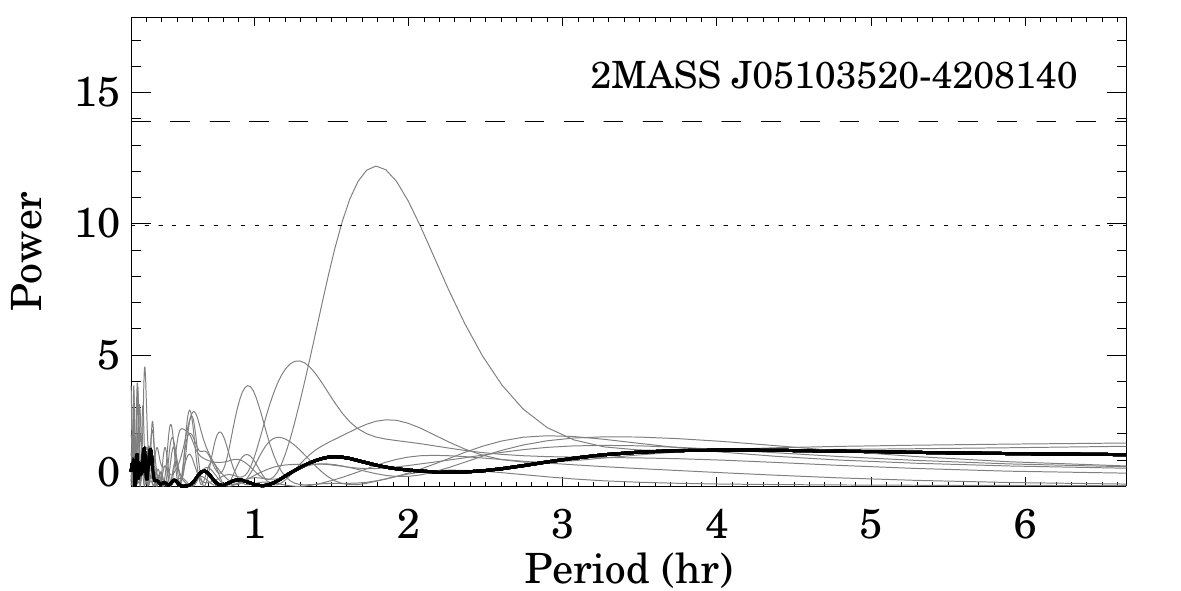} \\

\includegraphics[width=0.24\hsize]{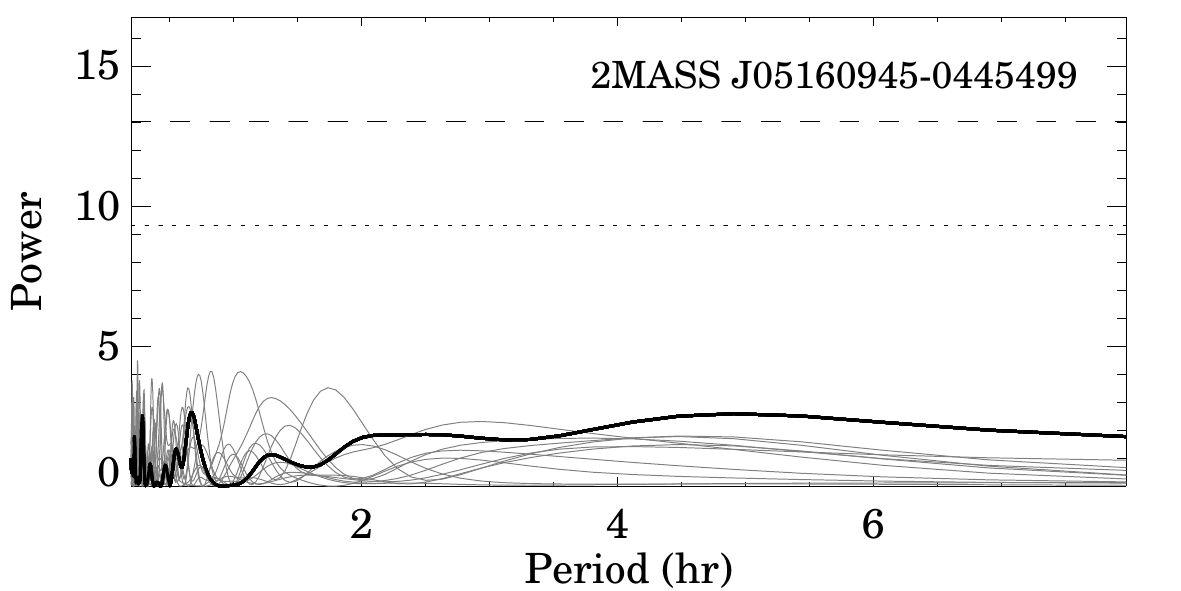} &
\includegraphics[width=0.24\hsize]{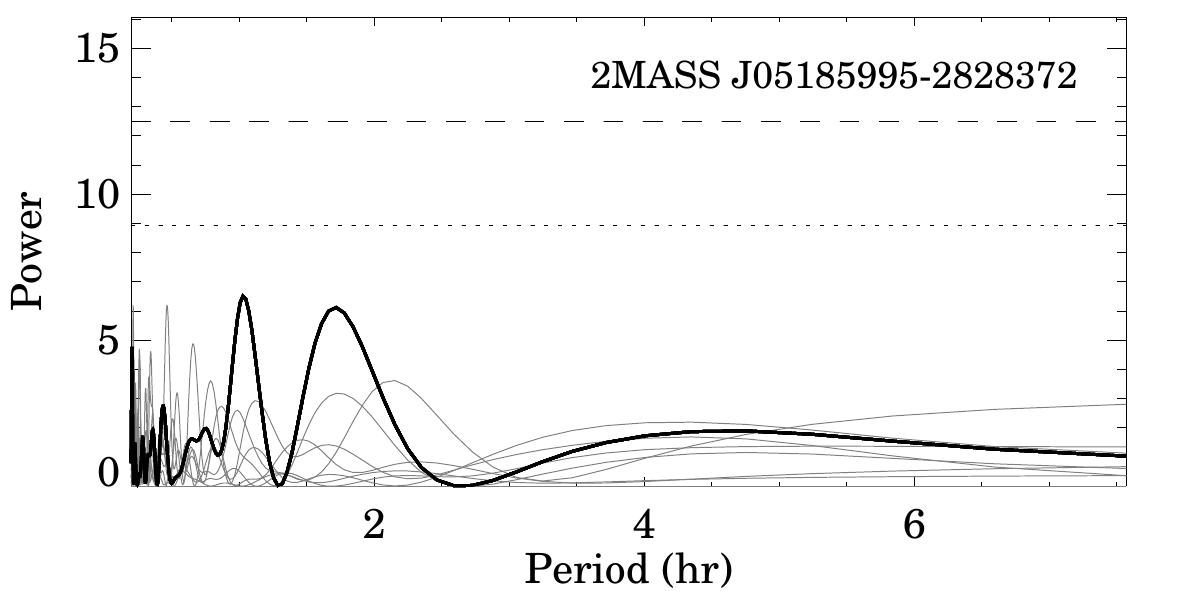} &
\includegraphics[width=0.24\hsize]{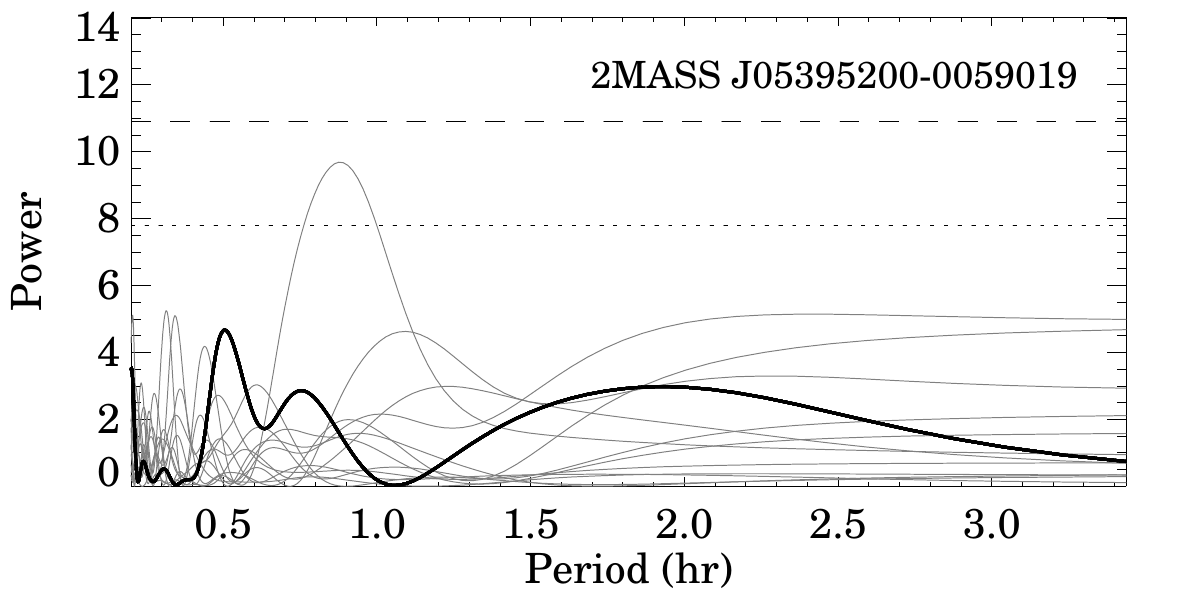} &
\includegraphics[width=0.24\hsize]{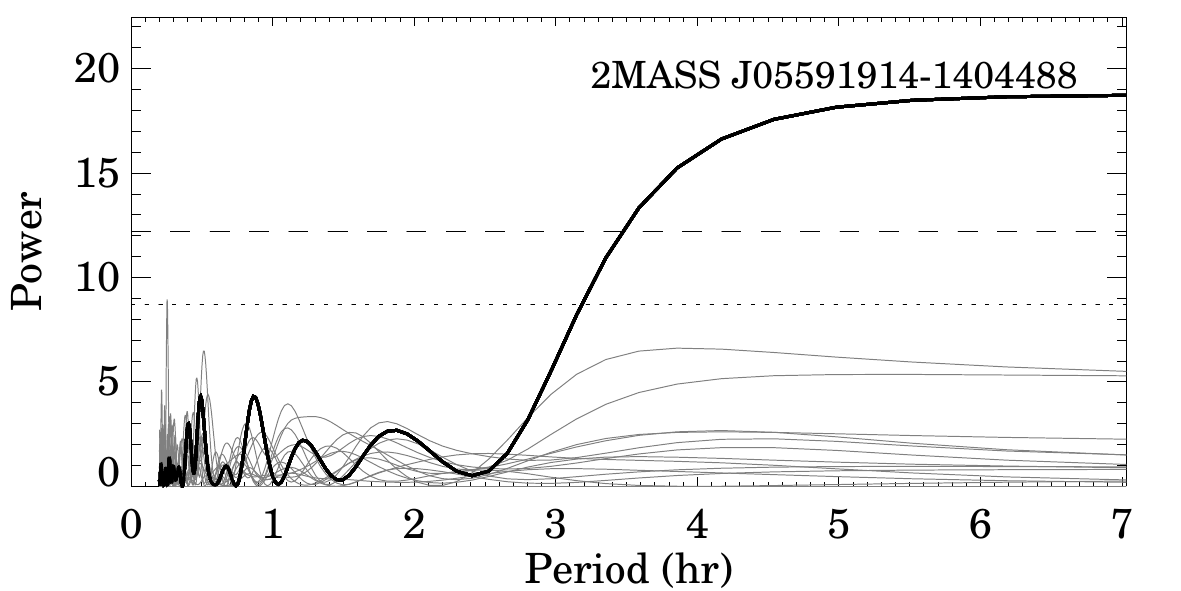} \\

\includegraphics[width=0.24\hsize]{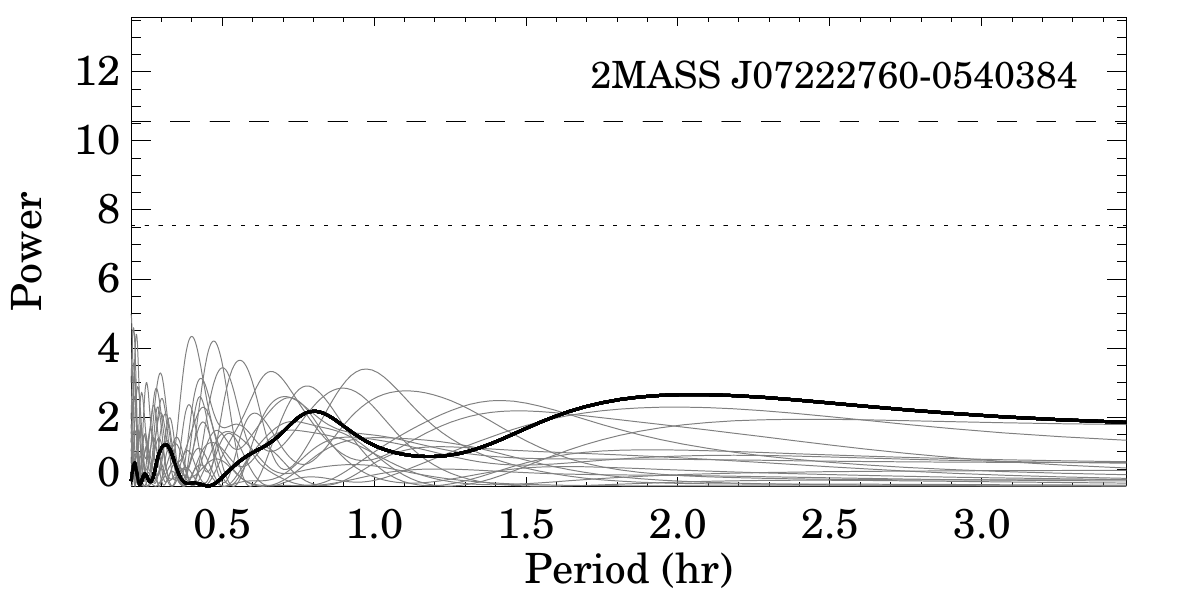} &
\includegraphics[width=0.24\hsize]{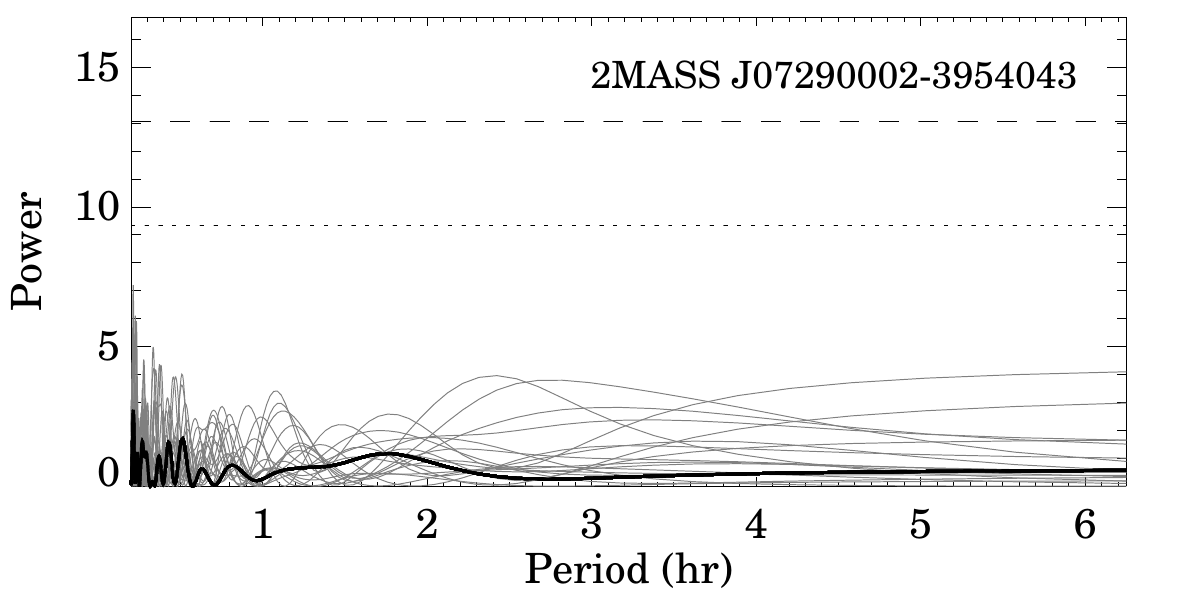} &
\includegraphics[width=0.24\hsize]{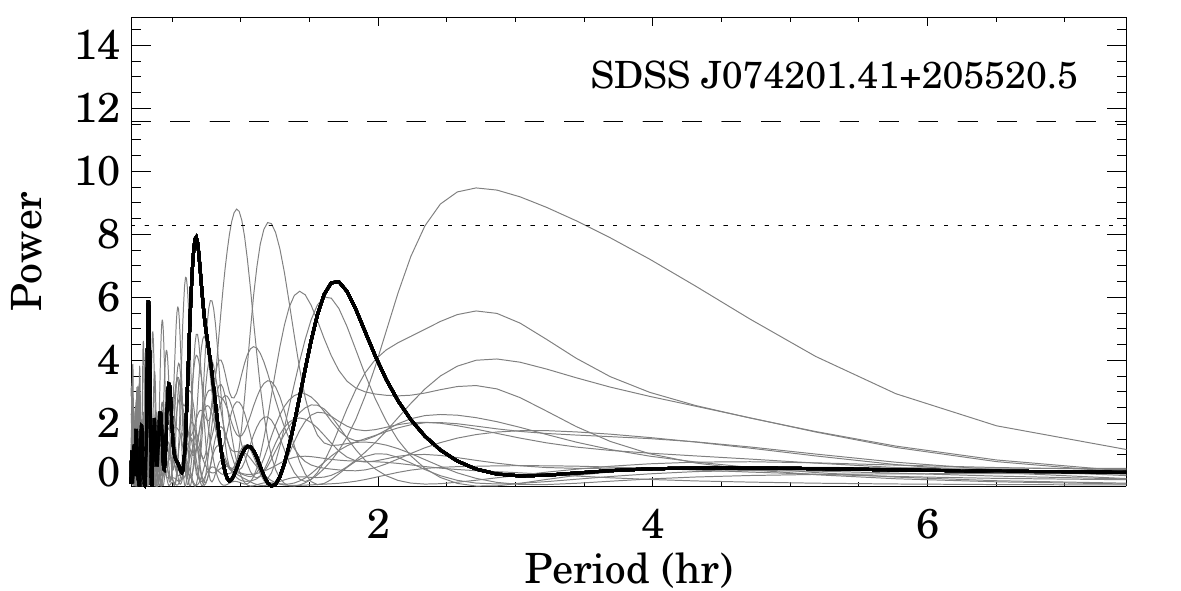} &
\includegraphics[width=0.24\hsize]{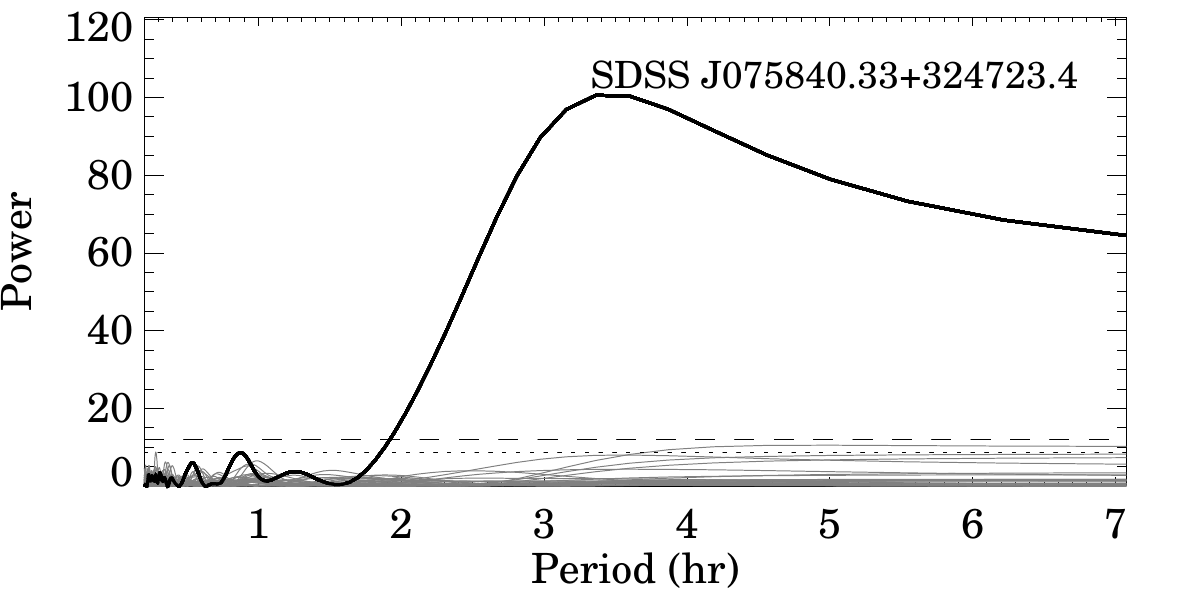} \\

\includegraphics[width=0.24\hsize]{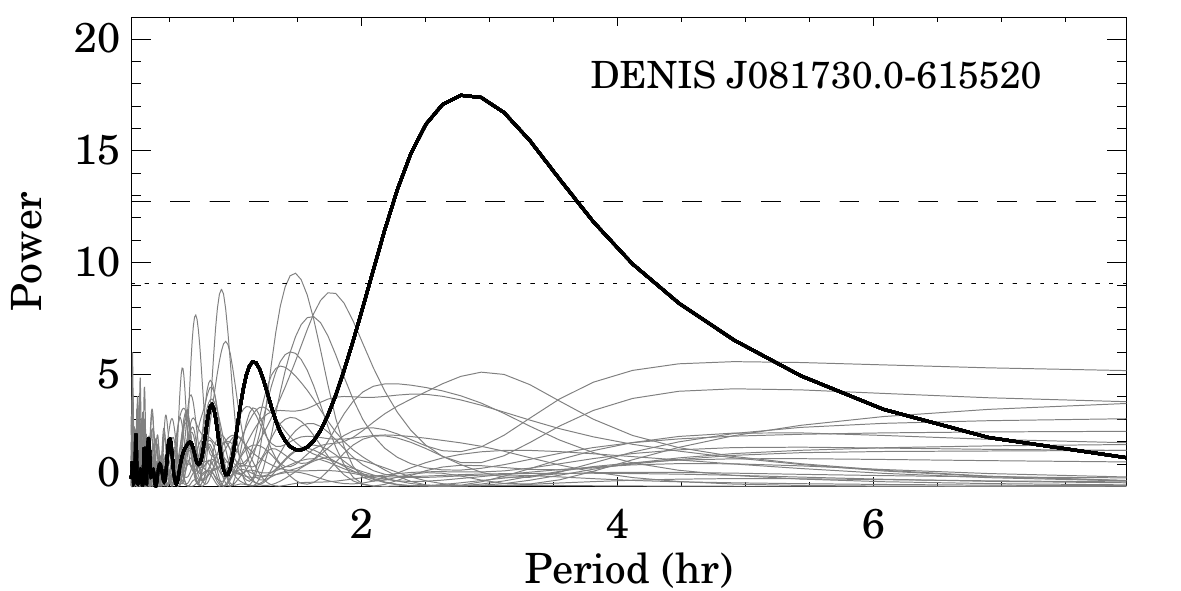} &
\includegraphics[width=0.24\hsize]{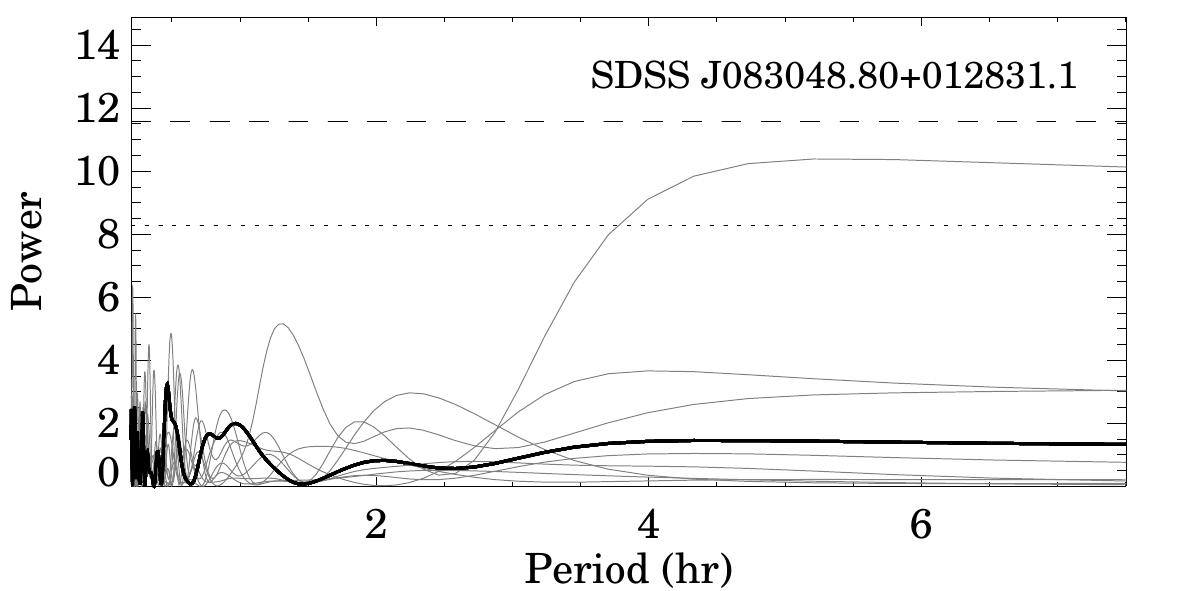} &
\includegraphics[width=0.24\hsize]{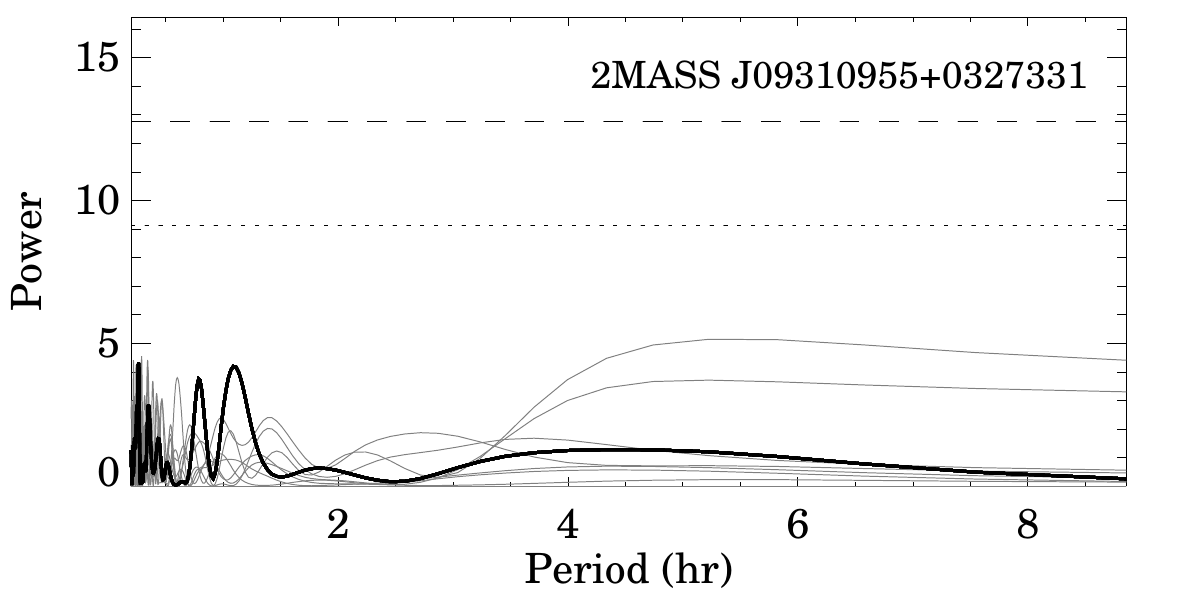} &
\includegraphics[width=0.24\hsize]{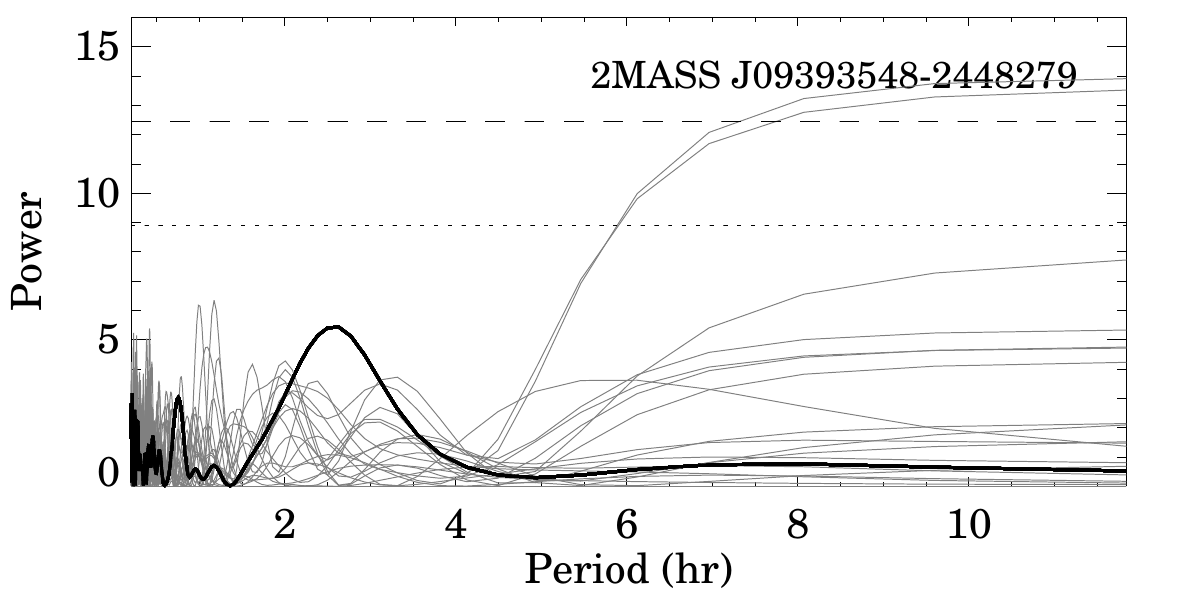} \\

\includegraphics[width=0.24\hsize]{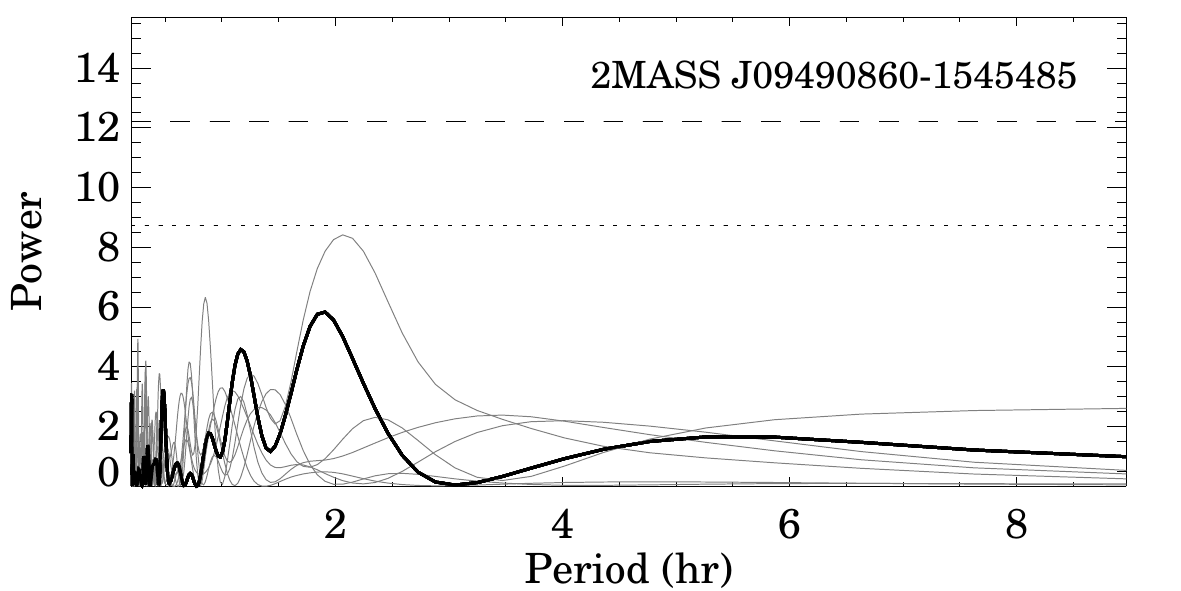} &
\includegraphics[width=0.24\hsize]{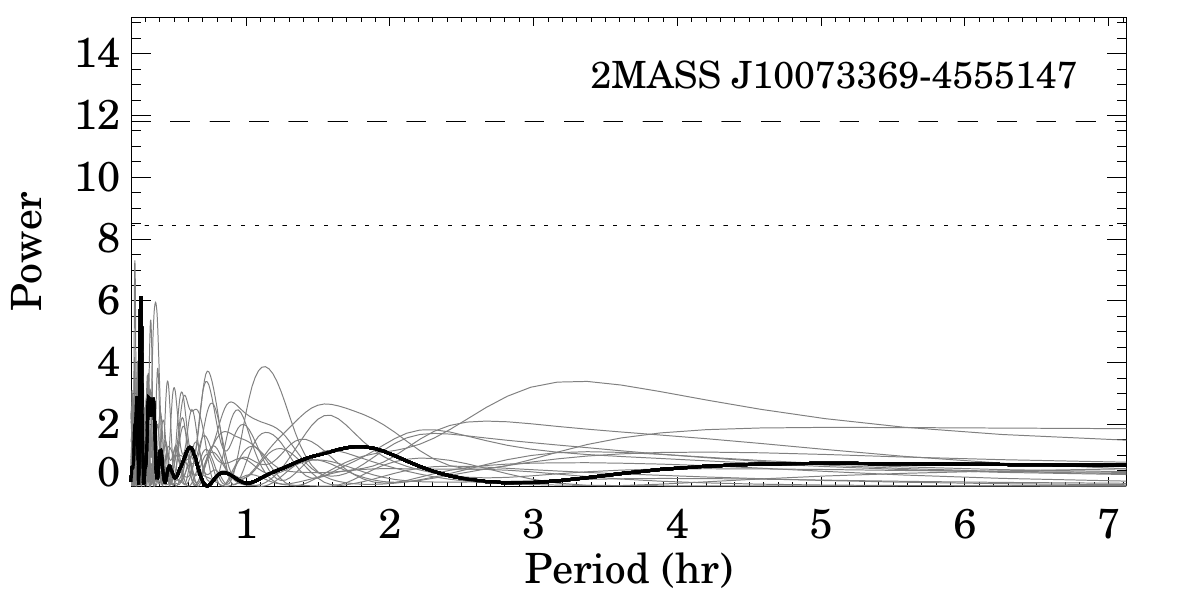} &
\includegraphics[width=0.24\hsize]{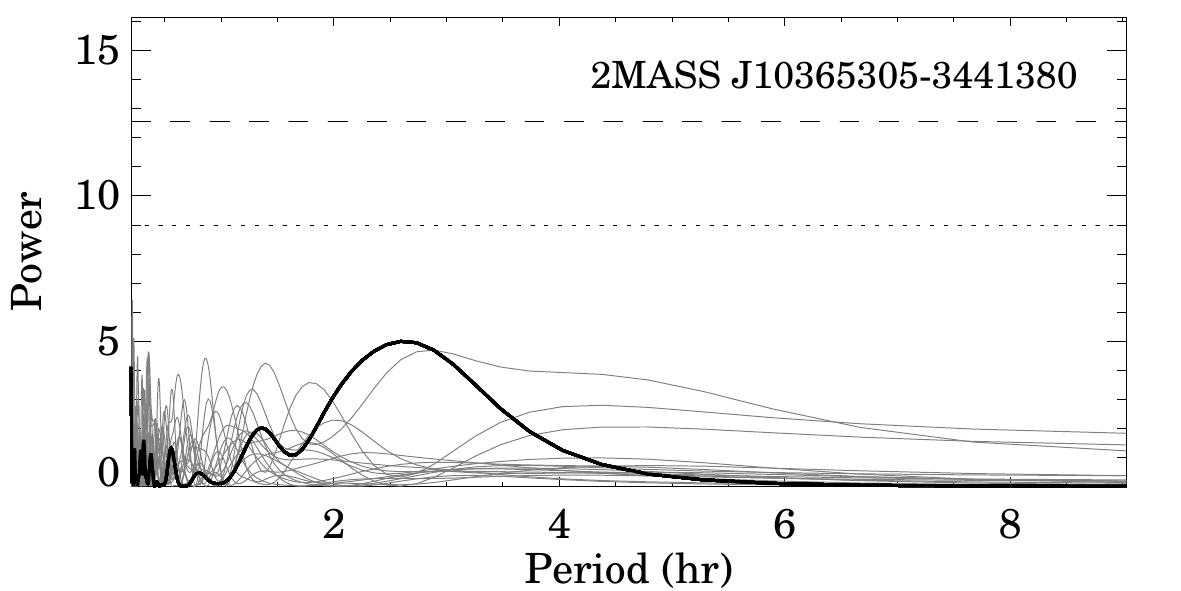} &
\includegraphics[width=0.24\hsize]{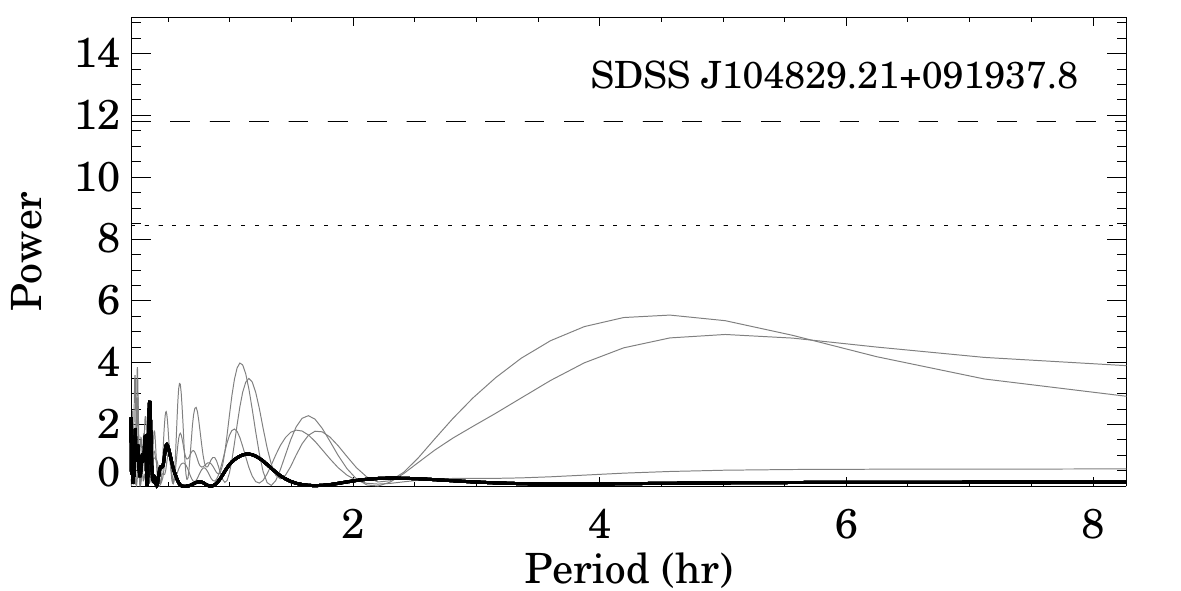} \\

\includegraphics[width=0.24\hsize]{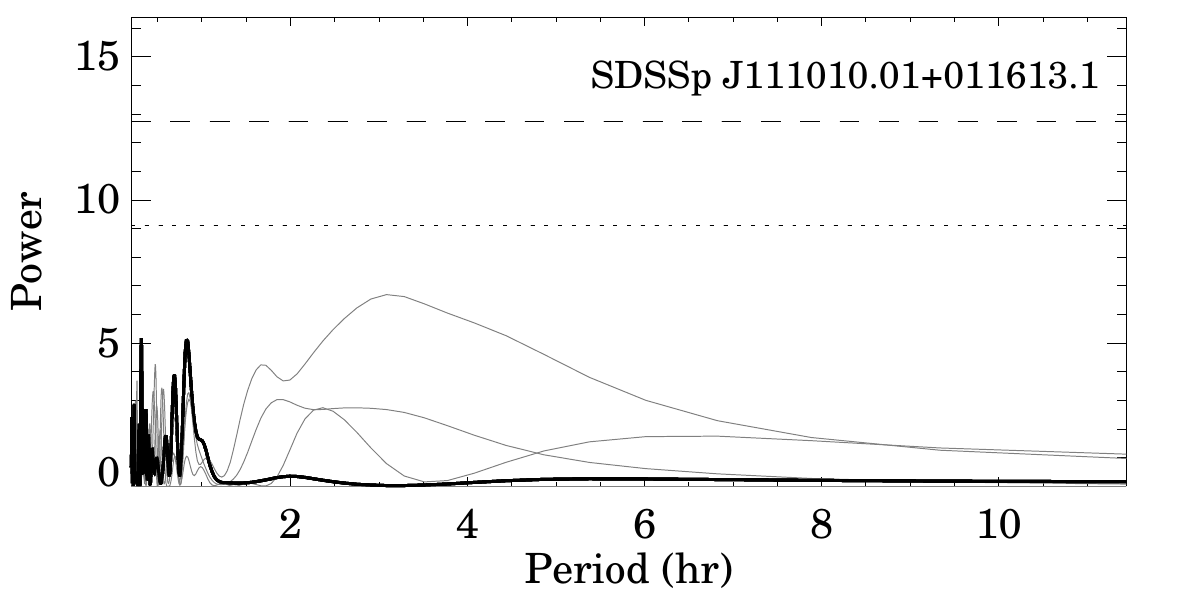} &
\includegraphics[width=0.24\hsize]{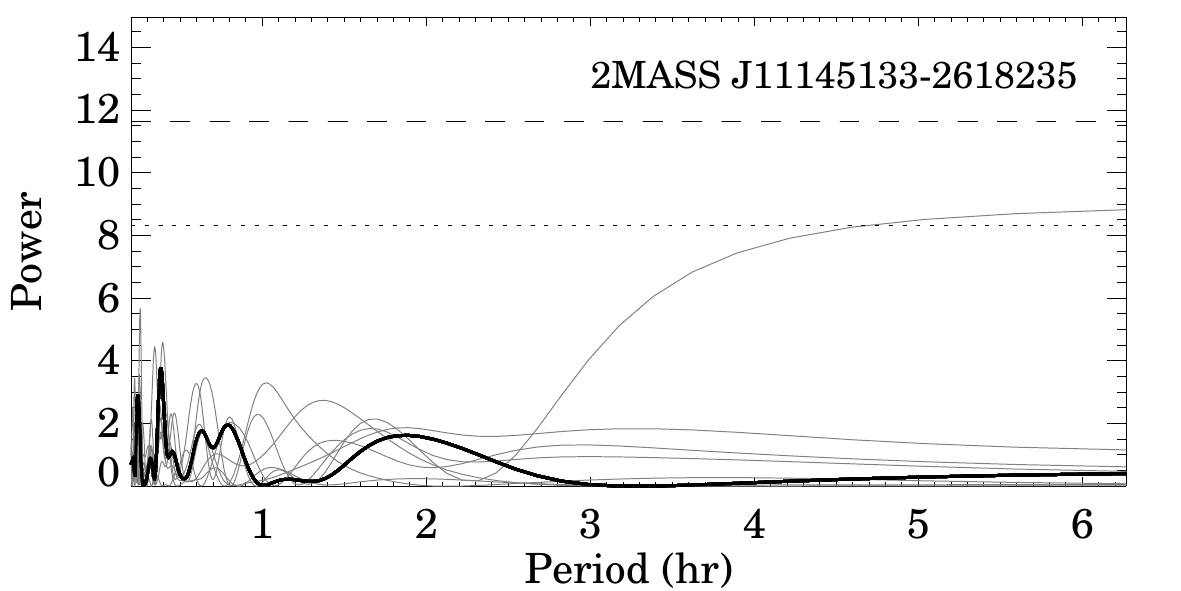} &
\includegraphics[width=0.24\hsize]{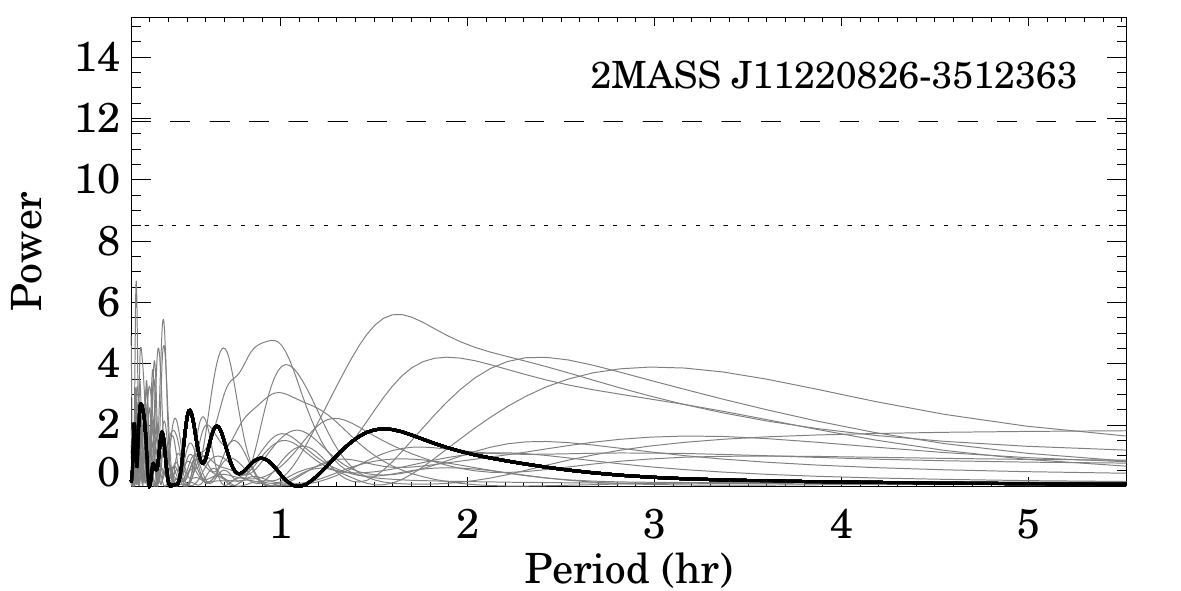} &
\includegraphics[width=0.24\hsize]{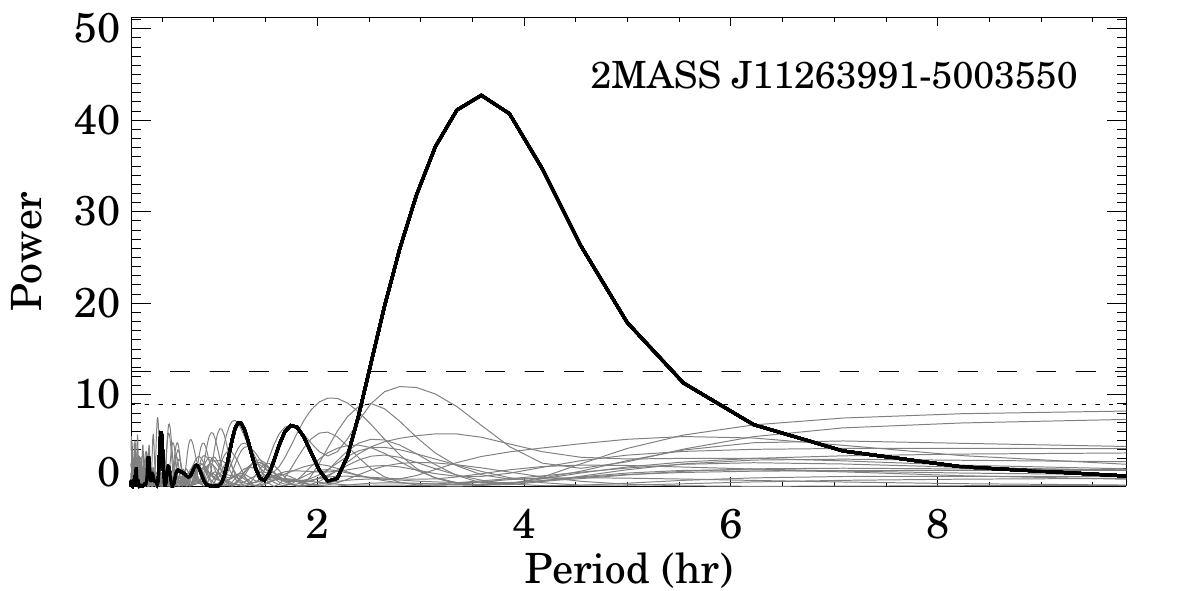} \\
\end{tabular}
\caption[]{Lomb Scargle periodograms for target (solid black line) and reference star (solid grey line) light curves.  The 0.01 FAP for the target, obtained from simulations, is shown as a dotted horizontal line.  A dashed line shows 1.4$\times$ the 0.01 FAP level (our criterion for a significant detection).  In cases where the target is flagged for quality (SDSSp J042348.57-041403.5 and SDSS J141624.08+134826.7), flagged reference stars are also plotted (dashed grey lines) to make a fair comparison. 
 \label{fig:lsper}}
\end{figure*}
\clearpage
\begin{figure*}[ht!]

\begin{tabular}{cccc}
\includegraphics[width=0.24\hsize]{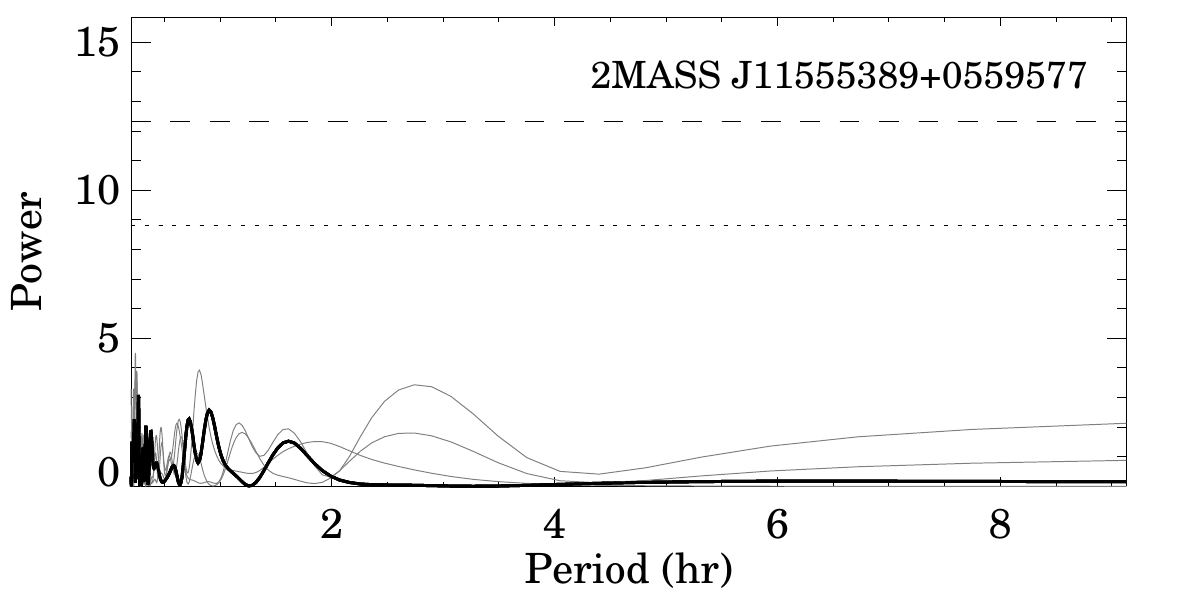} &
\includegraphics[width=0.24\hsize]{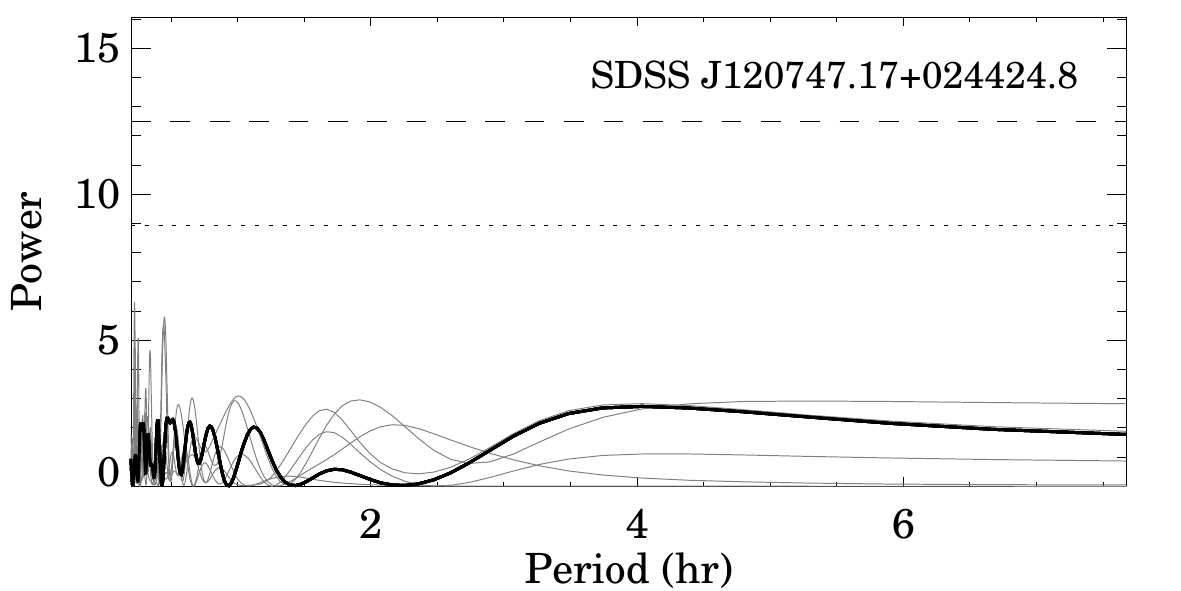} &
\includegraphics[width=0.24\hsize]{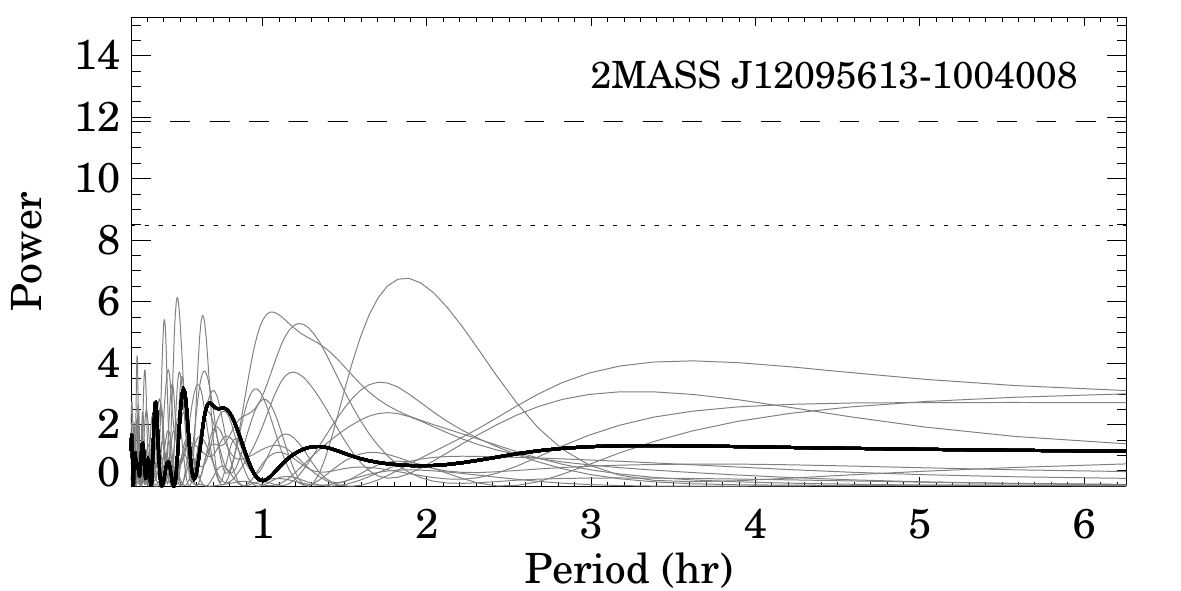} &
\includegraphics[width=0.24\hsize]{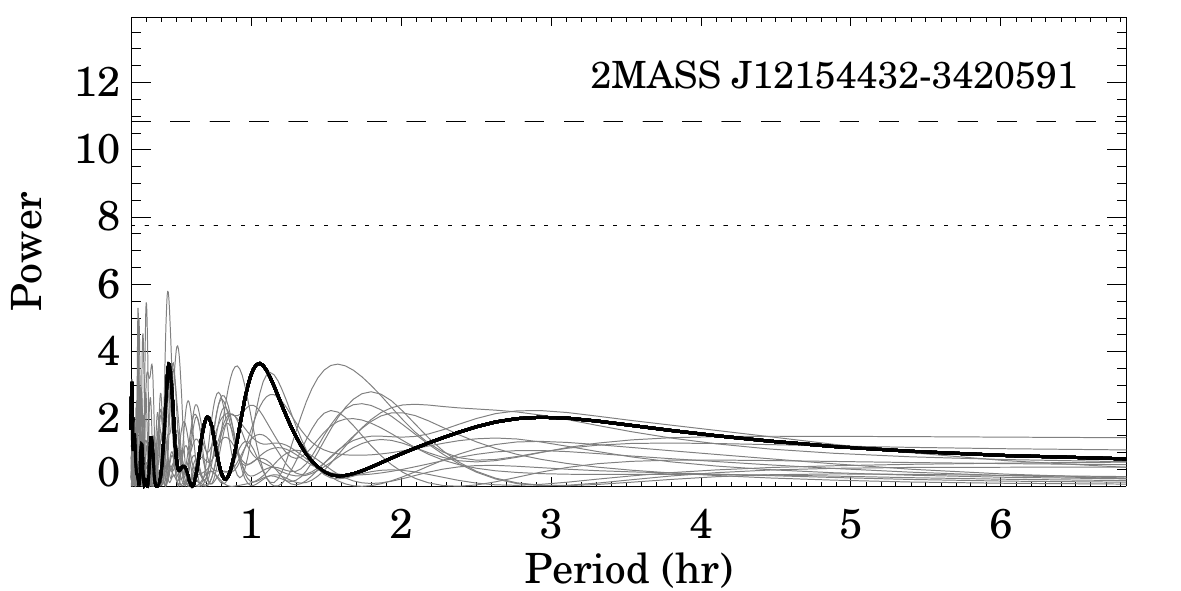} \\

\includegraphics[width=0.24\hsize]{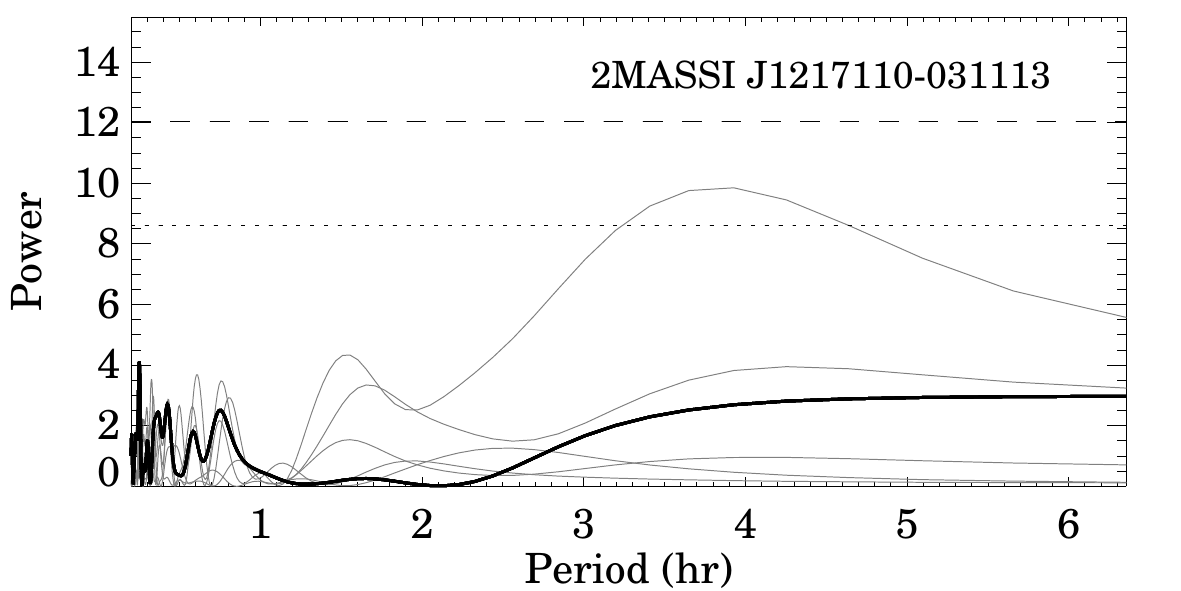} &
\includegraphics[width=0.24\hsize]{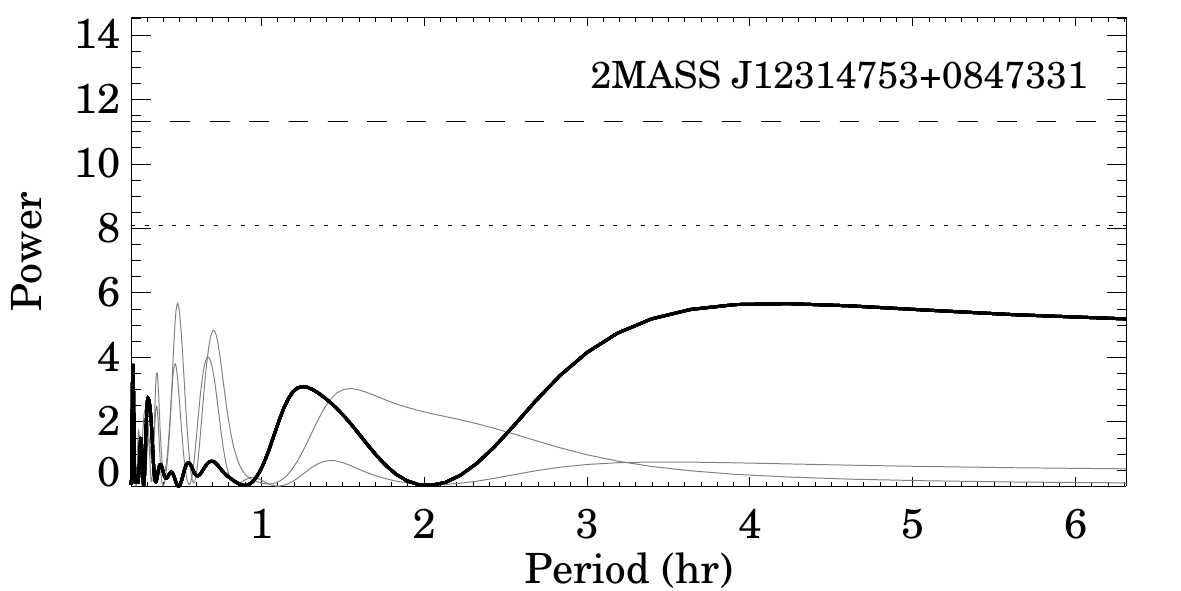} &
\includegraphics[width=0.24\hsize]{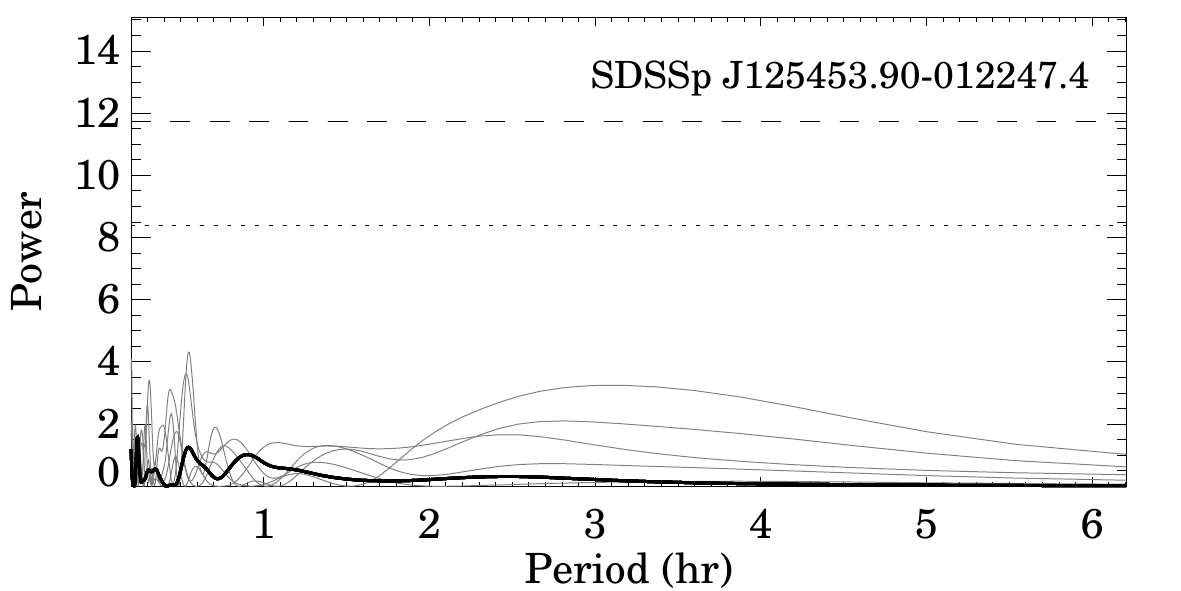} &
\includegraphics[width=0.24\hsize]{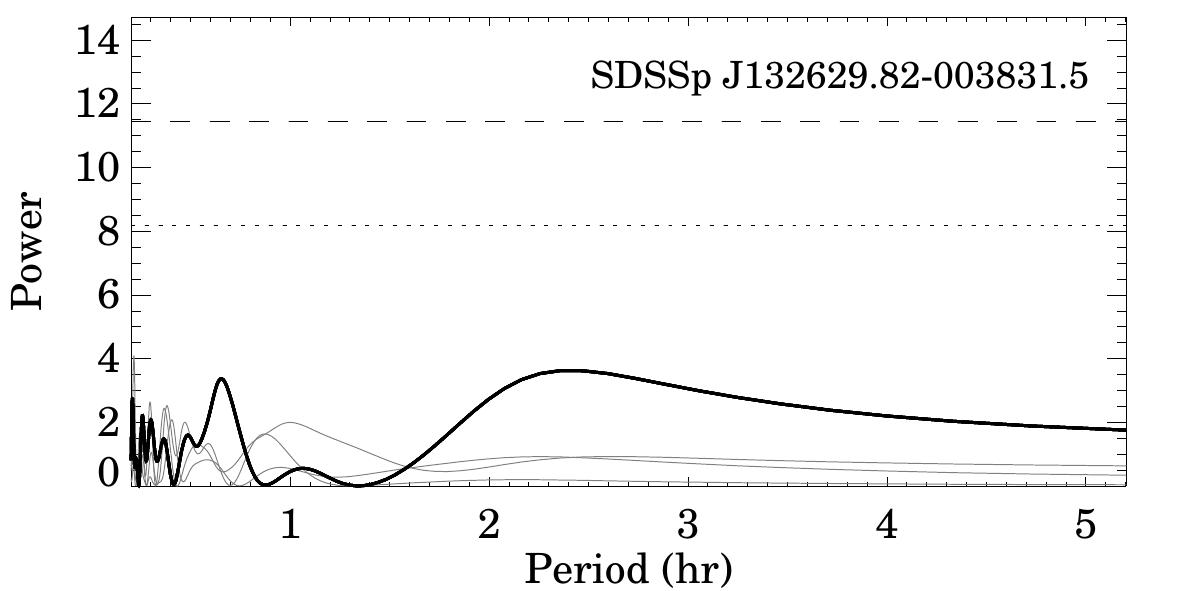} \\

\includegraphics[width=0.24\hsize]{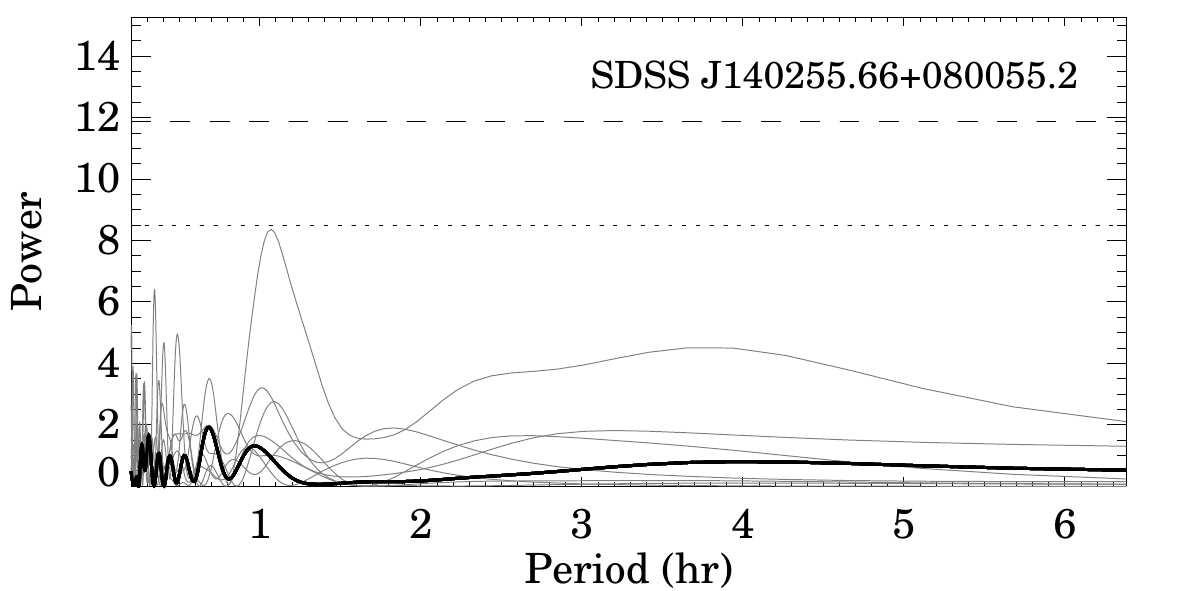} &
\includegraphics[width=0.24\hsize]{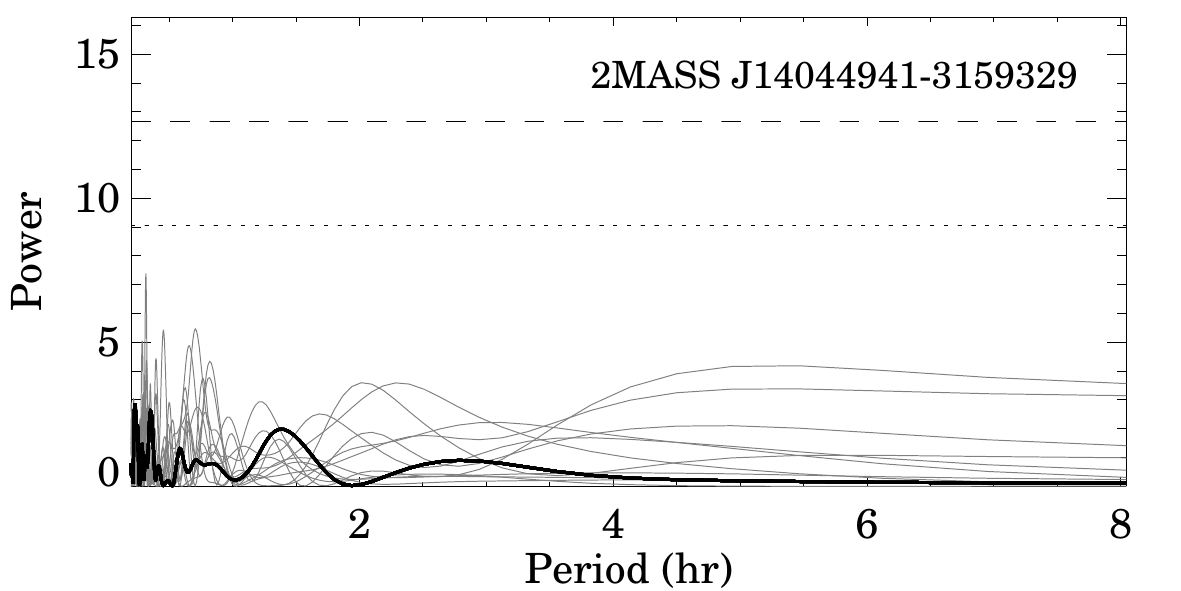} &
\includegraphics[width=0.24\hsize]{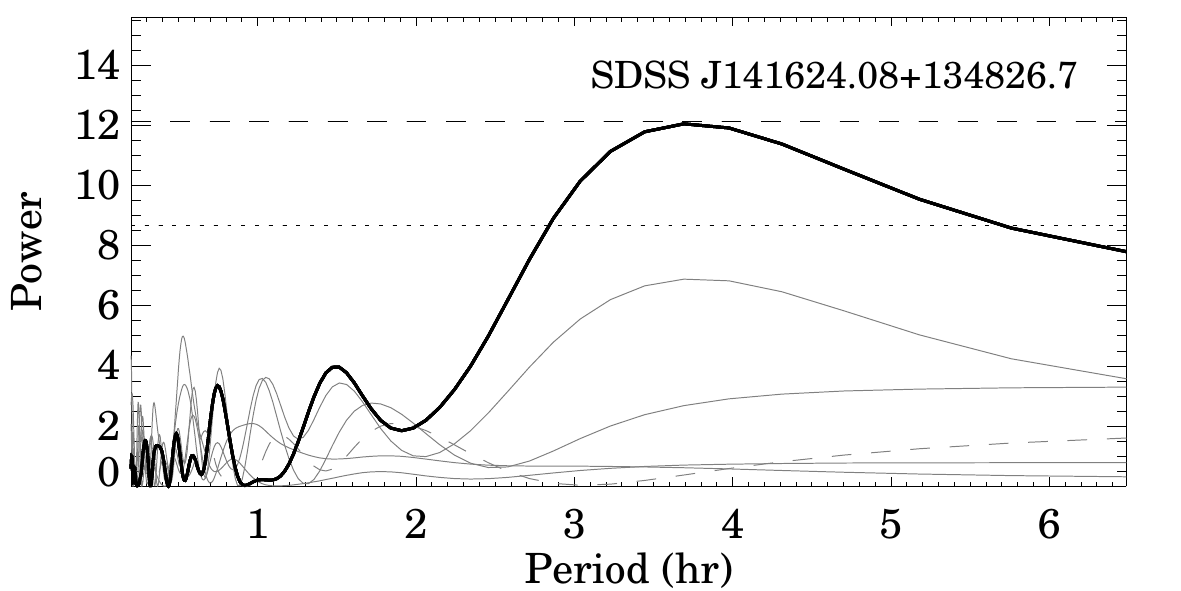} &
\includegraphics[width=0.24\hsize]{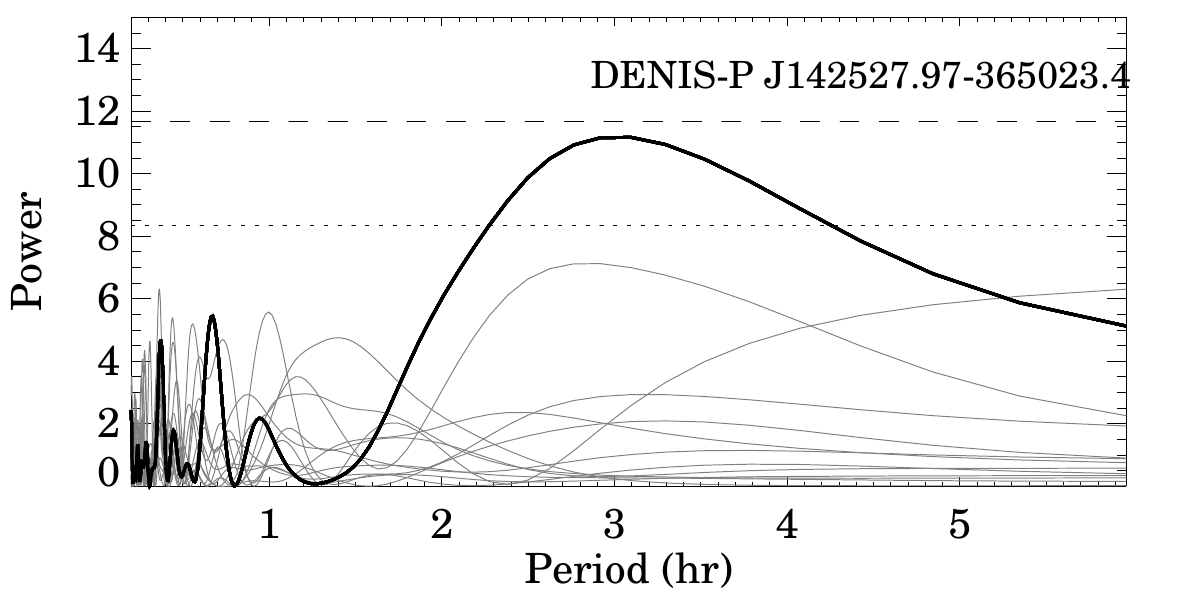} \\

\includegraphics[width=0.24\hsize]{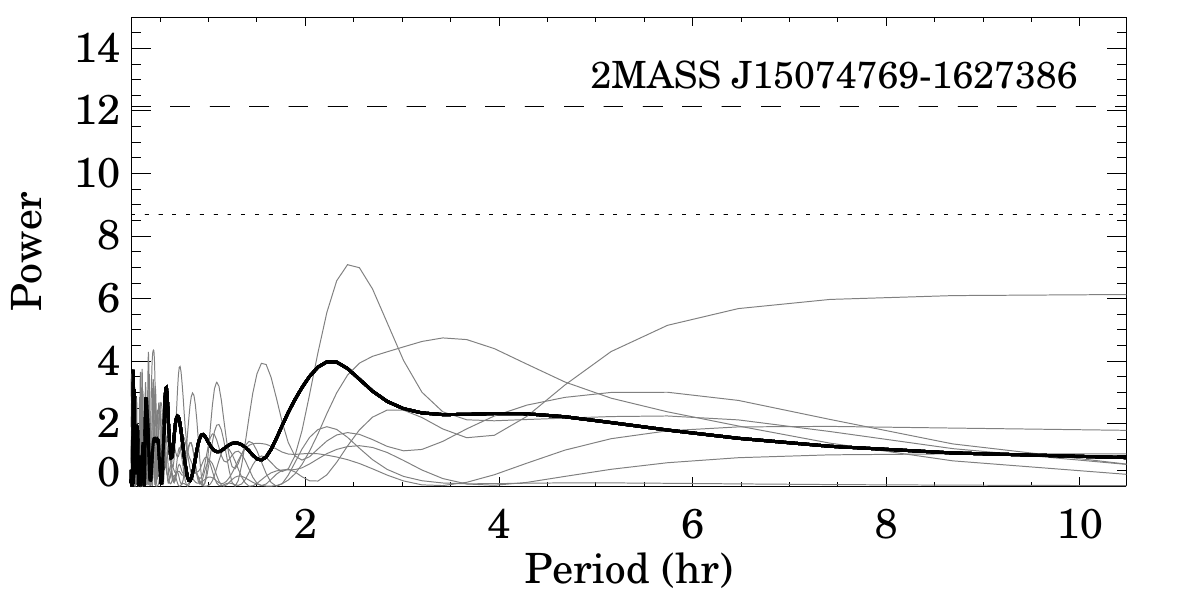} &
\includegraphics[width=0.24\hsize]{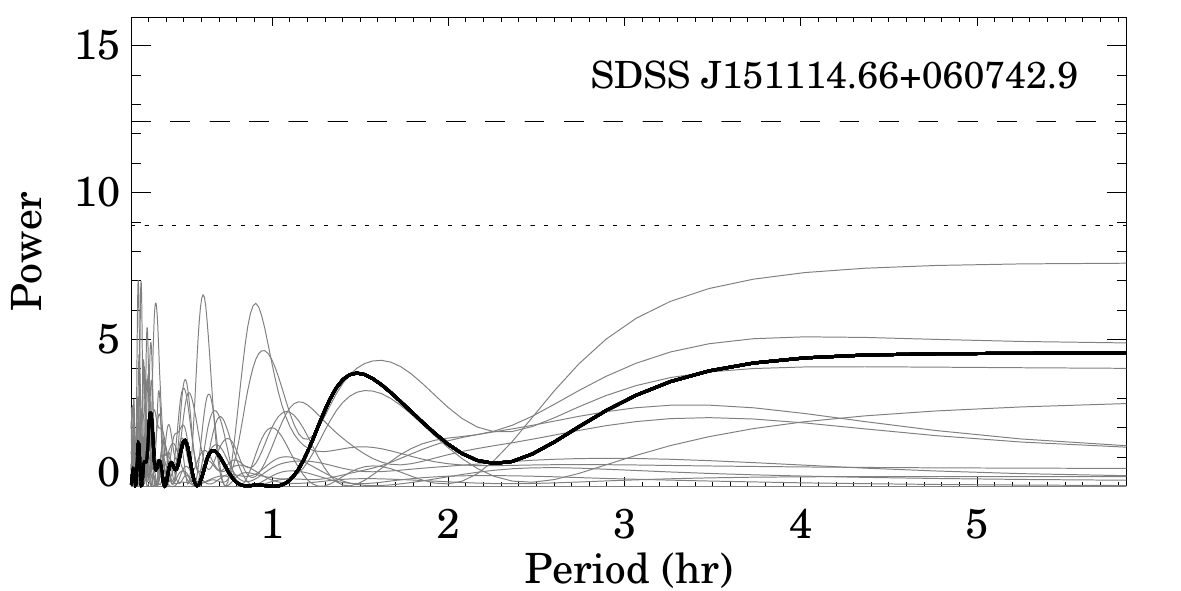} &
\includegraphics[width=0.24\hsize]{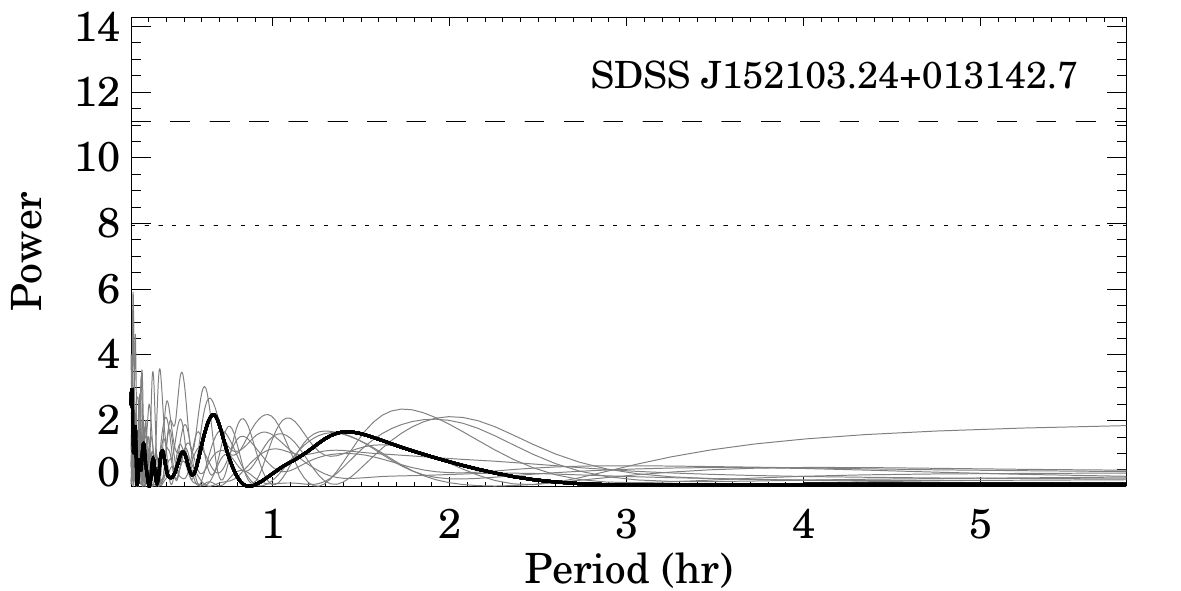} &
\includegraphics[width=0.24\hsize]{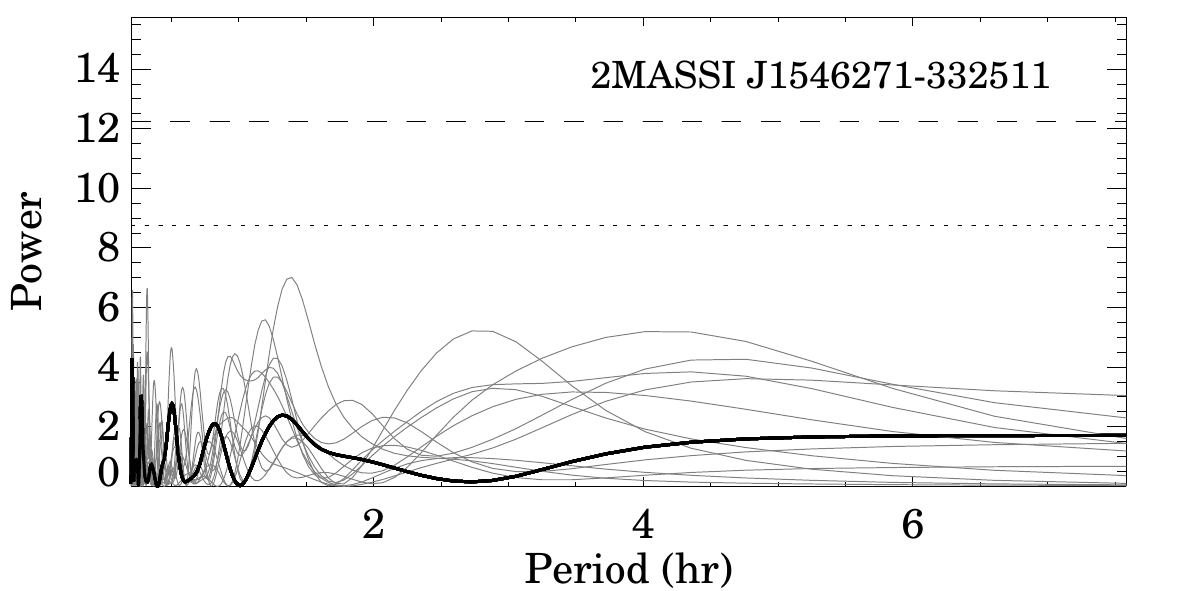} \\

\includegraphics[width=0.24\hsize]{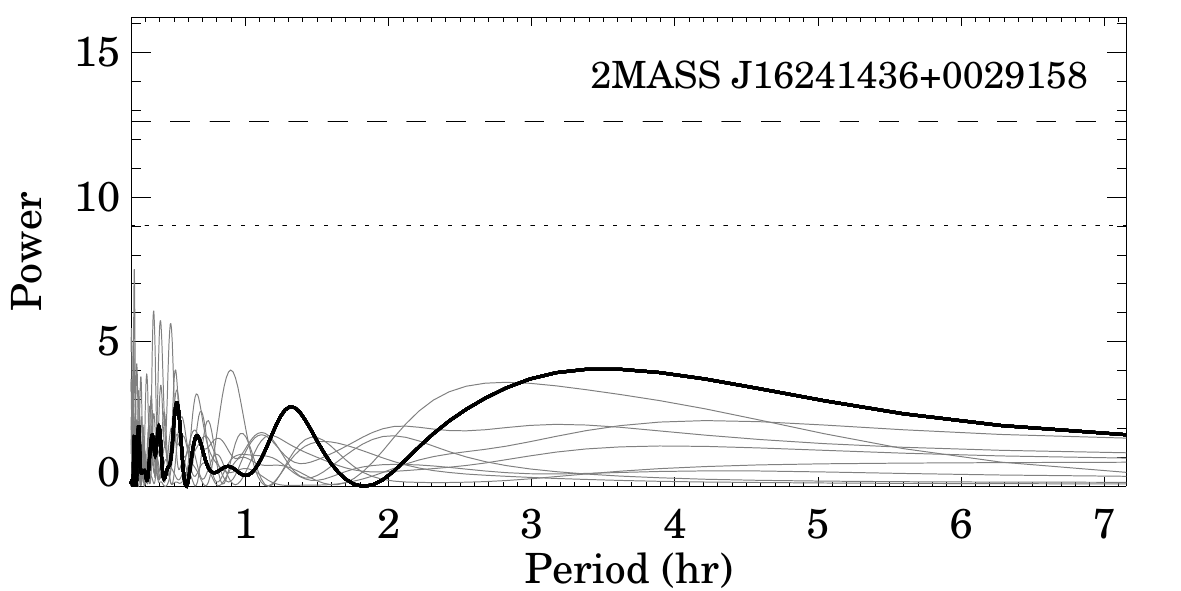} &
\includegraphics[width=0.24\hsize]{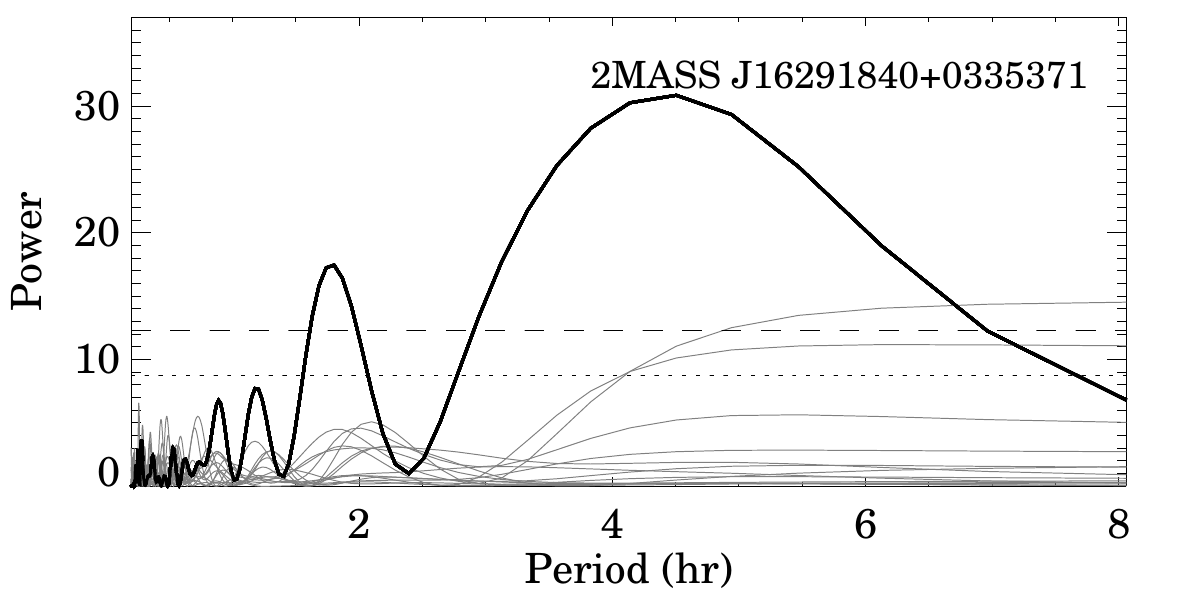} &
\includegraphics[width=0.24\hsize]{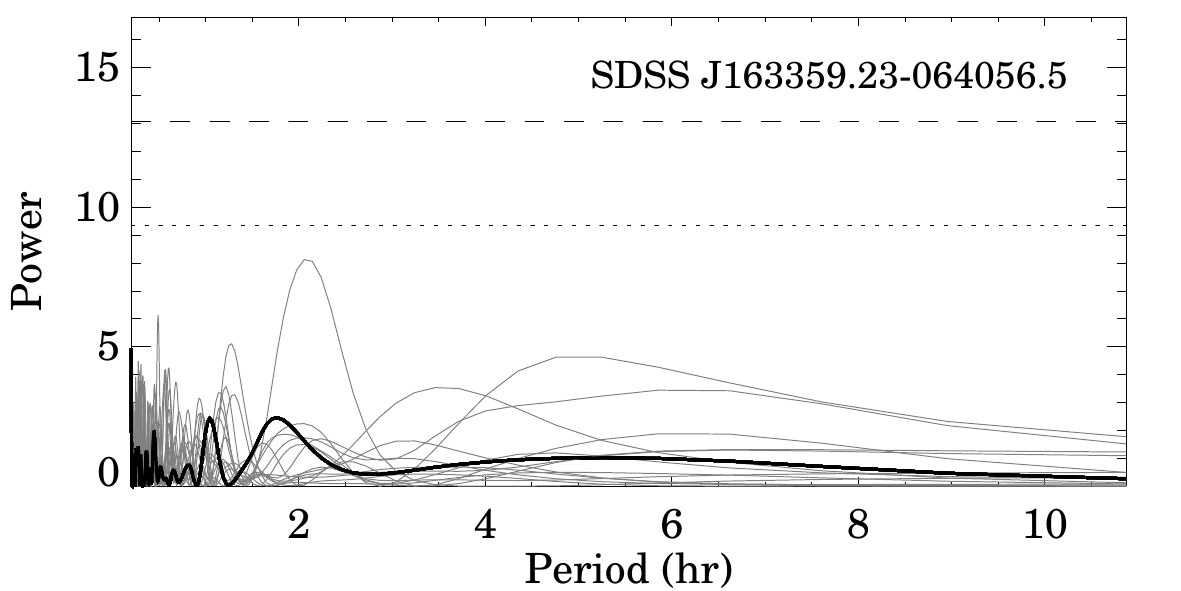} &
\includegraphics[width=0.24\hsize]{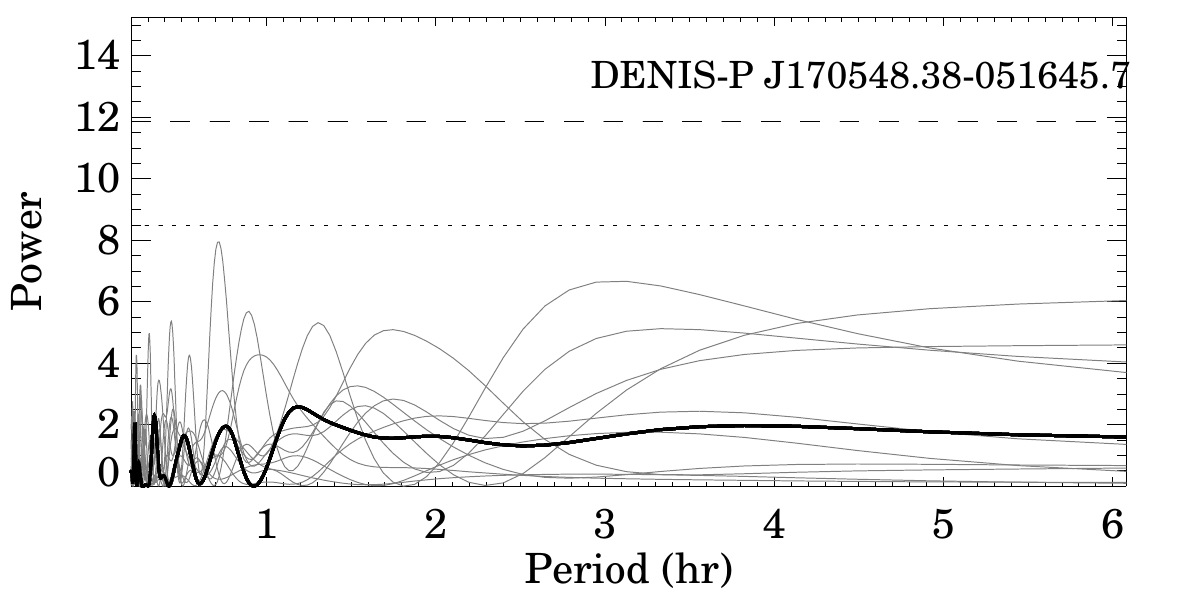} \\

\includegraphics[width=0.24\hsize]{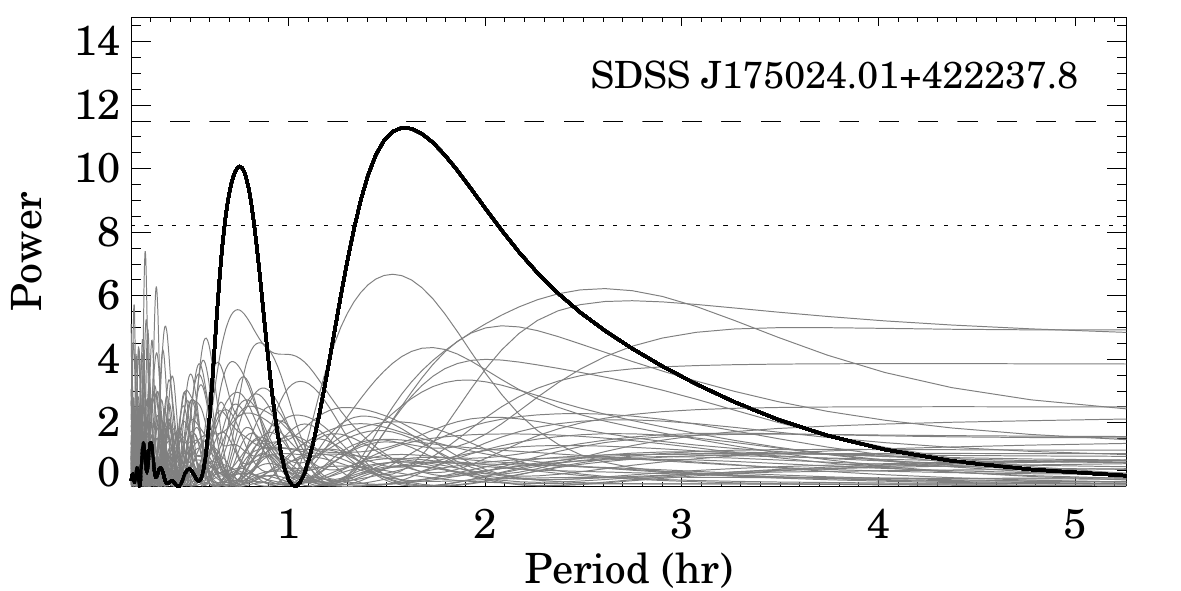} &
\includegraphics[width=0.24\hsize]{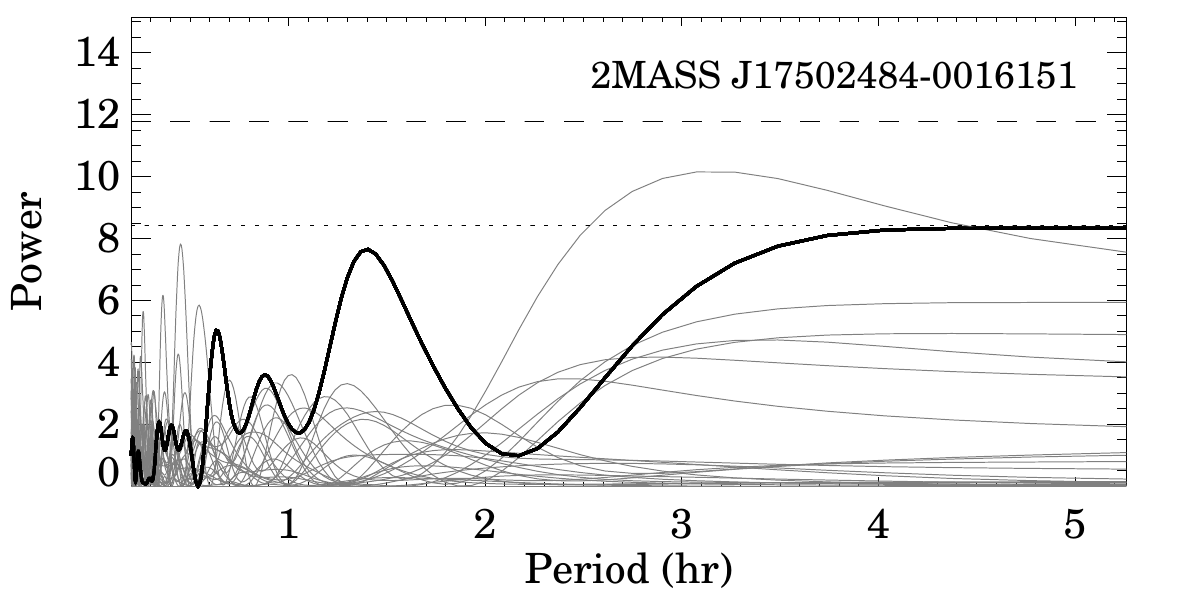} &
\includegraphics[width=0.24\hsize]{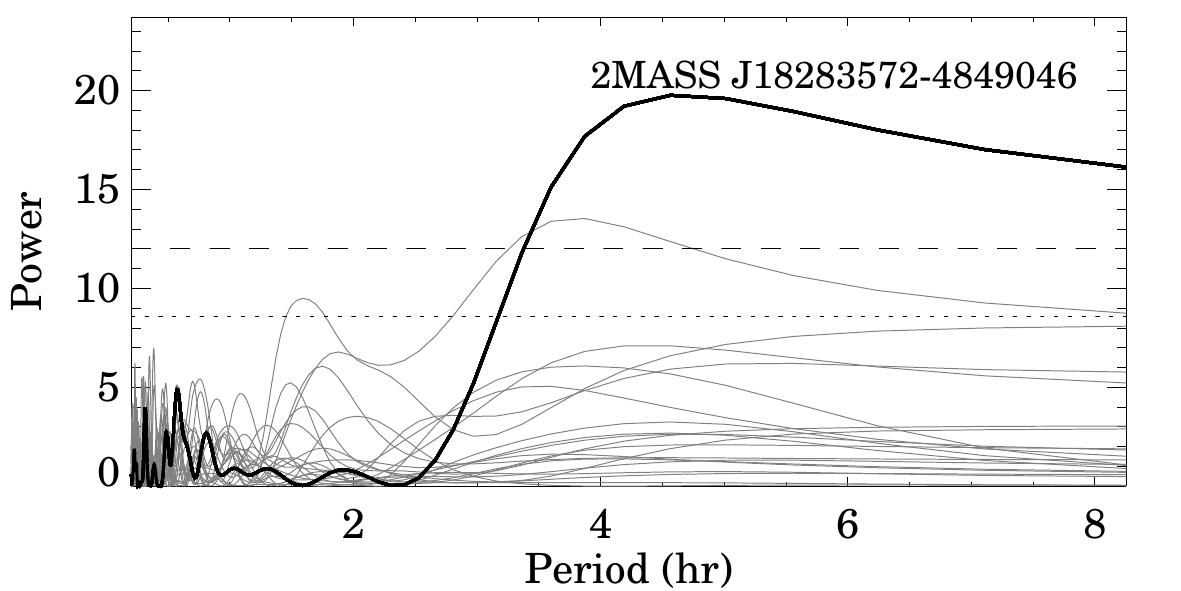} &
\includegraphics[width=0.24\hsize]{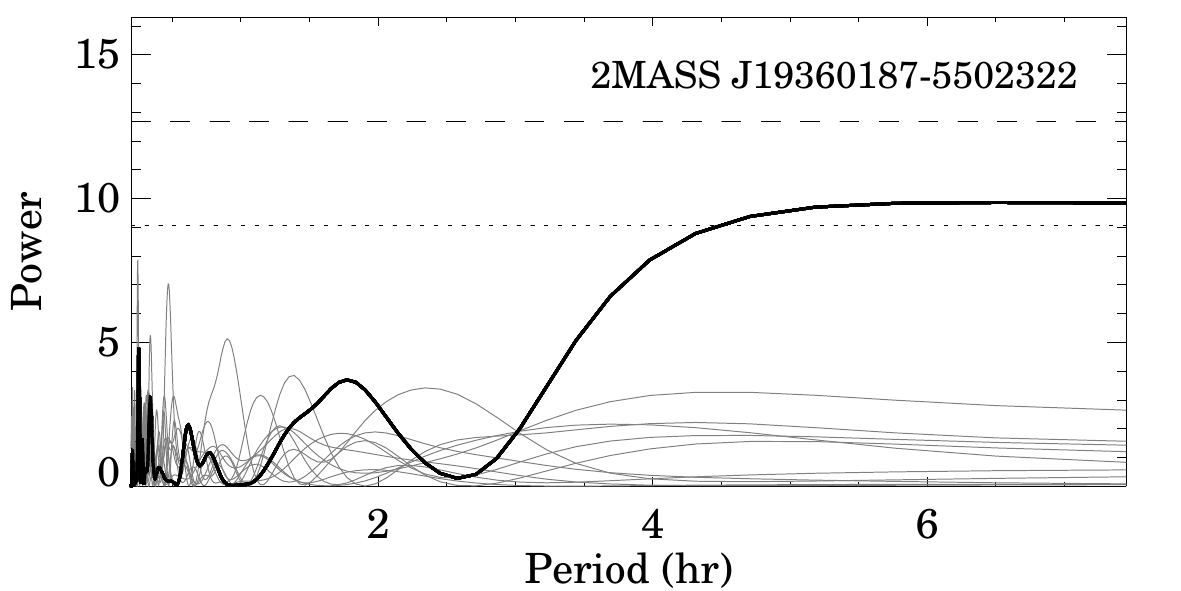} \\

\includegraphics[width=0.24\hsize]{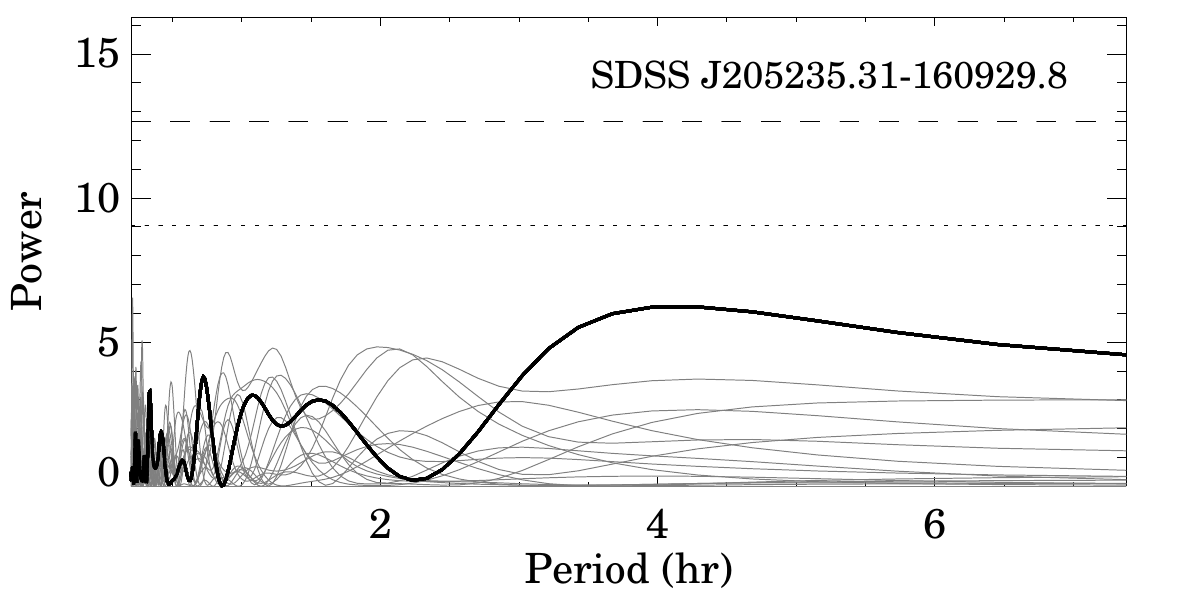} &
\includegraphics[width=0.24\hsize]{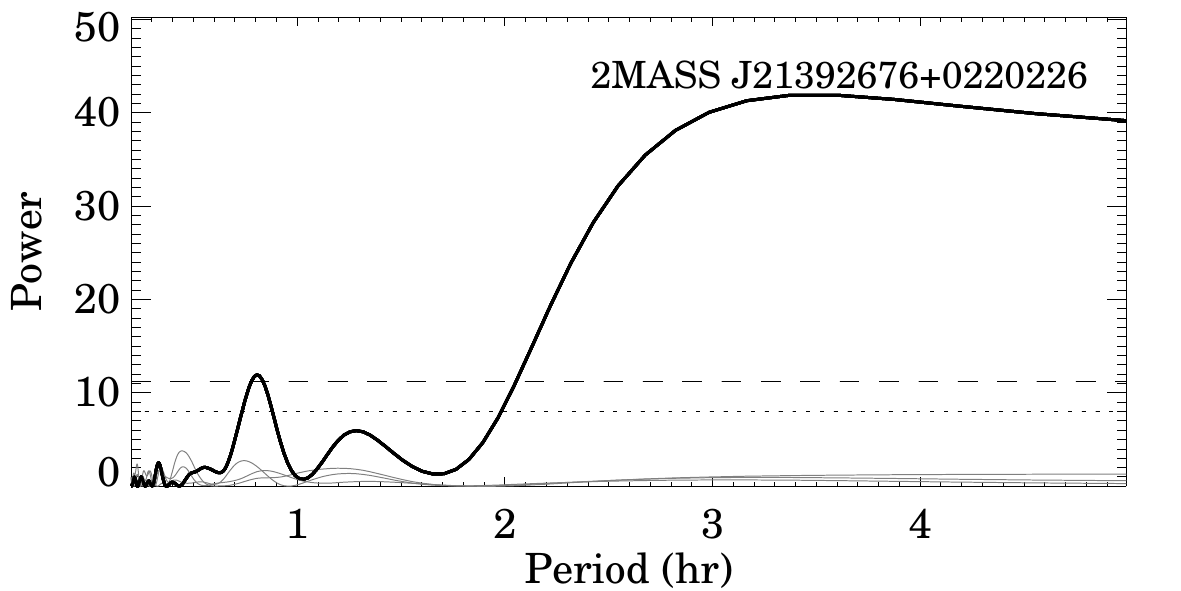} &
\includegraphics[width=0.24\hsize]{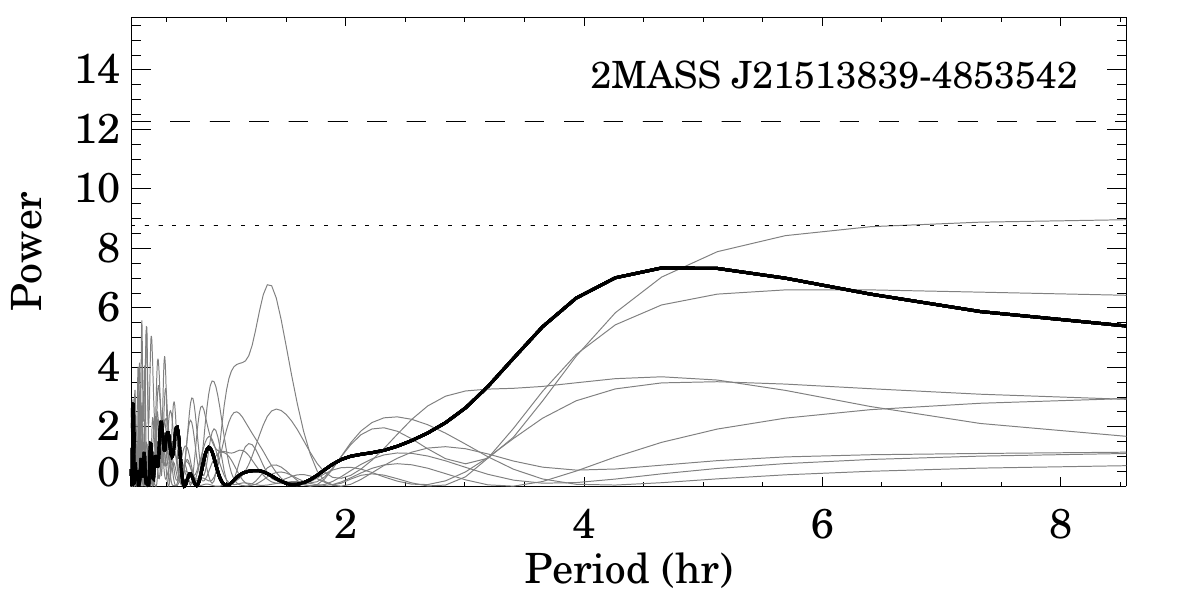} &
\includegraphics[width=0.24\hsize]{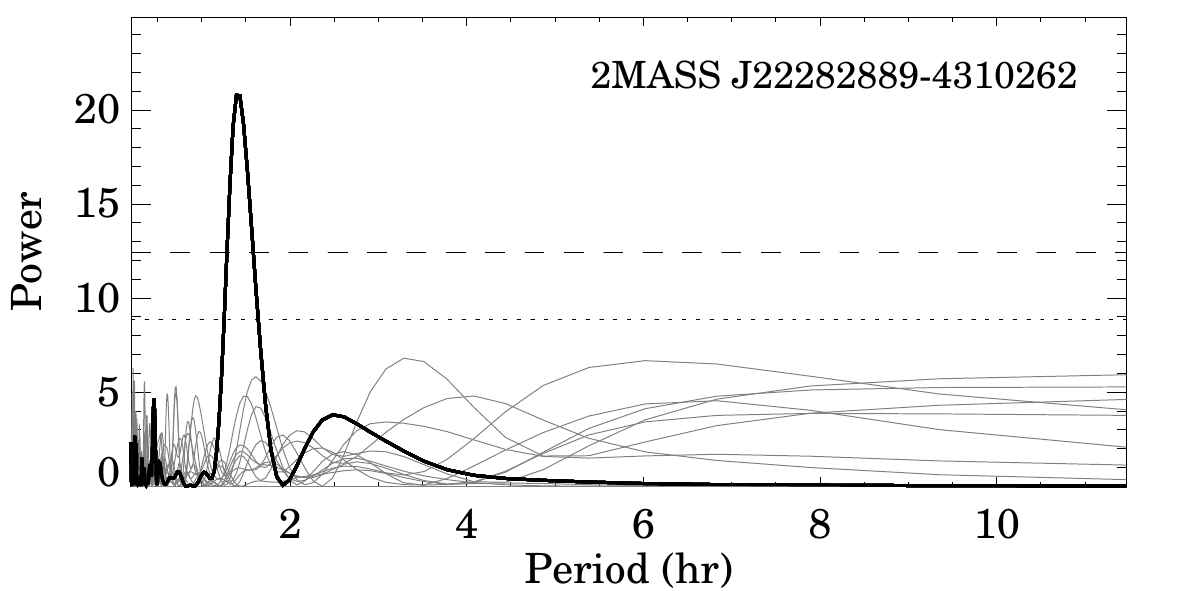} \\

\includegraphics[width=0.24\hsize]{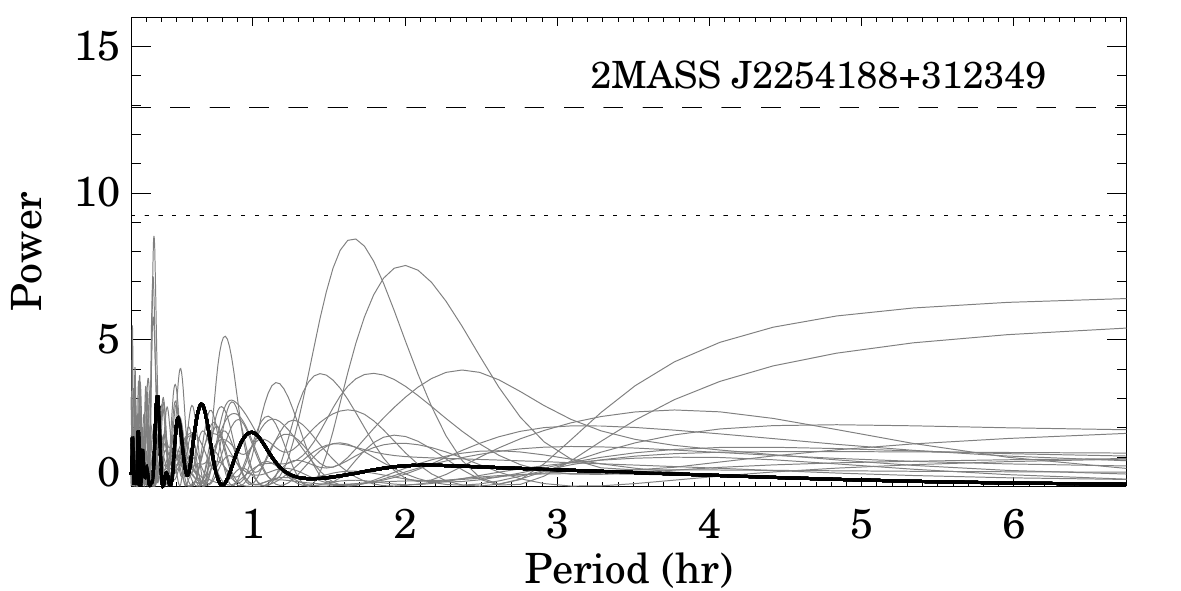} &
\includegraphics[width=0.24\hsize]{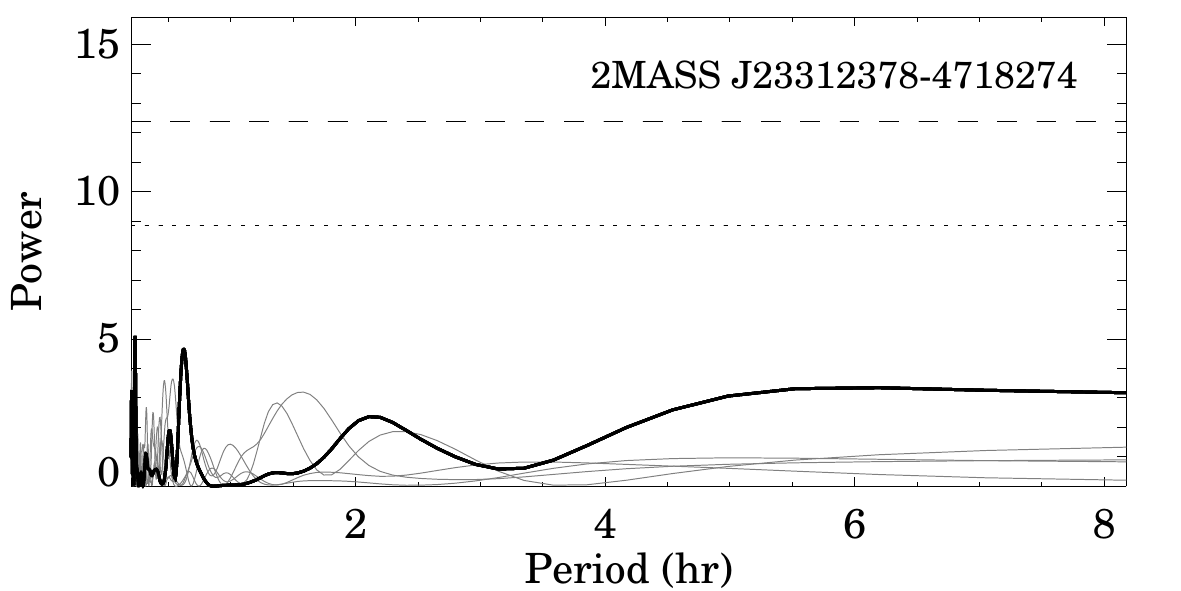}\\
\end{tabular}
\caption{Continued from figure \ref{fig:lsper}. \label{fig:lsper2}}
\end{figure*}

\section*{Acknowledgements}
JR and RJ wish to acknowledge support in the form of a Reasearch, Technology and Instrumentation grant from the Natural Sciences and Engineering Research Council of Canada to RJ, which made this work possible.  We thank Ian Thompson of Carnegie Observatories and the staff of the Las Campanas Observatory, especially Javier Fuentes, Oscar Duhalde and Herman Olivares, for their help in scheduling and carrying out the observations. This research has benefited from the SpeX Prism Spectral Libraries, maintained by Adam Burgasser, the M, L, T, and Y dwarf compendium housed at DwarfArchives.org, the Database of Ultracool parallaxes maintained by Trent Dupuy, and the SIMBAD database operated at CDS, Strasbourg, France. This publication makes use of data products from the Two Micron All Sky Survey, which is a joint project of the University of Massachusetts and the Infrared Processing and Analysis Center/California Institute of Technology, funded by the National Aeronautics and Space Administration and the National Science Foundation. JR acknowledges support from a Vanier Canada Graduate Scholarship from the National Sciences and Engineering Research Council of Canada, and is currently supported by a Giaconni Fellowship through the Space Telescope Science Institute, which is operated by the Association of Universities for Research in Astronomy, Incorporated, under NASA contract NAS5-26555.

\bibliographystyle{/home/radigan/manuscripts/astronat/apj/apj}
\bibliography{mylib}

\begin{deluxetable*}{lccccccccc}
\tabletypesize{\small}
\tablecolumns{9}
\tablewidth{0pt}
\tablecaption{Observing Log \label{tab:ch5_obslog}}
\tablehead{ \colhead{Name} & \colhead{Ref\tablenotemark{a}} & \colhead{Camera} & \colhead{Date} & \colhead{$\Delta t$}  & \colhead{${\rm t_{exp}}$} & \colhead{N$_{{\rm obs}}$} & \colhead{Median FWHM} & \colhead{$\sigma_{\rm raw}$\tablenotemark{b}} & \colhead{Type}\tablenotemark{c}}
\startdata
2MASSW J0030300-145033 & 2 & WIRC & 2009-09-26T04:24:40 & 4.49 & 60.0 & 188 & 0.99 & 0.01 & D \\
2MASS J00501994-3322402 & 3 & WIRC & 2009-08-02T07:37:18 & 3.26 & 60.0 & 175 & 0.73 & 0.10 & S \\
SIMP J013656.57+093347.3 & 4 & WIRC & 2009-07-31T08:44:56 & 2.32 & 40.0 & 177 & 1.42 & 0.03 & S \\
SDSS J015141.69+124429.6 & 5 & WIRCam & 2009-12-30T04:57:02 & 3.72 & 40.0 & 270 & 0.65 & 0.01 & S \\
SDSS J020742.48+000056.2 & 5 & WIRC & 2009-09-25T06:02:03 & 3.97 & 60.0 & 166 & 1.20 & 0.02 & D \\
2MASSI J0243137-245329 & 6 & WIRC & 2009-08-10T07:36:13 & 3.16 & 40.0 & 184 & 1.21 & 0.02 & D \\
2MASS J02572581-3105523 & 7 & WIRC & 2009-09-30T05:49:55 & 3.93 & 30.0 & 379 & 1.19 & 0.01 & S \\
2MASS J03185403-3421292 & 7 & WIRC & 2009-09-24T05:55:46 & 4.28 & 60.0 & 179 & 0.74 & 0.02 & D \\
2MASSI J0328426+230205 & 2 & WIRCam & 2009-12-26T05:14:20 & 3.69 & 40.0 & 270 & 0.38 & 0.02 & S \\
2MASSI J0415195-093506 & 6 & WIRC & 2009-09-23T06:07:36 & 3.79 & 60.0 & 203 & 0.84 & 0.01 & S \\
SDSSp J042348.57-041403.5 & 5 & WIRC & 2009-10-02T06:26:36 & 3.57 & 45.0 & 241 & 1.12 & 0.02 & S \\
2MASS J05103520-4208140 & 8 & WIRC & 2010-02-04T00:46:13 & 3.33 & 60.0 & 172 & 0.99 & 0.03 & D \\
2MASS J05160945-0445499 & 9 & WIRC & 2009-09-22T06:11:15 & 3.99 & 60.0 & 171 & 1.10 & 0.01 & D \\
2MASS J05185995-2828372 & 10 & WIRC & 2010-02-03T00:08:53 & 3.79 & 60.0 & 178 & 0.88 & 0.02 & S \\
SDSSp J053951.99-005902.0 & 11 & WIRC & 2010-02-07T00:27:30 & 1.72 & 45.0 & 119 & 1.29 & 0.01 & S \\
2MASS J05591914-1404488 & 12 & WIRC & 2010-02-01T00:31:38 & 3.52 & 45.0 & 242 & 1.03 & 0.07 & S \\
UGPS J072227.51-054031.2 & 13 & WIRC & 2010-04-22T23:24:07 & 1.74 & 60.0 & 71 & 1.05 & 0.02 & D \\
2MASS J07290002-3954043 & 8 & WIRC & 2010-02-02T01:02:15 & 3.13 & 60.0 & 138 & 0.82 & 0.26 & D \\
SDSS J074201.41+205520.5 & 1 & WIRCam & 2009-12-27T10:47:08 & 3.72 & 40.0 & 270 & 0.57 & 0.01 & S \\
SDSS J075840.33+324723.4 & 1 & WIRCam & 2009-12-26T10:15:20 & 3.54 & 15.0 & 476 & 2.31 & 0.02 & S \\
DENIS J081730.0-615520 & 14 & WIRC & 2010-04-23T23:25:45 & 3.99 & 45.0 & 215 & 0.77 & 0.04 & D \\
SDSS J083048.80+012831.1 & 1 & WIRC & 2010-02-08T00:24:18 & 3.75 & 60.0 & 166 & 1.13 & 0.01 & D \\
SDSS J093109.56+032732.5 & 1 & WIRC & 2010-03-30T00:05:51 & 4.43 & 60.0 & 197 & 0.82 & 0.02 & D \\
2MASS J09393548-2448279 & 3 & WIRC & 2010-04-02T23:29:53 & 5.92 & 60.0 & 262 & 0.79 & 0.07 & D \\
2MASS J09490860-1545485 & 3 & WIRC & 2010-03-26T01:04:35 & 4.48 & 40.0 & 333 & 0.83 & 0.02 & S \\
2MASS J10073369-4555147 & 8 & WIRC & 2010-01-31T03:34:10 & 3.56 & 30.0 & 244 & 0.76 & 0.02 & D \\
2MASSW J1036530-344138 & 15 & WIRC & 2010-03-28T01:43:22 & 4.52 & 60.0 & 179 & 1.05 & 0.02 & D \\
SDSS J104829.21+091937.8 & 16 & WIRC & 2010-03-31T01:14:03 & 4.14 & 60.0 & 184 & 0.78 & 0.02 & D \\
SDSSp J111010.01+011613.1 & 5 & WIRC & 2010-04-03T23:46:45 & 5.72 & 60.0 & 244 & 0.84 & 0.01 & D \\
2MASS J11145133-2618235 & 3 & WIRC & 2010-03-29T03:03:22 & 3.13 & 60.0 & 140 & 0.99 & 0.02 & D \\
2MASS J11220826-3512363 & 3 & WIRC & 2010-02-01T06:07:16 & 2.76 & 60.0 & 122 & 0.73 & 0.02 & D \\
2MASS J11263991-5003550 & 17 & WIRC & 2010-04-26T23:20:23 & 4.91 & 45.0 & 338 & 1.24 & 0.02 & S \\
SDSS J115553.86+055957.5 & 1 & WIRC & 2010-05-04T23:32:28 & 4.57 & 60.0 & 196 & 1.01 & 0.04 & D \\
SDSS J120747.17+024424.8 & 18 & WIRC & 2010-02-08T04:55:18 & 3.84 & 60.0 & 181 & 1.23 & 0.01 & S \\
2MASS J12095613-1004008 & 19 & WIRC & 2010-02-07T05:31:00 & 3.13 & 60.0 & 168 & 1.21 & 0.02 & S \\
2MASS J12154432-3420591 & 8 & WIRC & 2010-03-27T03:57:43 & 3.42 & 60.0 & 143 & 0.87 & 0.03 & D \\
2MASSI J1217110-031113 & 20 & WIRC & 2010-04-23T01:28:51 & 3.18 & 60.0 & 140 & 0.73 & 0.07 & D \\
2MASS J12314753+0847331 & 19 & WIRC & 2010-02-09T05:39:00 & 3.16 & 60.0 & 139 & 1.35 & 0.02 & D \\
SDSSp J125453.90-012247.4 & 21 & WIRC & 2010-02-02T05:43:40 & 3.10 & 45.0 & 143 & 0.95 & 0.45 & D \\
SDSSp J132629.82-003831.5 & 11 & WIRC & 2010-04-02T06:08:34 & 2.60 & 60.0 & 108 & 0.64 & 0.03 & D \\
SDSS J140255.66+080055.2 & 16 & WIRC & 2010-03-30T04:53:38 & 3.19 & 60.0 & 143 & 0.99 & 0.01 & D \\
2MASS J14044941-3159329 & 8 & WIRC & 2010-03-28T06:32:51 & 4.02 & 60.0 & 178 & 1.09 & 0.03 & D \\
SDSS J141624.08+134826.7 & 22 & WIRC & 2010-05-04T03:28:04 & 3.24 & 30.0 & 218 & 0.98 & 0.02 & D \\
DENIS-P J142527.97-365023.4 & 23 & WIRC & 2010-04-25T04:25:08 & 2.98 & 45.0 & 161 & 0.86 & 0.02 & D \\
2MASSW J1507476-162738 & 24 & WIRC & 2010-04-06T05:22:58 & 5.24 & 40.0 & 427 & 0.72 & 0.14 & D \\
SDSS J151114.66+060742.9 & 16 & WIRC & 2010-03-26T07:12:40 & 2.92 & 60.0 & 157 & 0.86 & 0.02 & S \\
SDSS J152103.24+013142.7 & 1 & WIRC & 2010-03-27T07:38:15 & 2.91 & 60.0 & 120 & 1.19 & 0.02 & D \\
2MASSI J1546291-332511 & 6 & WIRC & 2009-07-29T00:44:34 & 3.79 & 60.0 & 203 & 0.97 & 0.02 & S \\
SDSSp J162414.37+002915.6 & 25 & WIRC & 2009-08-09T23:33:14 & 3.58 & 60.0 & 192 & 1.38 & 0.20 & S \\
SIMP J162918.41+033537.0  & 29 & WIRC & 2009-07-29T23:47:39 & 4.03 & 60.0 & 216 & 0.96 & 0.01 & S \\
SDSS J163359.23-064056.5 & 16 & WIRC & 2010-04-22T05:13:23 & 5.44 & 60.0 & 249 & 0.86 & 0.01 & D \\
DENIS-P J170548.38-051645.7 & 23 & WIRC & 2010-04-25T07:47:48 & 3.04 & 25.0 & 235 & 0.90 & 0.02 & D \\
SDSS J175024.01+422237.8 & 1 & WIRCam & 2009-08-09T07:53:44 & 2.63 & 40.0 & 224 & 0.96 & 0.03 & S \\
2MASS J17502484-0016151 & 26 & WIRC & 2010-05-01T06:05:06 & 2.63 & 30.0 & 250 & 1.70 & 0.03 & S \\
2MASS J18283572-4849046 & 19 & WIRC & 2009-08-11T00:55:16 & 4.13 & 40.0 & 314 & 0.90 & 0.04 & S \\
2MASS J19360187-5502322 & 27 & WIRC & 2010-05-04T07:00:42 & 3.73 & 45.0 & 195 & 0.90 & 0.01 & D \\
SDSS J205235.31-160929.8 & 16 & WIRC & 2009-07-30T04:41:41 & 3.69 & 60.0 & 198 & 0.89 & 0.01 & S \\
2MASS J21392676+0220226 & 27 & WIRC & 2009-08-02T04:38:41 & 2.50 & 40.0 & 140 & 0.71 & 0.05 & D \\
2MASS J21513839-4853542 & 28 & WIRC & 2009-07-29T05:34:29 & 4.28 & 60.0 & 229 & 0.96 & 0.01 & S \\
2MASS J22282889-4310262 & 9 & WIRC & 2009-08-01T05:03:12 & 5.73 & 60.0 & 307 & 1.49 & 0.01 & S \\
2MASSI J2254188+312349 & 6 & WIRCam & 2009-11-03T05:28:16 & 3.37 & 40.0 & 231 & 1.60 & 0.01 & S \\
2MASS J23312378-4718274 & 19 & WIRC & 2009-09-30T00:02:44 & 4.09 & 60.0 & 212 & 1.31 & 0.02 & S
\enddata
\vspace{0.cm}
\tablenotetext{a}{Discovery references}
\tablenotetext{b}{\footnotesize RMS deviations of the normalized, median-combined raw light curves of reference stars on-chip.  Large values indicate variable or cloudy observing conditions.}
\tablenotetext{c}{{\footnotesize Method used to obtain light curve.  D=``dithered'', S=``staring''}}
\tablerefs{{\footnotesize 
(1)Knapp et al. (2004) (2)Kirkpatrick et al. (2000) (3)Tinney et al. (2005)
(4)Artigau et al. (2006) (5)Geballe et al. (2002) (6)Burgasser et al. (2002)
(7)Kirkpatrick et al. (2008) (8)Looper, Kirkpatrick, \& Burgasser (2007)
(9)Burgasser, McElwain, \& Kirkpatrick (2003) (10)Cruz et al. (2004)
(11)Fan et al. (2000) (12)Burgasser et al. (2000)b (13)Lucas et al. (2010)
(14)Artigau et al. (2010) (15)Gizis (2002) (16)Chiu et al. (2006)
(17)Folkes et al. (2007) (18)Hawley et al. (2002) (19)Burgasser et al. (2004)
(20)Burgasser et al. (1999) (21)Leggett et al. (2000) (22)Bowler, Liu, \& Dupuy
(23)Kendall et al. (2004) (24)Reid et al. (2000) (25)Strauss et al. (1999)
(26)Kendall et al. (2007) (27)Reid et al. (2008) (28)Ellis et al. (2005) (29)Artigau et al., in prep
}}
\end{deluxetable*}

\begin{deluxetable*}{lcccccl}[ht!]
\tabletypesize{\footnotesize}
\tablecolumns{7}
\tablewidth{0pt}
\tablecaption{Detections \label{tab:detections}}
\tablehead{ \colhead{Name} & \colhead{SpT} & \colhead{$J$} & \colhead{$J-K_s$} & \colhead{$A_{\rm sin}$ (\%)} & \colhead{$P_{\rm sin}$ (hr)} & \colhead{Notes}\\
\hline\hline\\
\multicolumn{7}{c}{Significant Detections ($p>$99\%)}}
\startdata
SIMP J013656.57+093347.3 & T2.5 & 13.45 & 0.893 & 2.9$\pm$0.3 & 2.4 & Period from \citet{artigau09} \\
2MASS J05591914-1404488 & T4.5 & 13.58 & 0.22 & $\sim$0.7 & $\sim$10 & Linear trend: amplitude/period highly uncertain  \\
SDSS J075840.33+324723.4 & T2 & 14.95 & 1.07 & 4.8$\pm$0.2 & 4.9$\pm$0.2 & \\
DENIS J081730.0-615520 & T6.5 & 13.61 & 0.09 & 0.6$\pm$0.1 & 2.8$\pm$0.2 & \\
2MASS J11263991-5003550 & L6.5 & 14.00 & 1.17 & 1.2$\pm$0.1 & 4.0$\pm$0.1  & \\
SIMP J162918.41+033537.0 & T2 & 15.06 & 0.22 & 4.3$\pm$2.4 &  6.9$\pm$2.4 & \\
2MASS J18283572-4849046 & T5.5 & 15.18 & 0.00 & 0.9$\pm$0.1 & 5.0$\pm$0.6 &  Similar trend at lower significance in a comparison star\\
2MASS J21392676+0220226 & T1.5 & 15.26 & 1.68 & 9$\pm$1 &  7.72 & Period from \citet{radigan12} \\
2MASS J22282889-4310262 & T6 & 15.66 & 0.37 & 1.6$\pm$0.3 & 1.42$\pm$0.03 & \\
\cutinhead{Marginal Detections ($p>$96\%)}
DENIS-P J142527.97-365023.4 & L5 & 13.75 & 1.94 & 0.6$\pm$0.1 & 3.7$\pm$0.8 &  \\
SDSS J175024.01+422237.8 & T2 & 16.47 & 0.98 & 1.5$\pm$0.3 & 2.7$\pm$0.2 & \\
2MASS J19360187-5502322 & L5 & 14.49 & 1.44 & \nodata & \nodata & Linear trend, amplitude/period not well defined
\enddata
\tablecomments{Periods and amplitudes were estimated from sinusoidal fits to the data unless otherwise noted. Amplitudes are expressed as peak-to-peak values.  Periods and uncertainties reported here are only accurate in the ideal case of a sinusoidal signal, and should be interpreted as rough estimates.  In the case of non-sinusoidal signals the reported uncertainties are almost certainly underestimated. See Table 4 for spectral type and photometry references.}
\end{deluxetable*}

\clearpage
\LongTables 
\turnpage
\begin{deluxetable}{lccccccccccccc}[h]
\tabletypesize{\footnotesize}
\tablecolumns{14}
\tablewidth{0pt}
\tablecaption{Target Information \label{tab:ch5_targ}}
\tablehead{ \colhead{Name} & \multicolumn{3}{c}{Spectral Type} & \multicolumn{2}{c}{2MASS} & \multicolumn{3}{c}{MKO} & \multicolumn{2}{c}{2MASS/SpeX\tablenotemark{d}} & \colhead{$\Pi$} & \colhead{Ref} \\
\colhead{} & \colhead{Opt} & \colhead{NIR} & \colhead{Ref} & \colhead{$J$ (mag)} & \colhead{$K_s$ (mag)} & \colhead{$J$ (mag)} & \colhead{$K$ (mag)} & \colhead{Ref\tablenotemark{e}} & \colhead{$J-K_s$} & \colhead{Ref} & \colhead{(mas)} & \colhead{}}
\startdata
2M0030$-$14 & L7 & \nodata & 33 & 16.28$\pm$0.11 & 14.48$\pm$0.10 & 16.39$\pm$0.03 & 14.49$\pm$0.03 & 36 &  2.06$\pm$0.07 & 15 &  37.4$\pm$4.5 & 49 \\
2M0050$-$33 & \nodata & T7 & 10 & 15.93$\pm$0.07 & 15.24$\pm$0.19 & 15.65$\pm$0.10 & 15.91$\pm$0.10 & 36 & 0.12$\pm$0.07 & 10 & \nodata & \nodata \\
SI\,0136$+$09 & \nodata & T2.5 & 2 & 13.45$\pm$0.03 & 12.56$\pm$0.02 & 13.24$\pm$0.03 & 12.62$\pm$0.03 & 20 & 0.89$\pm$0.04 & 20 & \nodata & \nodata \\
SD0151$+$12 & \nodata & T1 & 10 & 16.57$\pm$0.13 & 15.18$\pm$0.19 & 16.25$\pm$0.05 & 15.18$\pm$0.05 & 36 &  1.21$\pm$0.07 & 9 &  46.7$\pm$3.4 & 49 \\
SD0207$+$00 & \nodata & T4.5 & 10 & 16.80$\pm$0.16 & 16.39$\pm$0.10\tablenotemark{c} & 16.63$\pm$0.05 & 16.62$\pm$0.05 & 36 &  0.41$\pm$0.07 & 10 &  29.3$\pm$4.0 & \nodata \\
2M0243$-$24 & \nodata & T6 & 10 & 15.38$\pm$0.05 & 15.22$\pm$0.17 & 15.13$\pm$0.03 & 15.34$\pm$0.03 & 36 &  0.30$\pm$0.07 & 9 &  93.6$\pm$3.6 & 49 \\
2M0257$-$31 & L8 & \nodata & 34 & 14.67$\pm$0.04 & 12.88$\pm$0.03 & 14.52$\pm$0.04 & 12.86$\pm$0.03 & 20 &  1.79$\pm$0.05 & 20 & \nodata & \nodata \\
2M0318$-$34 & L7 & \nodata & 34 & 15.57$\pm$0.05 & 13.51$\pm$0.04 & 15.43$\pm$0.06 & 13.48$\pm$0.04 & 20 &  2.06$\pm$0.07 & 20 &  72.9$\pm$7.7 & 25 \\
2M0328$+$23 & L8 & L9.5 & 33,35 & 16.69$\pm$0.14 & 14.92$\pm$0.11 & 16.35$\pm$0.03 & 14.87$\pm$0.03 & 36 &  1.61$\pm$0.07 & 14 &  33.1$\pm$4.2 & 49 \\
2M0415$-$09 & T8 & T8 & 8,10 & 15.69$\pm$0.06 & 15.43$\pm$0.20 & 15.32$\pm$0.03 & 15.83$\pm$0.03 & 36 & -0.14$\pm$0.07 & 9 & \nodata & \nodata \\
SD0423$-$04 & L7.5 & T0 & 18,10 & 14.47$\pm$0.03 & 12.93$\pm$0.03 & 14.30$\pm$0.03 & 12.96$\pm$0.03 & 36 &  1.54$\pm$0.04 & 20 &  72.1$\pm$1.1 & 23 \\
2M0510$-$42 & \nodata & T5 & 39 & 16.22$\pm$0.09 & 16.00$\pm$0.28 & 15.95$\pm$0.09 & 16.10$\pm$0.28 & 20 &  0.24$\pm$0.07 & 39 & \nodata & \nodata \\
2M0516$-$04 & \nodata & T5.5 & 10 & 15.98$\pm$0.08 & 15.49$\pm$0.20 & 15.70$\pm$0.08 & 15.60$\pm$0.20 & 20 &  0.15$\pm$0.07 & 20 &  44.5$\pm$6.5 & 25 \\
2M0518$-$28 & L7 & T1p & 34,10 & 15.98$\pm$0.10 & 14.16$\pm$0.07 & 15.79$\pm$0.10 & 14.19$\pm$0.07 & 20 &  1.67$\pm$0.07 & 19 &  43.7$\pm$0.8 & 23 \\
SD0539$-$00 & L5 & L5 & 26,35 & 14.03$\pm$0.03 & 12.53$\pm$0.02 & 13.85$\pm$0.03 & 12.40$\pm$0.03 & 20 &  1.50$\pm$0.04 & 37 &  76.1$\pm$2.2 & 49 \\
2M0559$-$14 & T5 & T4.5 & 8,10 & 13.80$\pm$0.02 & 13.58$\pm$0.05 & 13.57$\pm$0.03 & 13.73$\pm$0.03 & 36 &  0.23$\pm$0.06 & 10 & \nodata & \nodata \\
UG0722$-$05 & \nodata & T9+ & 40 &  16.90$\pm$0.02\tablenotemark{a} & 16.91$\pm$0.08\tablenotemark{a} & 16.52$\pm$0.02 & 17.07$\pm$0.08 & 40 & -0.01$\pm$0.08 & 40 & \nodata & \nodata \\
2M0729$-$39 & \nodata & T8 & 39 & 15.92$\pm$0.08 & 16.51$\pm$0.10 \tablenotemark{b} & 15.57$\pm$0.08 & 16.66$\pm$0.10 & 20 & -0.59$\pm$0.07 & 39 & 126.3$\pm$8.3 & 25 \\
SD0742$+$20 & \nodata & T5 & 10 & 16.19$\pm$0.09 & 16.19$\pm$0.11\tablenotemark{c} & 15.60$\pm$0.03 & 16.06$\pm$0.03 & 35 &  0.00$\pm$0.07 & 15 &  66.5$\pm$8.6 & 25 \\
SD0758$+$32 & \nodata & T2 & 10 & 14.95$\pm$0.04 & 13.88$\pm$0.06 & 14.78$\pm$0.03 & 13.87$\pm$0.03 & 35 &  1.07$\pm$0.07 & 20 & \nodata & \nodata \\
DE0817$-$61 & \nodata & T6 & 3 & 13.61$\pm$0.02 & 13.52$\pm$0.04 & 13.32$\pm$0.02 & 13.64$\pm$0.04 & 20 & 0.09$\pm$0.05 & 20 & 203.0$\pm$13.0 & 3 \\
SD0830$+$01 & \nodata & T4.5 & 10 & 16.29$\pm$0.11 & 16.17$\pm$0.13\tablenotemark{c} & 15.99$\pm$0.03 & 16.38$\pm$0.03 & 36 &  0.12$\pm$0.07 & 15 &  43.1$\pm$6.1 & 25 \\
SD0931$+$03 & \nodata & L7.5 & 35 & 16.35$\pm$0.05\tablenotemark{b}  & 15.55$\pm$0.05\tablenotemark{b}  & 16.60$\pm$0.05 & 15.53$\pm$0.05 & 35 & 1.09$\pm$0.13 & 20 & \nodata & \nodata \\
2M0939$-$24 & \nodata & T8 & 10 & 15.98$\pm$0.11 & 16.85$\pm$0.13\tablenotemark{c} & 15.61$\pm$0.09 & 16.83$\pm$0.09 & 35 & -0.81$\pm$0.07 & 10 & 187.3$\pm$4.6 & 13 \\
2M0949$-$15 & \nodata & T2 & 10 & 16.15$\pm$0.12 & 15.23$\pm$0.17 & 15.94$\pm$0.12 & 15.27$\pm$0.17 & 20 &  0.85$\pm$0.07 & 10 &  55.3$\pm$6.6 & 25 \\
2M1007$-$45 & \nodata & T5 & 39 & 15.65$\pm$0.07 & 15.56$\pm$0.23 & 15.38$\pm$0.07 & 15.66$\pm$0.23 & 20 &  0.21$\pm$0.07 & 39 &  71.0$\pm$5.2 & 25 \\
2M1036$-$34 & L6 & \nodata & 29 & 15.62$\pm$0.05 & 13.80$\pm$0.04 & 15.49$\pm$0.05 & 13.77$\pm$0.04 & 20 &  1.82$\pm$0.06 & 20 &  61.5$\pm$9.1 & 25 \\
SD1048$+$09 & \nodata & T2.5 & 17 & 16.59$\pm$0.15 & 15.77$\pm$0.17\tablenotemark{c} & 16.39$\pm$0.03 & 15.87$\pm$0.03 & 36 &  0.83$\pm$0.07 & 15 & \nodata & \nodata \\
SD1110$+$01 & \nodata & T5.5 & 10 & 16.34$\pm$0.12 & 15.96$\pm$0.14\tablenotemark{c} & 16.12$\pm$0.05 & 16.05$\pm$0.05 & 36 & 0.38$\pm$0.07 & 10 &  52.1$\pm$1.2 & 23 \\
2M1114$-$26 & \nodata & T7.5 & 10 & 15.86$\pm$0.08 & 16.45$\pm$0.11\tablenotemark{c} & 15.52$\pm$0.05 & 16.54$\pm$0.05 & 36 & -0.59$\pm$0.07 & 10 & \nodata & \nodata \\
2M1122$-$35 & \nodata & T2 & 10 & 15.02$\pm$0.04 & 14.38$\pm$0.07 & 14.81$\pm$0.04 & 14.43$\pm$0.07 & 20 &  0.64$\pm$0.08 & 20 & \nodata & \nodata \\
2M1126$-$50 & L4.5 & L6.5 & 12 & 14.00$\pm$0.03 & 12.83$\pm$0.03 & 13.86$\pm$0.03 & 12.80$\pm$0.03 & 20 &  1.17$\pm$0.04 & 20 & \nodata & \nodata \\
SD1155$+$05 & \nodata & L7.5 & 35 & 15.66$\pm$0.08 & 14.12$\pm$0.07 & 15.63$\pm$0.03 & 14.09$\pm$0.03 & 35 &  1.67$\pm$0.07 & 15 &  57.9$\pm$10.2 & 25 \\
SD1207$+$02 & L8 & T0 & 30,10 & 15.58$\pm$0.07 & 13.99$\pm$0.06 & 15.38$\pm$0.03 & 14.16$\pm$0.03 & 35 &  1.59$\pm$0.09 & 39 &  44.5$\pm$12.2 & 25 \\
2M1209$-$10 & T3.5 & T3 & 34,10 & 15.91$\pm$0.07 & 15.06$\pm$0.14 & 15.55$\pm$0.03 & 15.17$\pm$0.03 & 36 &  0.71$\pm$0.07 & 9 &  45.8$\pm$1.0 & 23 \\
2M1215$-$34 & \nodata & T4.5 & 39 & 16.24$\pm$0.13 & 15.86$\pm$0.15\tablenotemark{c} & 15.98$\pm$0.13 & 15.95$\pm$0.15 & 20 &  0.37$\pm$0.07 & 39 &  39.8$\pm$8.9 & 25 \\
2M1217$-$03 & T7 & T7.5 & 8,10 & 15.86$\pm$0.06 & 15.69$\pm$0.09\tablenotemark{c} & 15.56$\pm$0.03 & 15.92$\pm$0.03 & 36 & 0.17$\pm$0.07 & 10 &  90.8$\pm$2.2 & 47 \\
2M1231$+$08 & \nodata & T5.5 & 10 & 15.57$\pm$0.07 & 15.22$\pm$0.19 & 15.14$\pm$0.03 & 15.46$\pm$0.03 & 36 &  0.04$\pm$0.07 & 9 & \nodata & \nodata \\
SD1254$-$01 & T2 & T2 & 8,10 & 14.89$\pm$0.04 & 13.84$\pm$0.05 & 14.66$\pm$0.03 & 13.84$\pm$0.03 & 36 &  1.06$\pm$0.06 & 20 &  84.9$\pm$1.9 & 21 \\
SD1326$-$00 & L8: & L5.5 & 26,35 & 16.10$\pm$0.07 & 14.21$\pm$0.07 & 16.21$\pm$0.03 & 14.17$\pm$0.03 & 36 &  1.9$\pm$0.09 & 20 &  50.0$\pm$6.3 & 49 \\
SD1402$+$08 & \nodata & T1.5 & 17 & 16.84$\pm$0.18 & 15.59$\pm$0.26 & 16.85$\pm$0.03 & 15.73$\pm$0.03 & 36 & 1.25$\pm$0.31 & 20 & \nodata & \nodata \\
2M1404$-$31 & T0 & T2.5 & 38,39 & 15.58$\pm$0.06 & 14.54$\pm$0.09 & 15.36$\pm$0.06 & 14.59$\pm$0.10 & 20 &  0.99$\pm$0.07 & 39 &  42.1$\pm$1.1 & 23 \\
SD1416$+$13 & L6 & L6p & 4 & 13.15$\pm$0.03 & 12.11$\pm$0.02 & 12.99$\pm$0.01 & 12.05$\pm$0.01 & 11 & 1.03$\pm$0.03 & 20 & 109.7$\pm$1.3 & 23 \\
DE1425$-$36 & L3: & L5 & 43,31 & 13.75$\pm$0.03 & 11.81$\pm$0.03 & 13.62$\pm$0.03 & 11.77$\pm$0.03 & 20 & 1.94$\pm$0.04 & 20 & \nodata & \nodata \\
2M1507$-$16 & L5 & L5.5 & 33,35 & 12.83$\pm$0.03 & 11.31$\pm$0.03 & 12.70$\pm$0.03 & 11.28$\pm$0.03 & 20 &  1.52$\pm$0.04 & 20 & 136.4$\pm$0.6 & 21 \\
SD1511$+$06 & \nodata & T0 & 17 & 16.02$\pm$0.08 & 14.54$\pm$0.10 & 15.83$\pm$0.03 & 14.52$\pm$0.03 & 36 &  1.56$\pm$0.07 & 15 &  36.7$\pm$6.4 & 25 \\
SD1521$+$01 & \nodata & T2: & 10 & 16.40$\pm$0.10 & 15.35$\pm$0.17 & 16.06$\pm$0.03 & 15.48$\pm$0.03 & 35 &  0.82$\pm$0.07 & 15 &  41.3$\pm$7.2 & 25 \\
2M1546$-$33 & \nodata & T5.5 & 10 & 15.63$\pm$0.05 & 15.48$\pm$0.18 & 15.35$\pm$0.05 & 15.75$\pm$0.10 & 20 & -0.01$\pm$0.07 & 14 &  88.0$\pm$1.9 & 47 \\
SD1624$+$00 & \nodata & T6 & 10 & 15.49$\pm$0.05 & 15.52$\pm$0.09\tablenotemark{c} & 15.20$\pm$0.05 & 15.61$\pm$0.05 & 36 & -0.08$\pm$0.07 & 10 &  90.9$\pm$1.2 & 47 \\
SI\,1629$+$03 & \nodata & T2 & 22 & 15.29$\pm$0.04 & 14.18$\pm$0.06 & 15.08$\pm$0.04 & 14.23$\pm$0.06 & 20 & 1.11$\pm$0.08& 20 & \nodata & \nodata \\
SD1633$-$06 & \nodata & L6 & 17 & 16.14$\pm$0.09 & 14.54$\pm$0.09 & 16.00$\pm$0.03 & 14.54$\pm$0.03 & 36 &  1.52$\pm$0.07 & 17 & \nodata & \nodata \\
DE1705$-$05 & L0.5 & L4 & 43,31 & 13.31$\pm$0.03 & 12.03$\pm$0.02 & 13.18$\pm$0.03 & 11.99$\pm$0.02 & 20 &  1.28$\pm$0.04 & 20 &  44.5$\pm$12.0 & 1 \\
SD1750$+$42 & \nodata & T2 & 10 & 16.47$\pm$0.10 & 15.48$\pm$0.17 & 16.12$\pm$0.03 & 15.31$\pm$0.03 & 35 &  1.08$\pm$0.07 & 10 & \nodata & \nodata \\
2M1750$-$00 & \nodata & L5.5 & 32 & 13.29$\pm$0.02 & 11.85$\pm$0.02 & 13.16$\pm$0.02 & 11.81$\pm$0.02 & 20 &  1.44$\pm$0.03 & 20 & 108.5$\pm$2.6 & 1 \\
2M1828$-$48 & \nodata & T5.5 & 10 & 15.18$\pm$0.06 & 15.18$\pm$0.14 & 14.89$\pm$0.06 & 15.29$\pm$0.14 & 20 &  0.16$\pm$0.07 & 9 &  83.7$\pm$7.7 & 25 \\
2M1936$-$55 & L5: & \nodata & 43 & 14.49$\pm$0.04 & 13.05$\pm$0.03 & 14.36$\pm$0.04 & 13.01$\pm$0.03 & 20 & 1.44$\pm$0.05 & 20 &  66.3$\pm$5.4 & 25 \\
SD2052$-$16 & \nodata & T1 & 17 & 16.33$\pm$0.12 & 15.12$\pm$0.15 & 16.04$\pm$0.03 & 15.00$\pm$0.03 & 36 &  1.32$\pm$0.07 & 15 &  33.9$\pm$0.8 & 23 \\
2M2139$+$02 & \nodata & T1.5 & 10 & 15.26$\pm$0.05 & 13.58$\pm$0.05 & 14.8$\pm$0.15 & 13.66$\pm$0.09 & 52 &  1.35$\pm$0.07\tablenotemark{f} & 10 & 101.5$\pm$2.0 & 51 \\
2M2151$-$48 & \nodata & T4 & 10 & 15.73$\pm$0.07 & 15.43$\pm$0.18 & 15.48$\pm$0.07 & 15.51$\pm$0.18 & 20 &  0.29$\pm$0.07 & 10 &  50.4$\pm$6.7 & 25 \\
2M2228$-$43 & \nodata & T6 & 10 & 15.66$\pm$0.07 & 15.30$\pm$0.21 & 15.37$\pm$0.07 & 15.41$\pm$0.21 & 20 &  0.25$\pm$0.07 & 9 &  94.0$\pm$7.0 & 25 \\
2M2254$+$31 & \nodata & T4 & 10 & 15.26$\pm$0.05 & 14.90$\pm$0.15 & 15.01$\pm$0.03 & 15.03$\pm$0.03 & 36 &  0.48$\pm$0.07 & 9 & \nodata & \nodata \\
2M2331$-$47 & \nodata & T5 & 10 & 15.66$\pm$0.07 & 15.39$\pm$0.20 & 15.56$\pm$0.10 & 16.40$\pm$0.10 & 36 &  0.38$\pm$0.07 & 9 & \nodata & \nodata
\enddata
\tablenotetext{a}{Photometry from \citet{lucas10}, converted to the 2MASS system using the relation of \citet{stephens04}.}
\tablenotetext{b}{Photometry from \citet{knapp04}, converted to the 2MASS system using the relation of  \citet{stephens04}.}
\tablenotetext{c}{These sources were undetected in the 2MASS $K_{\rm s}$ band.  Reported magnitudes were computed from synthetic $J-K_{\rm s}$ colors, using SpeX prism data.}
\tablenotetext{d}{2MASS $J-K_s$ magnitudes are provided except where uncertainties were larger than 0.1\,mag, in which case synthetic colors were computed from SpeX prism data if available.}
\tablenotetext{e}{MKO photometry with reference to \citet{cutri03} was obtained by converting 2MASS magnitudes to the MKO system using the relation of \citet{stephens04}.}
\tablenotetext{f}{The SpeX $J-K_s$ color of 2M2139$+$02 has been adopted.  The 2MASS color for this source is significantly redder than the color inferred from two subsequent SpeX epochs as well as the range of colors inferred from photometric monitoring \citep{radigan12,apai13}. }
\tablerefs{
(1)\citet{andrei11}; (2)\citet{artigau06}; (3)\citet{artigau10}; (4)\citet{bowler10}; (5)\citet{burgasser99}; (6)\citet{burgasser00b}; (7)\citet{burgasser02_spt1}; (8)\citet{burgasser03_opt}; (9)\citet{burgasser04}; (10)\citet{burgasser06}; (11)\citet{burningham10}; (13)\citet{burgasser08b}; (14)\citet{burgasser08c}; (15)\citet{burgasser10}; (16)\citet{burgasser03_2mass}; (17)\citet{chiu06}; (18)\citet{cruz03}; (19)\citet{cruz04}; (20)\citet{cutri03}; (21)\citet{dahn02}; (22)\citet{deacon11}; (23)\citet{dupuy12}; (24)\citet{ellis05}; (25)\citet{faherty12}; (26)\citet{fan00}; (27)\citet{folkes07}; (28)\citet{geballe02}; (29)\citet{gizis02}; (30)\citet{hawley02}; (31)\citet{kendall04}; (32)\citet{kendall07}; (33)\citet{kirkpatrick00}; (34)\citet{kirkpatrick08}; (35)\citet{knapp04}; (36)\citet{leggett10}; (37)\citet{leggett00}; (38)\citet{looper08}; (39)\citet{looper07}; (40)\citet{lucas10}; (41)\citet{marocco10}; (42)\citet{reid00}; (43)\citet{reid08}; (44)\citet{siegler07}; (45)\citet{stephens04}; (46)\citet{strauss99}; (47)\citet{tinney03}; (48)\citet{tinney05}; (49)\citet{vrba04}; (50)M. Cushing (unpublished); (51)\citet{smart13}; (52)\citet{radigan12}
}
\end{deluxetable}

    \turnpage
   \clearpage

\begin{deluxetable}{lccccccc}[h]
\tabletypesize{\footnotesize}
\tablecolumns{8}
\tablewidth{0pt}
\tablecaption{Light Curve Information \label{tab:lc_info}}
\tablehead{
\colhead{}  & \colhead{}   & \colhead{}   & \colhead{}   & \colhead{}   & \colhead{}   &\colhead{}    &   \colhead{High Ang. Res.} \\ 
\colhead{Object} & \colhead{$\delta t$} & \colhead{$\sigma_a$}\tablenotemark{a} & \colhead{$\beta$}\tablenotemark{b} & \colhead{A($f_{\rm det}=0.5$)\tablenotemark{c}} & \colhead{Detection?\tablenotemark{d}}& \colhead{Flag\tablenotemark{e}} & \colhead{Imaging Refs}
}
\startdata
2MASSW J0030300-145033 & 4.49 & 0.021 & 0.69 & 0.029 &  &   &  5\\ 
2MASS J00501994-3322402 & 3.26 & 0.011 & 0.58 & 0.018 &  &   &  \\ 
SIMP J013656.57+093347.3 & 2.32 & 0.008 & 3.69 & 0.019 & D &   & 6 \\ 
SDSS J015141.69+124429.6 & 3.72 & 0.006 & 0.62 & 0.013 &  &   &  4\\ 
SDSS J020742.48+000056.2 & 3.97 & 0.046 & 0.58 & 0.045 &  &   &  4\\ 
2MASSI J0243137-245329 & 3.16 & 0.011 & 0.40 & 0.019 &  &   &  5,10\\ 
2MASS J02572581-3105523 & 3.93 & 0.006 & 0.84 & 0.012 &  &   &  \\ 
2MASS J03185403-3421292 & 4.28 & 0.012 & 0.49 & 0.016 &  &   &  17\\ 
2MASSI J0328426+230205 & 3.69 & 0.006 & 0.52 & 0.013 &  &   &  \\ 
2MASSI J0415195-093506 & 3.79 & 0.007 & 0.67 & 0.013 &  &   &  4\\ 
SDSSp J042348.57-041403.5 & 3.57 & 0.006 & 1.76 & 0.013 & U & B,Q & 3\\
2MASS J05103520-4208140 & 3.33 & 0.019 & 0.14 & 0.031 &  &   &  \\ 
2MASS J05160945-0445499 & 3.99 & 0.014 & 0.28 & 0.019 &  &   &  4\\ 
2MASS J05185995-2828372 & 3.79 & 0.014 & 0.73 & 0.020 &  &  B & 4\\
SDSSp J053951.99-005902.0 & 1.72 & 0.004 & 0.60 & 0.020 &  &   &  5\\ 
2MASS J05591914-1404488 & 3.52 & 0.004 & 2.15 & 0.013 & D &   & 2 \\
UGPS J072227.51-054031.2 & 1.74 & 0.040 & 0.35 & 0.086 &  &   &  \\ 
2MASS J07290002-3954043 & 3.13 & 0.018 & 0.29 & 0.030 &  &   &  \\ 
SDSS J074201.41+205520.5 & 3.72 & 0.004 & 0.89 & 0.012 &  &   &  10\\ 
SDSS J075840.33+324723.4 & 3.54 & 0.008 & 11.60 & 0.014 & D &   & 10 \\ 
DENIS J081730.0-615520 & 3.99 & 0.005 & 1.93 & 0.012 & D &   & \\
SDSS J083048.80+012831.1 & 3.75 & 0.022 & 0.39 & 0.035 &  &   &  \\ 
SDSS J093109.56+032732.5 & 4.43 & 0.036 & 0.47 & 0.041 &  &   &  \\ 
2MASS J09393548-2448279 & 5.92 & 0.015 & 0.61 & 0.016 &  &   & 6 \\ 
2MASS J09490860-1545485 & 4.48 & 0.012 & 0.67 & 0.016 &  &   &  \\ 
2MASS J10073369-4555147 & 3.56 & 0.011 & 0.73 & 0.017 &  &   &  \\ 
2MASSW J1036530-344138 & 4.52 & 0.011 & 0.56 & 0.015 &  &   &  6\\ 
SDSS J104829.21+091937.8 & 4.14 & 0.036 & 0.33 & 0.043 &  &   &  \\ 
SDSSp J111010.01+011613.1 & 5.72 & 0.020 & 0.57 & 0.023 &  &   &  4\\ 
2MASS J11145133-2618235 & 3.13 & 0.018 & 0.45 & 0.030 &  &   &  \\ 
2MASS J11220826-3512363 & 2.76 & 0.008 & 0.32 & 0.018 &  &   &  \\ 
2MASS J11263991-5003550 & 4.91 & 0.005 & 4.77 & 0.009 & D &   &  \\ 
SDSS J115553.86+055957.5 & 2.87 & 0.013 & 0.25 & 0.024 &  &   &  \\ 
SDSS J120747.17+024424.8 & 3.84 & 0.016 & 0.31 & 0.023 &  &   &  \\ 
2MASS J12095613-1004008 & 3.13 & 0.014 & 0.37 & 0.024 &  &   B\tablenotemark{f} & 7 \\ 
2MASS J12154432-3420591 & 3.42 & 0.017 & 0.47 & 0.028 &  &   &  \\ 
2MASSI J1217110-031113 & 3.18 & 0.017 & 0.48 & 0.028 &  &   &  2,4\\ 
2MASS J12314753+0847331 & 3.16 & 0.014 & 0.70 & 0.024 &  &   &  \\ 
SDSSp J125453.90-012247.4 & 3.10 & 0.021 & 0.19 & 0.038 &  &   &  4\\ 
SDSSp J132629.82-003831.5 & 2.60 & 0.022 & 0.44 & 0.047 &  &   &  \\ 
SDSS J140255.66+080055.2 & 3.19 & 0.046 & 0.23 & 0.055 &  &   &  \\ 
2MASS J14044941-3159329 & 4.02 & 0.015 & 0.32 & 0.021 &  & B & \\
SDSS J141624.08+134826.7 & 3.24 & 0.007 & 1.39 & 0.015 & U &  Q & \\
DENIS-P J142527.97-365023.4 & 2.98 & 0.005 & 1.34 & 0.015 & M &   &  13\\ 
2MASSW J1507476-162738 & 5.24 & 0.006 & 0.46 & 0.010 &  &   &  11,12\\ 
SDSS J151114.66+060742.9 & 2.92 & 0.014 & 0.51 & 0.024 &  &   &  \\ 
SDSS J152103.24+013142.7 & 2.91 & 0.028 & 0.37 & 0.055 &  &   &  \\ 
2MASSI J1546291-332511 & 3.79 & 0.008 & 0.49 & 0.014 &  &   & 2 \\ 
SDSSp J162414.37+002915.6 & 3.58 & 0.017 & 0.45 & 0.027 &  &   & 4 \\ 
SIMP J162918.41+033537.0  & 4.03 & 0.008 & 3.51 & 0.013 & D &   &  \\ 
SDSS J163359.23-064056.5 & 5.44 & 0.015 & 0.53 & 0.016 &  &   &  \\ 
DENIS-P J170548.38-051645.7 & 3.04 & 0.008 & 0.31 & 0.016 &  &   &  14,17\\ 
SDSS J175024.01+422237.8 & 2.63 & 0.011 & 1.38 & 0.023 & M &   &  \\ 
2MASS J17502484-0016151 & 2.63 & 0.005 & 0.99 & 0.016 &  &   &  \\ 
2MASS J18283572-4849046 & 4.13 & 0.006 & 2.30 & 0.012 & D &   &  \\ 
2MASS J19360187-5502322 & 3.73 & 0.007 & 1.09 & 0.014 & M &  B? & 12,17 \\ 
SDSS J205235.31-160929.8 & 3.69 & 0.011 & 0.69 & 0.017 &  &  B &  16\\ 
2MASS J21392676+0220226 & 2.50 & 0.012 & 5.23 & 0.025 & D &   &  9\\ 
2MASS J21513839-4853542 & 4.28 & 0.009 & 0.84 & 0.013 &  &   &  \\ 
2MASS J22282889-4310262 & 5.73 & 0.014 & 2.34 & 0.016 & D &   & 4 \\ 
2MASSI J2254188+312349 & 3.37 & 0.008 & 0.34 & 0.015 &  &   &  4\\ 
2MASS J23312378-4718274 & 4.09 & 0.018 & 0.58 & 0.025 &  &   & 
\enddata
\tablenotetext{a}{White noise estimate for the relative-flux light curve, adjusted to a 67\,s cadence (see section 3.3)}
\tablenotetext{b}{Statistic used to indentify detections, described in section 3.2 ($p>$99\% when $\beta > 1.4$, and $p>$96\% when $\beta > 1.0$)}
\tablenotetext{c}{Peak-to-peak amplitude for which we could have detected variability 50\% of the time.  Values have been marginalized over an assumed distribution of rotation periods for BDs (see section 3.3).  For an idea of the level of variability detectable in the span of each observation, please refer to $\sigma_{\rm a}$.}
\tablenotetext{d}{D: significant detection ($p>$99\%), M:marginal detection ($p>$96\%); U:undetermined---marginal variability is detected, but the light curve has been flagged for quality, making the significance of the result uncertain.}
\tablenotetext{e}{B: resolved as a binary by high angular resolution imaging; Q: flagged for quality due to bad pixels in the photometry aperture}
\tablenotetext{f}{Highly unequal mass binary; this target was kept in the statistical sample. See section \ref{sect:stats}}
\tablerefs{(1)\citet{andrei11}; (2)\citet{burgasser03}; (3)\citet{burgasser05_0423}; (4)\citet{burgasser06_tbin}; (5)\citet{gizis03}; (6)\citet{goldman08}; (7)\citet{liu10}; (8)\citet{looper08}; (9)\cite{radigan12}; (10)\citet{radigan13}; (13)\citet{reid01}; (14)\citet{reid06}; (15)\citet{reid08}; (16)\citet{stumpf11}; (17)\citet{pope13}
}
\end{deluxetable}

\end{document}